\documentclass[a4paper, 12 pt, openany, twoside]{book}
\usepackage{a4wide}
\usepackage{german}
\usepackage[latin1]{inputenc}
\usepackage{float}
\usepackage[intlimits]{amsmath}
\usepackage{amsfonts}
\usepackage{amsbsy}
\usepackage{amssymb}
\usepackage[OT1]{fontenc}
\usepackage{booktabs}
\usepackage{slashed}
\usepackage{graphics}
\usepackage{graphicx}
\usepackage{feynmf}
\usepackage{epsfig}
\usepackage{rotating}
\usepackage[german]{varioref}
\usepackage[small,hang]{caption}
\usepackage[width=16cm, left=2.0cm]{geometry}

\title{{Eine untere Massenschranke für Neutralinos \\ [.4cm] 
 aus Supernova-Kühlung}}
\author{{Ulrich Langenfeld}}
\date{im März 2002}

\begin{document}
\frontmatter
\pagenumbering{roman}

\begin{titlepage}		
\begin{center}
    \font\Giant=cmr17 scaled\magstep2
	{\Giant U\kern0.8mm N\kern0.8mm I\kern0.8mm V\kern0.8mm %
	E\kern0.8mm R\kern0.8mm S\kern0.8mm I\kern0.8mm %
	T\kern0.8mm %
	\setbox0=\hbox{A}\setbox1=\hbox{.}%
	\dimen0=\ht0 \advance \dimen0 by -\ht1%
	\makebox[0mm][l]%
	{\raisebox{\dimen0}{.\kern 0.3\wd0 .}}A\kern0.8mm
	T\kern5mm  
	B\kern0.8mm O\kern0.8mm N\kern0.8mm N\kern0.8mm
	} \\[8mm]
    {\Giant
    	P\kern0.8mm h\kern0.8mm y\kern0.8mm s\kern0.8mm i\kern0.8mm 
	k\kern0.8mm a\kern0.8mm l\kern0.8mm i\kern0.8mm s\kern0.8mm
	c\kern0.8mm h\kern0.8mm e\kern0.8mm s\kern5mm
	I\kern0.8mm n\kern0.8mm s\kern0.8mm t\kern0.8mm
	i\kern0.8mm t\kern0.8mm u\kern0.8mm t\kern0.8mm
	} \\[3.5cm]
{\Large \bf
{{\textsl{Eine untere Massenschranke für Neutralinos \\ [.4cm] 
 aus Supernova-Kühlung}}}
 
}\\[20mm]
\vspace{5mm}
{\Large 
		von			\\[5mm]
Ulrich Langenfeld}\\[20mm]
\end{center}
\parindent 0pt

\parbox[t]{160mm}{\begin{center}
\large \bf{Abstract}
\end{center}

In my diploma thesis I calculate a lower mass bound of the lightest neutralino assuming an additional
supernova cooling mechanism. I consider elektron-positron annihilation and nucleon-nucleon-bremsstrahlung
as cooling processes. These cooling processes must not be too effective, because they would destroy the neutrino signal
(SN 1987a). From this requirement I derive an upper bound for the emitted energy and as a consequence I get a lower
bound for the mass of the lightest neutralino.   }\\[10mm]

Dieser Forschungsbericht wurde als
Diplomarbeit  
von der Mathematisch - Naturwissenschaftlichen Fakult\"at der Universit\"at 
Bonn angenommen. \hspace*{\fill}	\\[10mm]
\end{titlepage}

\tableofcontents
\listoffigures
\listoftables

\mainmatter
\pagenumbering{arabic}

\newcommand{\ie}{\ensuremath{\text{i}}}
\newcommand{\eu}{\ensuremath{\text{e}}}
\newcommand{\cw}[1][{}]{\ensuremath{\cos^{#1} \theta_W}}
\newcommand{\sw}[1][{}]{\ensuremath{\sin^{#1} \theta_W}}
\newcommand{\tw}[1][{}]{\ensuremath{\tan^{#1} \theta_W}}
\newcommand{\ctw}[1][{}]{\ensuremath{\cot^{#1} \theta_W}}
\newcommand{\vv}{\ensuremath{\bar{v}}}
\newcommand{\uu}{\ensuremath{\bar{u}}}
\newcommand{\real}{\ensuremath{\mathfrak{Re}}}
\newcommand{\imag}{\ensuremath{\mathfrak{Im}}}
\newcommand{\mink}[2]{\ensuremath{(#1 \cdot #2)}}
\newcommand{\dif}{\ensuremath{\mathrm{d}}}
\newcommand{\diff}[2]{\ensuremath{\frac{\mathrm{d}#1}{\mathrm{d}#2}}}
\newcommand{\cv}[1][{}]{\ensuremath{\cos^{#1} \beta}}
\newcommand{\sv}[1][{}]{\ensuremath{\sin^{#1} \beta}}
\newcommand{\tv}[1][{}]{\ensuremath{\tan^{#1} \beta}}
\newcommand{\M}{\ensuremath{\mathcal{M}}}
\newcommand{\Tr}{\ensuremath{\mathrm{Tr}}}
\chapter*{Überblick}
\addcontentsline{toc}{chapter}{Überblick}

In der vorliegenden Arbeit habe ich eine untere Massenschranke für das leichteste Neutralino berechnet.
Bei einer Supernova-Explosion wird der größte der Teil der freiwerdenden Energie in Form von Neutrinos
abgestrahlt. Dieses Signal ist durch simulationen bestätigt worden. Bei der Supernova SN 1987 A 
sind diese Neutrinos zum ersten Mal nachgewiesen worden. Ihr Signal dauerte ungefähr $10 \,\text{s}$. Ein 
zusätzlicher Kühlmechanismus darf nicht zu effektiv sein. Er würde die Energie, die die Neutrinos wegtragen 
müssen, schon vorher abführen. Dann würde man ein kürzeres Neutrinosignal messen. 
Folglich läßt sich aus der Forderung, dieses Signal nicht zu
zerstören, eine obere Schranke für die maximale Stärke eines zusätzlichen Kühlmechanismus ableiten, nämlich Kühlung
durch die Emission von Neutralinos. Dies geschieht ohne eine Supernova zu simulieren.

Ausgangspunkt sind zwei Produktionsmechanismen für Neutralinos im Supernova-Core\footnote{Ich verwende ``Core'' 
für das aktive Sternzentrum und ``Kern'' als Kurzform von Atomkern.},
nämlich Vernichtung von Elektronen und
Positronen sowie Nukleon-\-Nukleon-\-Brems\-strah\-lung. Das Core wird als durchsichtig für die Neutralinos angenommen.
Diese Annahme ist für einen Selektronenmasse im Bereich von mindestens $1\,\mathrm{TeV}$ erfüllt. 
Für den ersten Mechanismus ist ausgehend vom totalen
Wirkungsquerschnitt die abgestrahlte Energie in Abhängigkeit der Masse des leichtesten Neutralinos berechnet worden.
Für die Temperatur und die Entartung im Core der Supernova sind Profile aus einer gegebenen Simulation gewählt worden.
Diese Simulation hat die Neutralinos nicht berücksichtigt.
Die Masse des  Selektrons ist zwischen $100 \,\mathrm{GeV}$ und $1000 \,\mathrm{GeV}$ liegend angenommen worden.
Die Abstrahldauer der Neutralinos ist im Bereich von $0.002 - 2 \,\mathrm{s}$ variiert worden.   
Um die Massenschranke zu erhalten, wurde die abzustrahlende Energie nach oben begrenzt.
Die Mischungsparameter des leichtesten Neutralinos haben bei kleinen Selektronenmassen keinen Einfluß auf die
abgestrahlte Energie. Das leichtetste Neutralino besteht zu mindestens $98 \%$ aus Bino, der Rest verteilt sich 
hauptsächlich auf die beiden Higgsinos. 
Mit diesem Verfahren erhält man untere Schranken im Bereich von $50 \,\mathrm{MeV}$ (Selektronmasse groß, 
Abstrahldauer klein) bis $350 \,\mathrm{MeV}$ (Selektronmasse klein, Abstrahldauer groß).

Im zweiten Mechanismus ist die Emissivität direkt mit dem Raffelt-Kriterium - einer oberen Schranke für die Emissivität,
die durch Simulationen überprüft worden ist - verglichen worden.
Da die Strahlung als ``{\it soft}'' genähert wird, erhält man einen Ausdruck
für die Emissivität in Abhängigkeit der Neutralinomasse und der Coretemperatur, der für alle Teilchen verwendbar ist, 
die an den axialen Strom koppeln.  
Hier ist die Squarkmasse wie oben beim Selektron eine unbekannte Größe, für die ein ähnlicher Massenbereich wie für die 
Selektronenmasse zugrunde gelegt wird. Damit erhalte ich Massenschranken in derselben Größenordnung.

Ein leichtes Neutralino im Bereich von $34 \,\mathrm{MeV}$ kann somit nicht durch Supernova-Kühlung ausgeschlossen 
werden, sofern die Sfermionmassen hoch genug sind.

\chapter{Einführung}
\section{Einführung in die Themenstellung}

Das Ziel meiner Arbeit ist die Berechnung einer unteren Massenschranke für das als leichtestes supersymmetrisches
Teilchen angenommene Neutralino $\chi^0_1$ mittels Supernova-Kühlung. Dabei darf die Supernova nicht zu viel Energie in Form
von Neutralinos abstrahlen, damit das beobachtete zeitverzögerte Neutrinosignal nicht zerstört wird. 
Als 1987 der Stern Sanduleak 69202 in der Großen Magellanschen Wolke explodierte, konnten zum 
ersten Mal die bei einer Supernova entstehenden Neutrinos in den 
Wasser-Tscherenkow-Detektoren Irvine-Michigan-Brookhaven (IMB) und Kamiokande II beobachtet werden.
Diese waren eigentlich zur Messung des Protonzerfalls gebaut worden.
Neutrinos tragen wesentlich zur Kühlung
einer Supernova-Explosion bei. Sie führen hundert bis tausend Mal so viel Energie wie Photonen ab.
Ihr Signal dauerte etwa $10\,\text{sec}$. Jeder zusätzliche Kühlmechanismus darf nicht zu effektiv sein. Denn
sonst würde dieser Kühlmechanismus das Core so weit kühlen, daß die Neutrinos nicht mehr gefangen wären
und das Neutrinosignal kürzer wäre. Das steht jedoch im Widerspruch zu den Beobachtungen.  
Im folgenden nehmen wir an,
daß es einen weiteren Kühlmechanismus gibt, der Energie aus dem Super\-nova-Core nach außen abführen kann, nämlich 
die Produktion von Neutralinos.  
Neutralinos können auf folgende Arten erzeugt werden:
\noindent  
\begin{itemize}
\item Elektron-Positron-Annihilation in ein $Z^0$-Boson: $e^+ e^- \rightarrow Z^0 \rightarrow \chi^0_1 \chi^0_1$\\
      Elektron-Positron-Streuung: $e^+ e^- \rightarrow  \chi^0_1 \chi^0_1$ unter Austausch eines Selektrons	
\item Nukleon-Nukleon-Bremsstrahlung: $NN \rightarrow NN \chi^0_1 \chi^0_1$.\\
      Bei diesem Prozeß wird entweder ein $Z^0$ abgestrahlt, das in zwei Neutralinos zerfällt, oder ein Quark 
      geht kurzzeitig in ein Squark über und strahlt dabei zwei Neutralinos ab.
\end{itemize}
Für diese Mechanismen berechnen wir die Emissivität (Energieabgabe pro Zeit und Volumen oder Energieabgabe pro Zeit 
und Masse) in Abhängigkeit von der Neutralinomasse.
Im ersten Fall liefern
Volumen- und Zeitintegration die absolute freigesetzte Energie in Abhängigkeit von der 
Neutralinomasse. Diese Berechnungen erfolgen numerisch. 
Man kann aus der Anfangs- und der Endgröße des Cores abschätzen, wieviel Energie von einer Supernova maximal abgestrahlt 
wird. Dies sind etwa $3\cdot 10^{53}$ erg. Den Großteil davon führen Neutrinos ab.
Im Durchschnitt führt jede der sechs Neutrinoarten demzufolge $5\cdot 10^{52}\,\text{erg}$ ab.
Die Annahme, daß $10^{52}$ erg in Form von Neutralinos abgestrahlt werden, sollte nicht zu einer Zerstörung des Neutrinosignals
führen. 
In die Berechnungen gehen eine Reihe von Parametern ein:
Temperatur, chemisches Potential und Dichte des Sterncores, die Mischungsparameter des leichtesten Neutralinos,
die Selektronmasse und die Dauer der Energieabstrahlung. Für diese Parameter wurden geeignete Modelle
oder Werte zugrunde gelegt.

Im Falle der Nukleon-Bremsstrahlung ist es nicht notwendig, die Emissivität $\epsilon$ über das Volumen und die Zeit 
zu integrieren. Hier kann die Emissivität sofort mit einem
Kriterium von Raffelt verglichen werden: $\epsilon \le 10^{19}\,\mathrm{erg}/(\mathrm{g}\cdot \mathrm{s})$ 
(aus \cite{Stars:96}). Dieses Kriterium ist durch Simulationen bestätigt worden.
Das Endergebnis ist für andere Theorien verwendbar.

\section{Was sind Neutralinos?}
\subsection{Anschauliche Beschreibung}
Zuerst soll kurz erläutert werden, was Neutralinos überhaupt sind.
Supersymmterie ist eine Symmetrie zwischen Fermionen und Bosonen. Damit kann das Hierarchieproblem in der Teilchenphysik
gelöst werden(s. \cite{primer:97},\cite{susylow:93}). In der Supersymmetrie (kurz SUSY) wird 
jedem Boson ein Fermion als Superpartner zugeordnet und umgekehrt. Dadurch wird die Anzahl der Teilchen verdoppelt.
Die Eichbosonen  der Elektroschwachen Kraft sind nach spontaner Symmetriebrechung durch ein
komplexes Higgs-$SU(2)$-Dubletts die geladenen $W$-Bosonen, das neutrale $Z^0$-Boson
und das ebenfalls neutrale Photon (s. \cite{element:84}).
Von den vier Freiheitsgraden dieses Higgsdubletts werden drei durch den Higgsmechanismus ``aufgegessen''.
Ein Higgs-Teilchen bleibt übrig.
Zu den vier Austauschteilchen gibt es im ``Minimalen Supersymmetrischen Standardmodell'' (MSSM) vier fermionische 
Gauginos als Superpartner: die geladenen Winos und die neutralen Teilchen Zino und Photino (s. \cite{super:85}). 
In der Supersymmetrie muß man zwei komplexe Higgs-$SU(2)$-Dubletts einführen, damit u. a. sowohl up- als auch 
down-Quarks Masse erhalten (\cite{super:85},\cite{higgs:86}). Es bleiben am Ende nur fünf reelle Felder übrig, 
da drei Felder ``aufgegessen'' werden, um den schweren Vektorbosonen Masse zu geben.
Von diesen fünf Feldern sind zwei geladen und drei neutral. Die fermionischen Superpartner dieser Higgs-Felder heißen 
(nacd der Brechung von $SU(2)_L \times U(1)_Y \rightarrow U(1)_{\mathrm{em}}$) Higgsinos. Die zwei CP-geraden neutralen 
Higgsinos mischen mit den neutralen Gauginos
Photino und Zino sowie die geladenen Higgsinos mit den Winos. Die Eigenzustände der Schwachen
Wechselwirkung sind also nicht die propagierenden Masseneigenzustände. Die neutralen Masseneigenzustände werden 
Neutralinos genannt, die geladenen Charginos.
Um die Masseneigenzustände zu erhalten, muß man die Massenmatrix diagonalisieren.

Des weiteren gehe ich von erhaltener $R$-Parität aus (weswegen die Neutrinos nicht mit den Neutralinos mischen)
und nehme an, daß bei hohen Energien keine GUT-Symmetrie auftritt, also die 
$SU(3)_c \times SU(2) \times U(1)_Y$-Theorie nicht zur $SU(5)$-Theorie vereinigt wird.  

\subsection{Mathematische Beschreibung}

\subsubsection{Neutralinos}

Für die Einführung in die SUSY-Theorie verweise ich auf die Literatur:
Der Superfeld-Formalismus, mit dem man supersymmetrische Lagrangdichten konstruieren kann, wird z. B. in
\cite{Bailin:94} dargestellt.
Die Konstruktion supersymmetrischer Lagrangefunktionen für eine $SU(2)_L \times U(1)_Y$-Theorie wird in \cite{super:85}
beschrieben, der Higgs-Sektor wird in \cite{higgs:86} erläutert.   

Ich fasse hier die Resultate über die Neutralino-Massenmatrix kurz zusammen:
\begin{itemize}
\item{Massenterme:}
In der Lagrangedichte des MSSM findet man folgende Massenterme für die Superpartner der neutralen Eichbosonen
und neutralen Higgsfelder
unter Berücksichtigung des Symmetriebruchs von $SU(2) \times U(1)_Y$ zu $U(1)_{\text{em}}$
 und der {\textit{soft susy-breaking}} Terme (siehe dazu \cite{super:85}) (in Weylspinor-Darstellung):
\begin{eqnarray}
\mathcal{L}_M &\! =\! &    \frac{1}{2} \ie \sqrt{(g^2 +g'^2)} \lambda_Z (v_1 \psi_{H_{1}}^1 - v_2 \psi_{H_{2}}^2 )
	            + \frac{1}{2}(M \cw[2] + M' \sw[2])\lambda_Z \lambda_Z \notag \\  
		  & + & (M - M') \sw \cw \lambda_Z \lambda_{\gamma}
		    + \frac{1}{2} (M' \cw[2] + M \sw[2])\lambda_{\gamma}\lambda_{\gamma}
		    + \mu \psi_{H_{1}}^1  \psi_{H_{2}}^2  \notag \\
		 & + & \mathrm{h. c.} \enspace . \label{eq:lagrange}
\end{eqnarray}
Hier bedeuten $g, \enspace g'$: Kopplungskonstanten der $SU(2)\times U(1)_Y$-Theorie, $v_1,\enspace v_2$ sind die 
Vakuumserwartungswerte 
der zwei neutralen $CP$-geraden Higgsfelder, $\theta_W$ ist der elektroschwache Mischungswinkel, $M, \, M', \, \mu$ sind 
\mbox{Wino-} und Binomasse sowie der Higgsmischungsparameter aus dem Superpotential und  
$\lambda_\gamma,\, \lambda_Z, \, \psi_{H_{1}}^1,\enspace \psi_{H_{2}}^2$ bezeichnen \mbox{Photino-,} \mbox{Zino-} 
und die beiden Higgsinofelder.
\item{Basis:}
Anstatt der Photinobasis 
 $ \{  \lambda_{\gamma}, \lambda_{Z}, \psi_{H_{1}}^1,\psi_{H_{2}}^2 \} $ wird häufig auch die Bino-Basis
$\{\lambda', \lambda^3, \psi_{H_{1}}^1,\psi_{H_{2}}^2 \}$ verwendet. Dann sieht Gleichung (\ref{eq:lagrange}) so aus:
\begin{eqnarray}
\mathcal{L}_M & = &  \frac{1}{2}\ie g \lambda^3 (v_1 \psi_{H_{1}}^1 - v_2 \psi_{H_{2}}^2 ) 
		-  \frac{1}{2}\ie g' \lambda'(v_1 \psi_{H_{1}}^1 - v_2 \psi_{H_{2}}^2) \notag\\
	& &  + \frac{1}{2} M  \lambda^3  \lambda^3 + \frac{1}{2} M  \lambda'  \lambda' + 
			\mu  \psi_{H_{1}}^1\psi_{H_{2}}^2 
		 +  \mathrm{h. c.} \enspace . \label{eq:binolagrange}
\end{eqnarray}
\item{Mischungsmatrix:}
Schreibt man die Elemente der Bino-Basis als $\psi^0_j$, $j = 1 \ldots 4$, so kann man die letzte Gleichung
in die Form
\begin{equation}
\mathcal{L}_M  =  \frac{1}{2} (\psi^0_i)Y_{ij} \psi^0_j \label{eq:mlag1}
\end{equation}
bringen (über doppelt auftretende Indizes ist zu summieren) mit der (im allgemeinen komplexwertigen) 
Mischungsmatrix $Y$:
\begin{equation}
Y = 
\begin{pmatrix}
M'          & 0           & -m_Z \cv \sw &  m_Z \sv \sw \\
0           & M           &  m_Z \cv \cw & -m_Z \sv \cw \\
-m_Z\cv \sw & m_Z \cv \cw &  0           & -\mu         \\
m_Z \sv \sw & -m_Z \sv \cw& -\mu         &  0           \\
\end{pmatrix} \label{eq:neutralinomischung},
\end{equation} 
$\tv   =  \frac{v_2}{v_1}$ ist das Verhältnis der Vakuumserwartungswerte $v_1, v_2$ der Higgsfelder.

$M,\enspace M',\enspace \mu,\enspace \tv$ sind unabhängige Parameter im MSSM.
\item{Diagonalisierung der Mischungsmatrix:}
Die Mischungsmatrix (\ref{eq:neutralinomischung}) ist nicht diagonal. Um die propagierenden Masseneigenzustände $\chi^0_i$
zu erhalten, muß man die Mischungsmatrix diagonalisieren.
Dabei können die Eigenwerte  negativ werden. Deshalb nimmt man eine unitäre Diagonalisierungsmatrix $N$,
wobei man die freien Phasen so wählt, daß die Eigenwerte der Diagonalmatrix $N_D$ positiv sind (siehe \cite{neutrino:90}). Dann 
ist eine Interpretation der 
Eigenwerte als Massen möglich ist (\cite{super:85}). Die Vorzeichen der Massen können als $CP$-Eigenwerte gedeutet werden (\cite{cp:84}).
Es gilt:
\begin{eqnarray}
Y_{D} & = & N^* Y N^{-1}, \\
\chi^0_i & = & N_{ij}\psi^0_j, \enspace i,j = 1 \ldots 4 \quad \text{Neutralino-Masseneigenzustände} \enspace . 
\end{eqnarray}
Der Massenterm \ref{eq:mlag1} läßt sich dann als $ - \frac{1}{2}\sum_i M_i\overline{\chi_i}^0 \chi^0_i$schreiben, wobei $M_i$ 
die Eigenwerte der 
Matrix (\ref{eq:neutralinomischung}) sind. Man kann dies auch in vierkomponentige Majorana-Spinoren
umschreiben:
\begin{eqnarray}
\mathcal{L}_{M} = -\frac{1}{2}\sum_iM_i \overline{\tilde{\chi}}^0_i \tilde{\chi}^0_i; \quad
\tilde{\chi}^0_i = \begin{pmatrix}\chi^0_i \\[1mm] \overline{\chi}^0_i \end{pmatrix} \quad .
\end{eqnarray}
Die Eigenwerte der Massenmatrix bestimmt man am besten numerisch. Es gibt aber auch analytische Lösungen, siehe dazu 
\cite{analytisch:92}.
\end{itemize}

\subsubsection{Charginos}
Die Mischungsmatrix der Charginos hängt von $M,\enspace \mu$ und $\beta$ ab. 
Weil die Gauginos elektromagnetisch wechselwirken können und sie deshalb im Prinzip
einfacher zu entdecken sind als die elektrisch neutralen Neutralinos, müssen die Parameter der Mischungsmatrix
so gewählt werden, daß die Charginos schwerer als 120 GeV sind, der momentanen unteren Massenschranke, die durch die 
Beschleuniger gesetzt ist. 
Hier eine Zusammenfassung der Resultate für die Charginos:
\begin{itemize}
\item Der Chargino-Massenterm (Weyl-Spinor-Schreibweise) ist gegeben durch: 
\begin{eqnarray}
\mathcal{L}_M = \frac{\ie g}{\sqrt{2}}\left( v_1 \lambda^+ \psi^2_{H_{1}}+ v_2 \lambda^- \psi^1_{H_{2}} \right) 
              + M \lambda^+ \lambda^- -\mu\psi^2_{H_{1}} \psi^1_{H_{2}} + \text{h.c.} \label{eq:mchar}
\end{eqnarray}

\item Gl. (\ref{eq:mchar}) läßt sich durch eine Basistransformation umformen:
\begin{eqnarray}
\psi_j^+ & = & (-\ie \lambda^+,\psi^1_{H_{2}}), \enspace \psi_j^- = (-\ie \lambda^-,\psi^1_{H_{1}}) \\[2mm]
\mathcal{L}_M & = & - \frac{1}{2}\begin{pmatrix}\psi^+ & \psi^- \end{pmatrix}
                          \begin{pmatrix} 0 & X^T \\ X & 0 \end{pmatrix}
                          \begin{pmatrix}\psi^+ \\ \psi^- \end{pmatrix} + \text{h. c.},
\end{eqnarray}
wobei die Massenmatrix $X$ gegeben ist durch:
\begin{eqnarray}
X & = & \begin{pmatrix}M & m_w \sqrt{2} \sv \\ m_w\sqrt{2} \cv & \mu \end{pmatrix}; \\[2mm]
&&  m^2_w = \frac{1}{4}g^2(v_1^2 + v_2^2); \quad \tv = \frac{v_2}{v_1}
\end{eqnarray}

\item Die Masseneigenzustände erhält man durch Diagonalisieren der Matrix 
$\begin{pmatrix} 0 & X^T \\ X & 0 \end{pmatrix}$:  
\begin{eqnarray}
\chi_i^+ & = & V_{ij}\psi_j^+ , \quad\chi_i^- = U_{ij}\psi_j^- , \quad   
\mathcal{L}_M  = - (\chi_i (X_D)_{ij}\chi_j^+ + \text{h. c.})\\[2mm]
     && \text{mit}\enspace U,V \enspace \text{so, daß}\enspace U^* X V^{-1} = X_D \quad\text{diagonal}\notag
\end{eqnarray}
\item Aus den Weylspinoren lassen sich Diracspinoren $\tilde{\chi}_1$, $\tilde{\chi_2}$ konstruieren: 
\begin{eqnarray}
&& \tilde{\chi}_1 = \begin{pmatrix}\chi^+_1 \\[2mm]\overline{\chi}_1^{~-} \end{pmatrix}, \enspace
   \tilde{\chi}_2 = \begin{pmatrix}\chi^+_2 \\[2mm]\overline{\chi}_2^{~-} \end{pmatrix},\quad 
\mathcal{L}_M  =  -(M_+\overline{\tilde{\chi}}_1 \tilde{\chi}_1 + M_-\overline{\tilde{\chi}}_2\tilde{\chi}_2) \quad.
\end{eqnarray}
\paragraph{Anmerkung:} Die oberen Indizes $+,-$ an den Spinoren kennzeichnen in diesem Abschnitt die Ladung und 
                       nicht die hermitesche Konjugation.
\end{itemize}

\subsubsection{Anmerkung zu Majorana-Massen:}

Massenterme für Diracspinoren haben die Form $m \overline{\psi} \psi = m(\overline{\psi}_L \psi_R  + \text{h. c.})$, 
($\psi$: Diracspinor, $\psi_{L,R} = P_{L,R}\psi,\enspace P_{L,R} = \frac{1}{2}\left(1\mp \gamma_5 \right)$). 
$\psi_L$ und $\psi_R$ sind unabhängig voneinander und transformieren sich unter 
Lorentztransformationen verschieden. Solche Dirac-Massenterme mischen links- und rechtshändige Felder.

Majoranaspinoren $\Psi$ haben die definierende Eigenschaft $\Psi^c = C\overline{\Psi}^T = \Psi$, $C$: 
Ladungskonjugationsoperator.
Damit gilt $\Psi_L^T = \overline{\Psi}_R$ und Massenterme haben die Form $\frac{m}{2}(\psi_L^T \psi_L + \text{h. c.})$.
Es werden Komponenten gleicher Händigkeit gemischt. Dadurch ergibt sich ein anderes Transformationverhalten für 
Majoranaspinoren, da allgemeine Massenterme die Gestalt $\psi^T M \psi + \text{h. c.}$ haben, wobei 
$\psi = (\psi_1 \ldots \psi_n)$
Weyl-Spinoren sind. Transformiert man $\chi_i = N_{ij} \psi_j$, wobei $N$ unitäre Matrix ist, dann gilt 
$(N^{-1} \chi)^T M N^{-1}\chi = \chi^T N^* M N^{-1} \chi$, was das auf den ersten Blick ungewöhnliche Transformationverhalten
der Neutralinomassenmatrix erklärt. Denn um eine hermitesche Matrix $A$ mittels einer unitären Transformation $U$ 
auf Diagonalgestalt $A_D$ zu bringen, erwartet man das Transformationsgesetz $A_D = U A U^+$.

\chapter{Sternentwicklung}
\section{Bildung einer Gaswolke}
In den folgenden Abschnitten werde ich die Entwicklung von Sternen aus einer Wasserstoffwolke kurz 
dargestellen, um das 
mögliche Supernova-Ereignis in der Entwicklung eines Sterns verstehen zu können. Dabei werden
die thermodynamischen Zustandsgrößen zueinander in Beziehung gesetzt und erklärt, wie sie 
die Sternentwicklung steuern.  
Ich stütze mich auf \cite{stern:82}.

\subsection{Gravitation gegen kinetische Energie}
Um den Kollaps einer  Wasserstoffwolke geringer Dichte zu verstehen, betrachte man die
Energieverhältnisse in einer solchen Wolke:
\begin{eqnarray}
E   =  E_{\text{kin}} + E_{\text{pot}}
    = \frac{1}{2} M \overline{v^2} - \frac{3}{5} \frac{GM^2}{R}
    =   \frac{2}{3} \pi \rho \overline{v^2} R^3
        - \frac{16}{15} G \pi^2 \rho^2 R^5 
\end{eqnarray}
($M,\enspace R,\enspace \rho$: Masse, Radius und Dichte der Gaswolke, $\overline{v^2}$: mittlere quadratische 
Geschwindigkeit der Wasserstoffatome, $G$: Gravitationskonstante).
 
In einem gebundenem System muß die Gesamtenergie  $E$ negativ sein, das heißt, die
potentielle Energie muß betragsmäßig größer als die kinetische Energie sein. Da die
potentielle Energie stärker mit dem Radius als die kinetische wächst, kann die Wolke 
einen gebundenen Zustand erreichen. Das kann auf zwei Wegen 
geschehen: durch Kompression und durch Akkretion. 
Kompression bedeutet die Erzeugung von Dichtewellen, die durch die Wolke laufen. Diese werden
zum Beispiel durch Supernova-Explosionen oder durch vorbeiziehende Sterne erzeugt.

Akkretion heißt, daß die Wolke Materie aus ihrer Umgebung einsammelt und so sich ihr Durchmesser vergrößert. Dies 
geschieht in kosmologischen Zeiträumen, das heißt, der Vorgang benötigt etwa
eine Milliarde Jahre.
Den Radius dieser kollabierenden Wolke kann man leicht abschätzen, indem man $E$ zu Null setzt. 
Dann folgt für den kritischen Radius $R_{\text{k}}= \frac{6}{5} \frac{GM}{\overline{v^2}}$. 

Der Kollaps wird durch Gegenkräfte angehalten, die im folgenden diskutiert werden. 

\subsection{Quasihydrodynamisches Gleichgewicht}
Wir nehmen eine sphärische Massenverteilung an und betrachten eine Schale
der infinitesimalen\footnote{$\Delta x$ bezeichnet die infinitesimale Größe zu $x$}
 Dicke $\Delta r$ vom Radius $r$. Deren Volumen ist $4 \pi r^2_2 \Delta r$,
ihre Masse $\Delta M_{\text{S}} = 4 \pi r^2  \rho \Delta r$. Auf diese Schale wirkt die Kraft 
$\Delta F_S =  -\frac{GM_\text{S} \Delta M_\text{S}}{r_2^2}$. 
Die Gesamtkraft lautet demnach:
\begin{eqnarray}
\Delta F_{\text{ges}} =  4 \pi r^2 \Delta P - \frac{GM_r \Delta M_r}{r^2}
                    =  a_{\text{ges}} \Delta M_r \quad .
\end{eqnarray} 
Daraus ergibt sich die Gesamtbeschleunigung $a_\text{ges}$ zu
\begin{eqnarray}
 a_{\text{ges}} = \frac{1}{\rho} \frac{\Delta P}{\Delta r} - \frac{GM}{r^2} \label{eq:be} 
\quad \text{wobei} \enspace \Delta P = P_2 - P_1 \enspace .
\end{eqnarray} 
Hier können wir jetzt einige Fälle betrachten:
\begin{enumerate}
\item $a_{\text{ges}} = 0$: Gleichgewicht
\item $a_{\text{des}} \simeq 0$: Kleine Beschleunigung, sog. quasihydrostatisches Gleichgewicht QHE
                              (quasihydrostatic equilibrium)
\item $a_{\text{ges}} > 0$: expandierende Wolke
\item $a_{\text{ges}} < 0$: 
     \begin{enumerate}
      \item $\frac{\Delta P}{\Delta r} > 0$: nicht physikalisch
      \item $\left| \frac{1}{\rho} \frac{\Delta P}{\Delta r} \right| < 
            \left|\frac{GM}{r^2}\right|$: schlimmstenfalls freier Fall, Kollaps
     \end{enumerate}            
\end{enumerate}
Ein Stern (Materieansammlung) ist immer bestrebt, ins QHE zu gelangen, insbesondere ist eine Kontraktionsphase
dann beendet, wenn der Stern sich wieder im QHE befindet.
\subsection{Die Gasgleichung}
Um den Tatsachen Rechnung zu tragen, daß ein Stern erstens nicht nur aus einer Teilchensorte besteht,
zweitens die Teilchen manchmal ionisiert sind und daß drittens das Volumen keine dem Problem angemessene Größe ist,
wird die Gasgleichung $PV = NkT$ in eine passende Form gebracht.
Es seien $x,y,z$ die Massenanteile H, He und Metall ($=$ alles, was schwerer als He ist), also
$x = M_{\text{H}}/ M, \enspace M_{\text{H}}$: Gesamtmasse des Wasserstoffs usw. Dann kann man die Anzahl der 
Teilchen ausrechnen (nicht ionisiert):
\begin{eqnarray}
N_{\text{H}}  & = & \frac{M_{\text{H}}}{m_{\text{H}}} = \frac{x M}{m_{\text{H}}} \\
N_{\text{He}} & = & \frac{M_{\text{He}}}{m_{\text{He}}} = \frac{M_{\text{He}}}{4 m_{\text{H}}} =
                    \frac{y M}{4 m_{\text{H}}}\\
N_{\text{Me}} & = & \frac{M_{\text{Me}}}{m_{\text{Me}}} = \frac{M_{\text{Me}}}{A m_{\text{H}}} =
                    \frac{z M}{m_{A \text{H}}} \\
N_{\text{tot}}& = & \Bigl(\underbrace{x + \frac{1}{4}y + \frac{1}{A}z}_{\frac{1}{w}}\Bigr)
                    \frac{M}{m_{\text{H}}} \quad .
\end{eqnarray} 
Im Falle vollständiger Ionisierung liefert ein Atom mit $A$ Nukleonen $Z$ Elektronen, damit haben wir $Z+1$ 
Teilchen. Daher müssen die vorhergehenden Gleichungen um entsprechende Vorfaktoren verändert
werden:
\begin{eqnarray}
N_{\text{H}}  & = &  2 x \frac{M}{m_{\text{H}}},\\
N_{\text{He}} & = & \frac{3}{4} y \frac{M}{m_{\text{H}}},\\
N_{\text{Me}} & = & \frac{Z+1}{A} z \frac{M}{m_{\text{H}}} \simeq  \frac{1}{2} z \frac{M}{m_{\text{H}}}\\
N_{\text{tot}}& = & (2x + \frac{3}{4}y + \frac{1}{2}z) \frac{M}{m_{\text{H}}} 
                    = \left( \frac{1}{w}\right)_{\text{ion}}\frac{M}{m_{\text{H}}} \quad . 
\end{eqnarray} 
Jetzt kann die Gasgleichung umgeschrieben werden:
\begin{eqnarray}
PV =  \frac{1}{w} \frac{M}{m_{\text{H}}} k T, \quad \text{mit} \quad \rho = \frac{M}{V} \Rightarrow \enspace
  P = \frac{\rho k T}{w m_{\text{H}}} \label{eq:gas}
\end{eqnarray} 
Aus dieser Gleichung kann man schließen, daß ein Druckgradient drei Ursachen haben kann:
\begin{alignat}{5}
\frac{\Delta P}{\Delta r} & \propto & \frac{\Delta T}{\Delta r} & : \quad &
         \text{Temperaturgradient} \label{eq:tgrad} \\
\frac{\Delta P}{\Delta r} & \propto & \frac{\Delta \rho}{\Delta r} & : \quad & 
           \text{Dichtegradient}\label{eq:dgrad} \\
\frac{\Delta P}{\Delta r} & \propto & \quad \frac{\Delta w^{-1}}{\Delta r} & : \quad &
           \text{Teilchengradient}\label{eq:ngrad}
\end{alignat}
Später werden wir sehen, wie all diese Gradienten dazu beitragen, daß ein Stern nicht kollabiert.

\section{Thermonukleare Prozesse}

\subsection{Eine weitere Energiequelle: Kernfusion}
Sterne wie die Sonne beziehen ihre Energie aus der Verschmelzung von Wasserstoff zu Helium
oder ähnlichen Prozessen. Aus dem Verlauf der Kurve der mittleren Bindungsenergie pro Nukleon 
in Abhängigkeit von der Massenzahl kann man folgern, daß man durch Fusion
von Nukliden, die leichter als die eisenartigen Metalle sind, Energie gewinnen kann.
Für den Nettoprozeß
\begin{equation}
 6p \longrightarrow {}^4\text{He} + 2p +2e^+ + \overline{\nu} + E
\end{equation}
sind Temperaturen von $10^7 \enspace \text{K}$ \footnote{Die Geschwindigkeit der Protonen unterliegt einer Maxwellverteilung mit einer 
mittleren Geschwindigkeit von $10^7 \enspace \text{K}$. Eine solche energie würde allein nicht ausreichen, um die Coulomb-Abstoßung zu 
überwinden, aber die Anzahl der Teilchen im Schwanz der Verteilung mit einer entsprechenden Geschwindigkeit ist groß genug, um die Fusion 
aufrecht zu erhalten.}nötig. Gravitation und Kernfusion sind jetzt zwei 
gegeneinander wirkende Mechanismen: Durch die Fusion erhöht sich die Sterntemperatur im Core und 
infolge dessen auch der Druck. Dieser Druck wirkt der Gravitation entgegen.

\subsection{Entwicklung zum Roten Riesen}

\subsubsection{Prozesse im Inneren eines Sterns} 
Der Stern besteht aus zwei Teilen: einem Core, in dem Wasserstoff zu Helium fusioniert
wird, und einer äußeren, ``kälteren'' Hülle. Im Core reichert sich mit der Zeit immer
mehr Helium an, dabei wird die Teilchenanzahl kleiner, nach der Gasgleichung nimmt damit auch der
Druck ab, und der Kern zieht sich zusammen.  

An der Grenzfläche stimmen Druck und Temperatur von Hülle und Core überein, aus der Gasgleichung 
folgt dann $\frac{\rho_1}{w_1} = \frac{\rho_2}{w_2}$. Da $w_1$ mit der Zeit wächst, nimmt
die Dichte des Core zu. Seine Masse bleibt aber konstant (den Massendefekt
kann man vernachlässigen), so daß das Volumen des Sterns schrumpft. Bei der Kontraktion des Cores 
nimmt jedoch die Gravitationsenergie
weiter ab, die freigesetzte Gravitationsenergie erhöht die  Gesamtleuchtkraft des Sterns, und der Temperaturgradient im
Reststern wird größer. Aus Gl. (\ref{eq:be}) und der Gasgleichung folgt, daß die Hülle
sich ausdehnen muß: Wir haben ein kontrahierendes Core und eine expandierende Hülle.
Der Stern kann sich dabei nicht aufheizen, da die Energie zur Expansion verwendet wird.

Es gibt aber noch einen anderen Prozeß. An der Oberfläche des Sterns wächst mit der Expansion die potentielle
Energie eines Teilchen. Weil die Energie aber konstant ist, muß seine kinetische Energie abnehmen,
infolgedessen auch seine Oberflächentemperatur. 
Währenddessen nimmt im Core die Anzahl der Protonen ab. 
Da sich die Emissivität wie $\epsilon_{pp} = x^2 T^4$   
verhält, nimmt die Fusionsrate ab ($x$ wird kleiner) damit einher geht eine verminderte Kerntemperatur und 
Leuchtkraft. Dies wiederum führt zu einer Kontraktion des Sterns, wodurch sich ein neues 
Gleichgewicht einstellen kann. 
Es gibt also zwei gegeneinander wirkende Prozesse, welcher davon die Oberhand gewinnen kann,
hängt von der Masse des Sterns ab. Die weitere Entwicklung des Sterns hängt von seiner Masse ab.

\subsubsection{Alterung des Sterns}
Die Alterung beginnt mit der Bildung eines festen homogenen Heliumcores. Die Temperatur ist für ein 
Zünden des nächsten möglichen Prozesses, nämlich $3 \alpha \rightarrow {}^{12}\text{C}$, zu niedrig
($10^8$ K\footnote{siehe eben} sind nötig, $\alpha = {}^{4}\text{He}^{2+}$). Das Core ist somit tot, seine Leuchtkraft Null, 
man sagt, daß das Core
jetzt isotherm sei (da $\frac{\text{d} T}{\text{d} r} = 0$). Der Stern würde gerne im QHE bleiben,
im Core gibt es keinen Temperaturgradienten, also auch keinen daraus resultierenden Druckgradienten. Aber der Stern
kann einen Dichtegradienten entwickeln:
\begin{equation}
 \frac{\text{d} P}{\text{d} r} = 
\frac{\text{d} \rho}{\text{d} r}\frac{k T}{w m_{\text{H}}} = -\rho \frac{G M}{r^2} \quad .
\end{equation}
Dies geschieht so: Das isotherme Core kann die Hülle nicht länger tragen, der Stern kontrahiert,
dabei wird Gravitationsenergie frei, und es besteht die Möglichkeit, daß dadurch eine Schale um das Core
herum eine Temperatur zum Fusionieren von Wasserstoff erreicht.

Diese Schale sorgt jetzt für den Dichtegradienten und die Entwicklung zum Roten Riesen:
Sollte der Radius der Schale schrumpfen, bewegt sie sich in Richtung größerer Dichte und 
Temperatur. Da die Emissivität proportional mit der Dichte wächst, wird die Energieproduktion
so lange größer, bis Energieabgabe ($\propto L$) und Energieproduktion ($\propto \epsilon_{pp}$)
sich ausgleichen. Bei der Gleichgewichtstemperatur ist der Druck nach außen größer, so daß die
nach innen gerichtete Kraft kompensiert werden kann. Eine Störung dieser Gleichgewichtslage führt 
zu einer gedämpften Schwingung um diese Position und deshalb bleibt der Schalenradius konstant.
Da $r_{\text{S}}$ sich nicht ändert, ist auch das eingeschlossene Corevolumen und die Coremasse
konstant, folglich auch die mittlere Dichte.
Der Strahlungsstrom aus der Schale richtung Core drückt diesen zusammen. Dadurch wird seine Dichte
größer, die Dichte der Schale muß kleiner werden. Es gibt deshalb einen Dichtegradienten, der den
Einsturz des Sterns verhindern kann. 

Aber es passiert noch mehr: Da die Masse der Hülle sich ebenfalls nicht ändert, die Dichte
an der Oberfläche Null ist und die mittlere Dichte abnimmt, dehnt der Stern sich aus, solange 
das Core zusammengedrückt wird. Damit steuert die Schale zugleich das Core und die Hülle.
Durch das Ausdehnen der Hülle wird die Oberfläche des Sterns kühler. Ab einem bestimmten Punkt wird die 
Energiezufuhr in die Hülle von Konvektion bestimmt. Das stabilisiert die Oberflächentemperatur.
Der Stern erscheint rötlich, er ist zu einem Roten Riesen geworden.  

\subsection{Entartung der Elektronen}

\subsubsection{Entartung der Materie}

Entartung heißt, daß die mittlere kinetische Energie der Fermionen -- in unserem Falle sind das die Elektronen -- kleiner
als die Fermi-Energie ist.
Fast alle Energiezustände unterhalb der Fermi-Energie sind besetzt. Aus der Temperatur und der 
Teilchendichte läßt sich abschätzen, ob ein Gas entartet ist oder nicht. 
Fermi-Impuls und  Fermi-Energie sind gegeben durch:
\begin{eqnarray}
p_{\mathrm{F}} & = & \hbar \left(3 \pi^2 \frac{N}{V}\right)^{1/3}, \quad \label{eq:fermiimpuls}\\
E_{\mathrm{F}} & = & \frac{\hbar^2}{2m}\left(3 \pi^2 \frac{N}{V}\right)^{2/3}, \quad 
                                            \text{(nichtrelativistisch)}\label{eq:fermienergien}\\
 E_{\mathrm{F}}&  \approx&  c \hbar\left(3 \pi^2 \frac{N}{V}\right)^{1/3}, \quad \phantom{\frac{1}{2m}}
           \text{(relativistisch)} \quad .
\end{eqnarray}

Betrachten wir als Beispiel die Sonne. 
Ihre mittlere Dichte beträgt $\rho \approx 1.4 \cdot 10^3 \,\mathrm{kg}/\mathrm{m}^3$. Ihre Masse
wird hauptsächlich von  Protonen getragen, deren Masse $1.67\cdot 10^{-27}\, \mathrm{kg}$ beträgt. Somit
befinden sich $10^{30}$ Protonen in einem $\mathrm{m}^3$, was auch der Elektronendichte entspricht, da die 
Sonne elektrisch neutral ist.
Die Fusion findet bei $10^7\,\mathrm{K} = 800 \, \mathrm{eV}$ statt, man kann also nichtrelativistisch rechnen.
Der Fermi-Impuls beträgt $\approx 6200 \,\mathrm{eV}$, und damit die Fermi-Energie $38\,\mathrm{eV}$ für die Elektronen
und $2\cdot 10^{-2} \,\mathrm{eV}$ für die Protonen. Beide Gase sind also nicht entartet.
Beim Übergang eines Sterns wie der Sonne zum Roten Riesen kann der Fermi-Druck einen beträchtlichen Beitrag zum 
Gesamtdruck leisten: Wir nehmen einen Coreradius von $\frac{1}{70} R_{\odot}
= 10^7\,\mathrm{m}$ an, eine Masse von $0.2\, M_{\odot}$ und eine Temperatur von $2 \cdot 10^7\, \text{K}
= 2 \cdot 10^3 \,\text{eV}$.
Daraus folgt eine Teilchendichte (Helium) von $ n_{\text{He}} = 16 \cdot 10^{33}\, \text{m}^{-3}$
und $ n_e = 32 \cdot 10^{33}\, \text{m}^{-3}$. Der Fermi-Impuls ist $p_{\text{F},e} = c h 
\left(\frac{3 n_e}{8 \pi}\right)^{1/3} \approx 10^5\, \text{eV}$, woraus eine Fermi-Energie von 
$10^4\, \text{eV}$ folgt. Folglich sind die Elektronen teilweise entartet. 

Es hängt jetzt von der Masse
des Sterns ab, ob der Fermi-Druck der Elektronen (oder der Neutronen) den Kollaps des Sterns verhindern kann.
Wenn der Stern schrumpft, werden die Elektronen zum Atomkern hin gedrückt, d. h., der ihnen zur Verfügung
stehende Raum wird kleiner. Nach Gl. (\ref{eq:fermiimpuls}) und (\ref{eq:fermienergien}) wächst ihre kinetische 
Energie und damit auch der Fermi-Druck. 

\subsection{Weitere Entwicklung}
Gemäß Arnett \cite{super:96} kann man sich die weitere Entwicklung des Heliumcores näherungsweise so wie die
Entwicklung des Wasserstoff fusionierenden Sterns  vorstellen.  Der nächste
Prozeß nach der Fusion von Wasserstoff ist der $3\alpha$-Prozeß: 
\begin{eqnarray}
3 \alpha & \longrightarrow & {}^{12}\text{C} + \gamma \\
 {}^{12}\text{C} + \alpha & \longrightarrow & {}^{16}\text{O}
\end{eqnarray}
Aber zuerst muß die Entartung des Cores überwunden werden.
Wie oben erwähnt besteht das Core eines gealterten H-fusionierenden Sterns aus festem Helium. Das Core wird aber 
von den darüber liegenden Materieschichten zusammengedrückt und heizt sich dabei auf. Es kann sich aber nicht wie 
ein Gas ausdehnen.  
Sobald eine Temperatur von $1.5 \cdot 10^{8}$ K erreicht ist, setzt der $3\alpha$-Prozeß ein.
Die zusätzliche Energie kann nicht schnell genug abgeführt werden, obwohl die Wörmeleitfähigkeit eines entarteten
Elektronengases gut ist: Es kommt zu einer Explosion, dem sogenannten
Helium-Flash. Dadurch wird die Coredichte und die Temperatur vermindert, die Elektronen sind nicht mehr entartet.
Die Gravitation ist aber so stark, daß sie die Explosion so weit dämpfen kann, daß kein Material ausgeschleudert wird.
Es setzt eine weitere Phase der Kontraktion ein. Dadurch kann im Core des Sterns, der zur Zeit keine nukleare 
Energiequelle
besitzt, wieder die Temperatur für den  $3\alpha$-Prozeß erreicht werden. Die Kernfusion läuft diesmal kontrolliert ab,
weil das Core gasförmig und nicht fest ist.    
Die bei der Wasserstoffusion entstandene Asche ist also in einem Folgeprozeß zur Energiegewinnung genutzt worden.

\begin{figure}
\centering
\scalebox{0.45}{\includegraphics{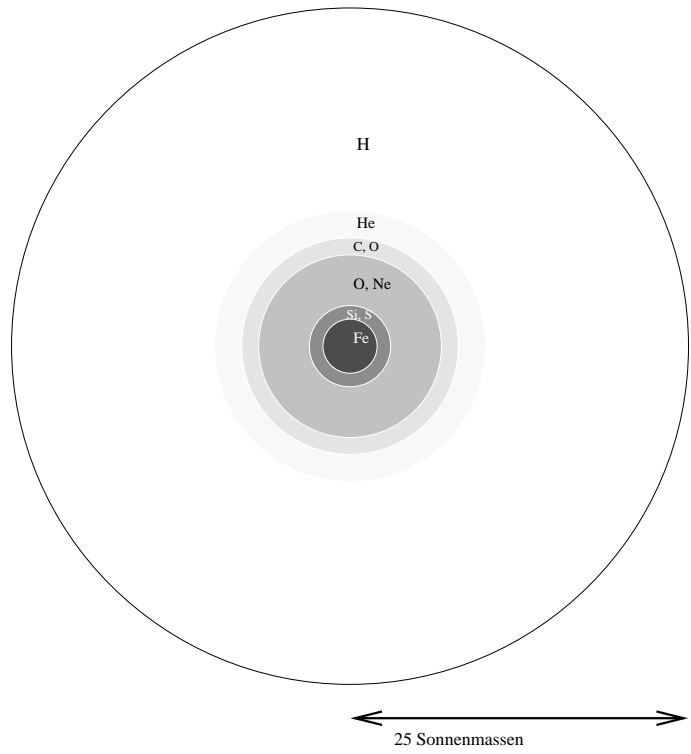}}
\caption[Sternaufbau]{Aufbau eines Sterns, der seinen gesamten Brennstoff aufgebraucht hat, nach \cite{How:85}}
\label{fig:aufbau}
\end{figure}

Insgesamt gibt es sechs solcher Fusionsprozesse. Dies führt dazu, daß ein ausgebrannter
Stern einen schalenförmigen Aufbau (ähnlich einer Zwiebel) besitzt, weil die Asche des vorausgegangenen Prozesses
der Brennstoff des nächsten ist. Das funktioniert solange, bis die mittlere Bindungsenergie pro Nukleon ein Maximum erreicht hat.
Dieser Fall tritt bei den eisenartigen Metallen ($\text{Ni},\,\text{Co},\,\text{Fe}$) ein.
Die Hauptprozesse sehen im einzelnen so aus (siehe \cite{super:96},\cite{How:85}):
\begin{enumerate}
\item  $4p \longrightarrow {}^4\text{He} + 2e^+ +2 \overline{\nu}$ bei $T \approx 10^7$ K
\item $ 3 \alpha  \longrightarrow  {}^{12}\text{C} + \gamma , \,
          {}^{12}\text{C} + \alpha  \longrightarrow  {}^{16}\text{O} $
                   bei $T \approx 1,5\cdot 10^8$ K
\item $ {}^{12}\text{C} + {}^{12}\text{C}  \longrightarrow {}^{20}\text{Ne} + \alpha, 
                                      \enspace $oder $ {}^{24}\text{Mg}
        {}^{16}\text{O} + {}^{4}\text{He}  \longrightarrow {}^{20}\text{Ne}$
                   bei $T \approx 1\cdot 10^9$ K
\item ${}^{20}\text{Ne} + \alpha \longrightarrow {}^{24}\text{Mg} + $ bei $T \approx 1,5\cdot 10^9$ K
\item ${}^{16}\text{O} + {}^{16}\text{O} \longrightarrow {}^{32}\text{S} \enspace
         \text{oder} 
         \enspace {}^{28}\text{Si} + \alpha $ bei $T \approx 2\cdot 10^9$ K
\item ${}^{28}\text{Si} + {}^{28}\text{Si} \longrightarrow {}^{56}\text{Ni}
                                   \longrightarrow {}^{56}\text{Fe} +
           2 e^+ +  2 \overline{\nu} $ bei $T \approx 3\cdot 10^9$ K
\end{enumerate}
In Abb. \ref{fig:aufbau} ist der schalenförmige Aufbau eines Sterns von $25 \,M_{\odot}$ skizziert, 
der alle Fusionsprozesse durchlaufen hat. 

Diese Reaktionen sind nur die wesentlichen Prozesse, daneben laufen noch viele weitere Reaktionen
ab, die für die Nukleosynthese wichtig sind. Das setzt aber voraus, daß der Stern ausreichend
Masse hat, so daß die Gravitation den Stern genügend zusammenquetschen kann.
Der innerste Prozeß erzeugt Eisen, das Element
mit der höchsten Bindungsenergie pro Nukleon. Die weitere Entwicklung hängt jetzt von der Masse
und der Elektronenanzahl pro Nukleon dieses Eisenkerns ab. Die Grenzmasse zwischen Stabilität und Instabilität ist gerade die
Chandrasekhar-Grenze, wird sie überschritten, ist Stabilität nicht möglich.
Wenn der Stern es schafft, in einer Supernova-Explosion genügend Masse abzustoßen,
bleibt ein kleiner, extrem dichter Neutronenstern übrig (für diesen gilt aber ein analoges Stabilitätsargument 
wie für den Weißen Zwerg), ansonsten ein Schwarzes Loch.

\subsection{Die Chandrasekhar-Grenze}
Im letzten Abschnitt habe ich die Chandrasekhar-Grenze erwähnt. Nun soll an dieser Stelle ein
Ausdruck für diese Grenzmasse hergeleitet werden. Dabei stütze ich mich auf \cite{Weinberg:71}.
 
\subsubsection{Differentialgleichung für die Sternstruktur}
Wir nehmen an, die Sternmaterie verhalte sich wie eine ideale, isotrope Flüssigkeit, d. h. keine 
Raumrichtung ist ausgezeichnet. Die mittlere
freie Weglänge im Inneren des Sterns sei kleiner als die Längenskala eines Beobachters, d. h. die innere
Struktur der Flüssigkeit spiele keine Rolle.
Außerdem sei die Flüssigkeit für den Beobachter in Ruhe. Dann können wir den Energie-Impuls-Tensor hinschreiben:
\begin{equation}
T^{00} = u, \, T^{0i} = T^{i0} = 0, \, T^{ij} = P \delta^{ij} \quad, \label{eq:spannung}  
\end{equation}
wobei $u$ die Energiedichte und $P$ den Druck bezeichnen.
 Die zugrundeliegende Metrik soll statisch und isotrop sein ($\tau$: Eigenzeit):
\begin{equation}
\text{d} \tau^2 = F(r)\text{d}t^2 - 2E(r)\text{d}t r \text{d}r
                                  -  D(r)r^2 \text{d}r^2
                                  -  C(r)(\text{d}r^2 + r^2 \sin^2 \theta \text{d} \varphi^2
                                          + r^2 \text{d} \theta^2) \quad .
\end{equation}
Durch eine geeignete Transformation der Zeit kann man den in der Raumzeit gemischten Anteil
zum Verschwinden bringen: $ t \mapsto t + \Phi(r)$, außerdem transformiert man $r \mapsto \sqrt{C(r)}r$.
Man erhält nach einer kurzen Rechnung:
\begin{eqnarray}
\text{d} \tau^2 & = & B(r)\text{d}t^2 - A(r)  \text{d}r^2 - r^2(\sin^2 \theta\text{d} \varphi^2
                                                             + \text{d} \theta^2) \enspace
\text{mit} \\
A(r) & = & \left( 1 + \frac{G(r)}{C(r)} \right) \left( 1 + \frac{1}{2}
\frac{rC'(r)}{C(r)} \right)^{-2}, \enspace
B(r) = F(r) \\
 & & \text{und der Nebenbedingung} \quad \Phi'(r) = -\frac{r E(r)}{F(r)} \quad .
\end{eqnarray} 
Hier ist $f'$ die Abkürzung für $\diff{f}{r}$.
Damit sieht unsere Metrik wie folgt aus:
\begin{eqnarray}
g_{\mu\nu} =
\begin{pmatrix}
-B(r)   &  0    &  0  &    0  \\
  0     &  A(r) &  0  &    0  \\
  0     &  0    & r^2 &    0  \\
  0     &  0    &  0  & r^2\sin^2 \theta 
\end{pmatrix}, \enspace
g^{\mu\nu} =
\begin{pmatrix}
- \frac{1}{B(r)} &           0       &  0            &    0  \\
  0               & \frac{1}{A(r)} 0  &                    0  \\
  0               &  0                & r^{-2}        &    0  \\
  0               &  0                &  0            & \frac{1}{r^2\sin^2 \theta }
\end{pmatrix} \quad .
\end{eqnarray} 
Das Ziel ist es, die Einstein-Gleichungen für ein solches System zu lösen.
Allgemein lauten sie:
\begin{eqnarray}
R_{\mu \nu} - \frac{1}{2} g_{\mu \nu} R = - 8 \pi G T_{\mu \nu} \quad .
\end{eqnarray}
Der Energie-Impuls-Tensor auf der rechten Seite der Gleichung ist bekannt, siehe Gleichung
 (\ref{eq:spannung}){}.
Ricci-Tensor und -Skalar  ergeben sich aus:
\begin{eqnarray}
R_{\mu \nu} & = & {R^{\lambda}}_{{\mu \lambda \nu}} =
                           \partial_{\nu}{\Gamma^{\lambda}}_{\mu \lambda}
                          - \partial_{\lambda} {\Gamma^{\lambda}}_{\mu \nu}
                          + {\Gamma^{\rho}}_{\mu \lambda} {\Gamma^{\lambda}}_{\nu \rho}
                          - {\Gamma^\rho}_{\mu \nu} {\Gamma^\lambda}_{\lambda \rho}
                            \enspace,\\
R & = & g^{\mu\nu}R_{\mu\nu} \quad.     
\end{eqnarray}
Die Christoffelsymbole ${{\Gamma_{\lambda}}^\sigma}_\mu$ sind wie folgt definiert:
\begin{eqnarray}
{{\Gamma_{\lambda}}^\sigma}_\mu = g^{\nu \sigma} \left(
                                  \partial_\lambda g_{\mu \nu} +
                                  \partial_\mu g_{\lambda \nu} -
                                  \partial_\nu g_{\mu \lambda} \right) \quad . 
\end{eqnarray}
Nach einer längeren Rechnung folgen dann die Einstein-Gleichungen:
\begin{eqnarray}
R_{rr} & = & \frac{B''}{2B} -\frac{B'}{4B} \left( \frac{A'}{A} + \frac{B'}{B} \right) -
 \frac{A'}{rA} =
                  -4 \pi G(\rho - P) A \\
R_{\theta \theta} & = & -1 + \frac{r}{2A}\left( -\frac{A'}{A} + \frac{B'}{B} \right) + 
\frac{1}{A} =
                   -4 \pi G(\rho - P) r^2  \label{eq:rtheta}\\
R_{tt} & = & -\frac{B''}{2A} + \frac{B'}{2A}\left( \frac{A'}{A} + \frac{B'}{B} \right) -
 \frac{B'}{rA} =
                -4 \pi G(\rho -3 P) B \\
R_{\phi \phi} & = & R_{\theta\theta}\\
R_{\mu \nu} & = & 0, \enspace  \text{falls} \enspace \mu \ne \nu
\end{eqnarray}
mit der Dichte $\rho$.
Zusätzlich benötigen wir eine Gleichung für hydrodynamisches Gleichgewicht. Aus $\partial_\beta{T^{\alpha \beta}} = 0$
folgt
\begin{eqnarray}
\frac{B'}{B} = - \frac{2P'}{P + \rho} \quad .
\end{eqnarray} 
Bildet man $\frac{R_{rr}}{2A} + \frac{R_{\theta \theta}}{r^2} + \frac{R_{tt}}{2B}$, so erhält 
man die Gleichung:
\begin{eqnarray}
\left(\frac{r}{A} \right)' = 1 - 8 \pi G \rho r^2 \enspace \Rightarrow
 A =\Bigl( 1 - \frac{2G}{r} \underbrace{\int^r_0 4 \pi x^2 \rho \text{d}x}_{M(r)} \Bigr)^{-1} . 
\end{eqnarray}
Dies setzen wir in Gleichung (\ref{eq:rtheta}) ein,  und es folgt nach einigen Umformungen
\begin{eqnarray}
- r^2 \diff{{}}{r}P(r) = G M(r)\rho (r) \left( 1 + \frac{\rho(r)}{P(r)} \right)
                             \left(1 + \frac{4 \pi r^3 P(r)}{M(r)} \right)
                             \left(1 - \frac{2 G M(r)}{r} \right)^{-1} \enspace \label{eq:Gleichung}.
\end{eqnarray}

\subsubsection{Newtonsche Sterne}

Wir betrachten jetzt Sterne, deren innere Energie und Druck viel kleiner als ihre Restmassendichte
 sind, wo relativistische Effekte keine Rolle spielen. Dann vereinfacht sich Gleichung 
(\ref{eq:Gleichung}) zu (siehe \cite{Weinberg:71}):
\begin{eqnarray}
\frac{\text{d}}{\text{d}r} \frac{r^2}{\rho(r)}\frac{\text{d}}{\text{d}r}P(r) 
= -4 \pi G r^2 \rho(r) \quad .
\end{eqnarray}
Eine Variablentransformation
\begin{eqnarray} 
\rho = \rho(0)\Theta^{1/(\gamma-1)},\enspace p = K \rho(0)^\gamma \Theta^{\gamma/( \gamma-1)},
\enspace r =\left( \frac{K \gamma}{4 \pi G (\gamma - 1)} \right) \rho(0)^{(\gamma - 2)/2} x 
\end{eqnarray} 
führt zur sogenannten Lane-Emden-Gleichung:
\begin{eqnarray}
\frac{1}{x^2} \frac{\text{d}}{\text{d}x} x^2  \frac{\text{d}}{\text{d}x} \Theta +
 \Theta^{1/(\gamma - 1)} ,
\Theta(0) = 1, \enspace \Theta'(0) = 0 \enspace .
\end{eqnarray}
Für $\gamma > \frac{6}{5}$ hat die Lösungsfunktion eine Nullstelle bei $x_1$, d. h. Druck und Dichte verschwinden bei $x_1$. 
Damit folgen der  Sternradius $R$ und die Sternmasse $M$ zu (mit den Bezeichnungen aus \cite{Weinberg:71})
\begin{eqnarray}
R & = & \sqrt{\frac{K \gamma}{4 \pi G (\gamma - 1)}} \rho(0)^{(\gamma - 2)/2}x_1, \enspace \\
M & = & 4\pi \rho(0) ^{(3\gamma -4)/2} \left(\frac{K \gamma}{4\pi G(\gamma - 1)} \right)^{3/2}
             x_1^{-(3\gamma - 4)/(\gamma - 2)} x_1^2 |\Theta'(x_1)|^2 \enspace. \label{eq:masseradius}
\end{eqnarray}
\subsubsection{Bedingung für Stabilität}
Nun können wir eine Bedingung angeben, wann ein Stern stabil ist. Dazu betrachten wir seine 
Energie 
\begin{eqnarray}
E = T + V = \underbrace{\frac{KM}{\gamma - 1}}_{a} \rho^{\gamma -1} - 
           \underbrace{\frac{3}{5} \left( \frac{4}{3}\pi \right)^{1/3} G M^{5/3}}_{b} \rho^{1/3}
              \enspace .
\end{eqnarray} 
Damit ein Stern instabil werden kann, muß er in Bezug auf seine zentrale Dichte 
bei konstanter Entropie pro Nukleon und konstanter chemischer Zusammensetzung
gemäß Theorem 1 aus \cite{Weinberg:71} stationär werden, das heißt, es muß
\begin{eqnarray}
\frac{\partial E(\rho(0))}{\partial\rho(0)} = 0, \enspace 
\frac{\partial N(\rho(0))}{\partial\rho(0)} = 0
\end{eqnarray} 
gelten mit $N$ der Nukleonenzahl. Wendet man dieses Kriterium auf den Ausdruck oben an, so liegt für
$\gamma > \frac{4}{3}$ ein Minimum von $E$vor, das heißt stabiles Gleichgewicht:
\begin{equation}
\rho = \left(\frac{b}{3a(\gamma-1)} \right)^{1/(\gamma-4/3)}\quad.
\end{equation}
 $\gamma < \frac{4}{3}$ liefert instabiles Gleichgewicht. Der Fall
$\gamma = \frac{4}{3}$ erfordert $a=b$, damit verschwindet die Energie.

\subsubsection{Anwendung auf Weiße Zwerge}

Die relativistische kinetische Energie eines Teilchens ist gegeben durch
\begin{eqnarray}
E_{\mathrm{kin}} = \sqrt{p^2 + M^2} - M \quad.
\end{eqnarray}
Wir nehmen an, daß die Temperatur gegenüber der kinetischen Energie des Teilchens vernachlässigbar sei.
Die Dichte der kinetischen Energie $e$ und der Druck der Elektronen $p$ in einem Weißen Zwerg sind gegeben durch
(Rechnungen ab sofort in natürlichen Einheiten, d.h. $\hbar=c=1$)
\begin{eqnarray}
e & = & \frac{8\pi}{(2\pi)^3}\int_0^{k_F} \left(\sqrt{k^2 + m^2_e} - m_e\right)k^2 \dif k \\
p & = & \frac{8\pi}{3(2\pi)^3}\int_0^{k_F}\frac{k^2}{\sqrt{k^2 + m^2_e}} k^2 \dif k\\
k_F & = & (3 \pi^2 n)^{1/3} =  \left(\frac{3\pi^2\rho}{m_N \mu}\right)^{1/3}, \enspace 
         \rho = n\underbrace{{\frac{n_B}{n}}}_{n_{e}} m_N \\
k_F & : & \text{Fermi-Impuls}, \enspace n_e: \text{Anzahl der Baryonen pro Elektron} \quad.
\end{eqnarray}
Zwei Extremfälle sind leicht zu diskutieren: $k_F \ll m_e$ und $k_F \gg m_e$.
Im ersten Fall gilt:
\begin{eqnarray}
p = \frac{8\pi}{15 m_e(2\pi)^3}k_F^5 = \frac{1}{15 m_e \pi^3}\left(\frac{3\pi^2}{m_N n_e}\right)^{5/3}\rho^{5/3}; 
     \enspace e = \frac{3}{2}p \quad.
\end{eqnarray}
Das ergibt unter Verwendung von Gl. (\ref{eq:masseradius}) eine Masse von
\begin{eqnarray}
M & = & 2.79 \left(\frac{\rho(0)}{\rho_c}\right)^{1/2} \frac{M_{\odot}}{m_e^2}\\
\rho_c & : & \text{Dichte, wo der Fermi-Impuls gleich der Elektronmasse ist}
\end{eqnarray}
Im zweiten Fall kann die Elektronmasse überall vernachlässigt werden:
\begin{eqnarray}
p & = & \frac{8\pi}{12(2\pi)^2}k_F^4 = \frac{1}{12\pi^2}\left(\frac{3\pi^2}{m_N\mu}\right)^{4/3}\rho^{4/3}; \enspace
e = 3p \\
M & = & 5.87 \frac{M_{\odot}}{n_e^2} \quad \text{{\bf{Chandrasekhar-Grenze}}} \quad. \label{eq:cha}
\end{eqnarray}
In beiden Fällen haben wir es mit Polytropen zu tun: Im ersten Fall mit einem Adiabatenexponenten von
 $\gamma = 5/3$, im zweiten Fall ist  $\gamma = 4/3$. Weiße Zwerge können unter diesen Voraussetzungen der   
Gravitation standhalten und sind bis zu der Grenzmasse in Gl.(\ref{eq:cha}) stabil.

Materie verhält sich aber etwas anders. Ab einer gewissen Größe des Fermi-Impulses wird Elektroneneinfang energetisch
günstiger, so daß die Anzahl der Baryonen pro Elektron wächst. Dies wiederum reduziert die Grenzmasse. 
Man kann zeigen, daß an diesem Punkt Stabilität in Instabilität umschlägt (s.\cite{Weinberg:71}).

\section{Endzustände der Sternentwicklung}

In diesem Abschnitt werden Endzustände der Sternentwicklung erläutert. Es gibt genau drei Endstadien der
Sternentwicklung:
\begin{itemize}
\item Weißer Zwerg
\item Neutronenstern
\item Schwarzes Loch
\end{itemize}
Der Endzustand eines Sterns hängt von seiner Anfangsmasse ab und von den äußeren Lebensbedingungen.
So können Materieströme von einem Begleiter des Sterns seine Masse erhöhen und dafür sorgen, daß
die Entwicklung weitergeht. 


\subsection{Weiße Zwerge}
Irgendwann ist der gesamte Brennstoff eines Sterns aufgebraucht. Der thermische Druck kann der Gravitation
nicht weiter standhalten. Bei kleinen Massen verhindert die elektrostatische Abstoßung der Elektronen einen
Kollaps. Ab einer Dichte von $80 \, \mathrm{g} \cdot \mathrm{cm}^{-3}$ tritt wie beim Roten Riesen 
Entartung der Materie ein. Der Stern kühlt ab.
Ist die Masse des Cores kleiner als die Chandrasekhar-Grenze, und gibt es auch keine allzu schwere Hülle,
so entwickelt sich der Stern zu einem Weißen Zwerg. Sein Radius beträgt einige tausend Kilometer. 
Die Elektronen im Inneren des Cores sind relativistisch entartet und können
dadurch den Kollaps des Sterns verhindern. Die äußeren Schichten des Cores sind weniger dicht, und die
Elektronen sind nicht entartet. Diese äußere Schicht wirkt als Isolation gegenüber dem Außenraum. Durch
das Schrumpfen gewinnt der Stern Energie aus dem Gravitationsfeld und strahlt diese Energie in langen
Zeiträumen ab, der Weiße Zwerg strahlt weiß-bläulich.

Hat der Stern dagegen eine schwere Hülle, so kann der Kern sich so weit aufheizen, daß Helium zu Kohlenstoff 
verbrannt wird.

\subsection{Neutronensterne}
Ist das Core des Sterns schwerer als die Chandrasekhar-Grenze, wird der Stern instabil.
Elektroneneinfang wird jetzt energetisch günstiger, wodurch sich 
die Zustandsgleichung der Materie ändert und ein neues Gleichgewicht erreicht wird.
In Gl. (\ref{eq:cha}) kann für leichte Neutronensterne  die Elektronenmasse $m_e$ durch die Neutronenmasse $m_n$,
ersetzt werden, und $n_e$ wird gleich $1$ gesetzt. Bei schwereren Neutronensternen bricht die Analogie zum Weißen Zwerg
zusammen, weil die Materie eines  Weißen Zwergs aus nichtrelativistischen Nukleonen besteht, während die Nukleonen 
eines Neutronensterns (von einer Sonnenmasse) eine kinetische Energie im Bereich ihrer Ruhmasse haben.   
Neutronen sind gegen $\beta$-Zerfall stabil, wenn der Fermi-See der Elektronen im Core mindestens bis zum maximalen
Elektronenimpuls der beim $\beta$-Zerfall entstehenden Elektronen aufgefüllt ist. Dieser maximale Impuls beträgt
etwa $1.2\,\text{MeV}$ (\cite{Weinberg:71}). Im Core sind bei diesen hohen Dichten die Elektronen entartet.
Ihr chemisches Potential beträgt am Anfang etwa $6\,\text{MeV}$ (\cite{state:79}). Es bildet sich ein
$\beta$-Gleichgewicht aus, so daß am Ende das chemische Potential der Elektronen gleich der Differenz der
chemischen Potentiale von Neutronen und Protonen ist: $\mu_e = \mu_n-\mu_p$.

Im letzten Schritt wird bei massiven Sternen Silizium zu Eisen verbrannt. Dies geschieht innerhalb eines Tages.
Wenn der thermische Druck wegfällt, schrumpft der Stern. Dabei werden die Elektronen in die Atomkerne gepreßt.
Ab einer bestimmten Dichte wird Elektroneneinfang für den Stern energetisch günstiger.
Dabei werden Neutrinos freigesetzt. Es werden aber nicht alle Protonen in Neutronen umgewandelt,
sonst wäre der Stern instabil gegenüber $\beta$-Zerfall. In dieser Umgebung können sogar Myonen
und andere exotische Teilchen wegen des Pauli-Blocking stabil sein!  
Elektroneinfang reduziert den Fermi-Impuls und somit den Druck  und leitet einen Kollaps des Kerns ein, der von einer Explosion
begleitet sein kann. Dabei wird die äußere Hülle weggesprengt. Dieses Ereignis ist eine Typ II-Supernova. 
Übrig bleibt ein extrem dichter Neutronenstern, sofern der Stern genügend Materie losgeworden ist.
Dieser wird von den nichtrelativistisch entarteten Neutronen am weiteren Kollaps gehindert. Die Rolle der Elektronen
haben die Neutronen übernommen. Die Rechnung von Chandrasekhar kann im wesentlichen übernommen werden, man muß nur
Elektronen durch Neutronen ersetzen. Dies führt zu einer Grenzmasse in derselben Größenordnung. Der resultiertende
Stern ist kleiner und dichter. 

Außen ist der Stern von einer festen Kruste aus Protonen in fester Phase umgeben. Die Elektronen sind relativistisch 
entartet und werden von den Atomkernen nicht beeinflußt. Im Inneren des Sterns befindet sich eine
``Neutronenflüssigkeit''.

\subsection{Schwarze Löcher}
Ist der Stern zu massiv und schafft er es nicht, genug Materie los zu werden, kann der Entartungsdruck der Fermionen
der Gravitation nichts mehr entgegensetzen. Der Stern kollabiert zu einem Schwarzen Loch.

\section{Sternkollaps und Neutrinokühlung}

In diesem Abschnitt wird die Explosion (Typ II-Supernova) eines massiven ($\approx 20-25 M_{\odot}$) gealterten Sterns 
genauer betrachtet (nach \cite{How:85}).

\subsubsection{Beginn des Kollaps}
Wir beginnen  mit dem Zeitpunkt, wo die Verbrennung von Silizium zu Eisen einsetzt. Dieser Prozeß dauert etwa einen
Tag, während die anderen Prozesse in längeren Zeiträumen stattgefunden haben. Innerhalb dieses Tages bildet sich ein
Core aus Eisen, der die Chandrasekhar-Grenze übersteigt. Dieses Core kann nicht mehr durch den Elektronendruck
stabilisiert werden, es implodiert zuerst, um dann zu explodieren.

Wenn die Implosion einsetzt, wird das Core komprimiert. Das führt zu einer Temperaturerhöhung, aber nicht unbedingt zu 
einem höheren Druck. Der Grund ist folgender: Der Druck hängt ab von der Anzahl der Teilchen und deren mittlerer Energie
(ideales Gas: Druck ist proportional zur Teilchendichte und zur Temperatur). Im Core befinden sich Elektronen und Eisenkerne. 
Da ein Eisenkern aus $56$ Nukleonen aufgebaut ist und weniger
Eisenkerne als Elektronen zur Verfügung stehen, ist der Beitrag der Elektronen zum Gesamtdruck wesentlich größer als 
der Beitrag der Eisenkerne. Durch das Aufheizen werden einige Eisenkerne in kleinere Bruchstücke zerlegt, dieser Vorgang
benötigt soviel Energie, wie bei deren Produktion freigesetzt wurde. Diese Energie wird den Elektronen entzogen, wodurch
deren Druck sinkt. Durch das Aufspalten dieser Eisenkerne steigt der Druck der Nukleonen, aber bei weitem
nicht so stark, daß der Druckabfall durch die kühleren Elektronen kompensiert werden kann. Die Folge dieses Druckabfalls
ist eine Beschleunigung des Kollapses.

\subsubsection{Die Entropie}
An dieser Stelle ist eine Bemerkung zur Entropie angebracht. Der Kollaps ist kein chaotischer Prozeß, wie es auf
den ersten Blick scheint. In einer Wasserstoffwolke ist die Entropie hoch, die ja ein Maß für die Unordnung ist.
In dem Core eines kollabierenden Sterns sind aber $56$ Nukleonen zu einem Eisenkern geordnet, folglich ist die Entropie
pro Nukleon klein. Photonen und Neutrinos haben im Laufe der Zeit die fehlende Entropie abgeführt.
Da die Zeitskala des Kollaps in der Größenordnung von Millisekunden liegt, Starke Prozesse aber eine Zeitskala von
$10^{-20} \, \mathrm{s}$ haben, werden Störungen sofort beseitigt, weil der Kollaps im Vergleich zu der Zeitskala 
starker Prozesse ``ewig'' dauert. 

Bedingt durch die hohe Dichte und der damit einhergehenden Entartung der Elektronen tritt vermehrt Elektroneneinfang ein. 
Es enstehen dabei Neutrinos, die bei ihrer
Emission Energie und Entropie abführen. Der Gesamteinfluß auf die Entropie ist unsicher. 

Die verminderte Elektronenzahl reduziert weiter den Druck und beschleunigt so den Kollaps. Da das 
Verhältnis der Anzahl von Elektronen zu Baryonen kleiner wird, wird auch die Chandrasekhar-Grenze kleiner,
welche ja von diesem Verhältnis abhängt.

\subsubsection{Neutrinoopazität}
Bei einer Dichte von $4 \cdot 10^{11} \, \mathrm{g}\cdot\text{cm}^{-3}$ tritt ein Effekt auf, der sonst nur
unmittelbar nach dem Urknall zu beobachten gewesen wäre: Die Sternmaterie wird für Neutrinos undurchsichtig.
Die mittlere Weglänge $\overline{\lambda}$ eines $30\,\text{MeV}$-Neutrinos beträgt im Core etwa $300 \,\text{cm}$.
Die Diffusionszeit (Zeit, die ein Teilchen braucht, um das Core zu verlassen) eines solchen Neutrinos beträgt damit 
etwa $1\,\text{s}$ (nach \cite{Stars:96}). 
Die Neutrinos können das Core nicht verlassen und es nicht kühlen. Hier ändert sich auch die Rolle der 
Chandrasekhar-Grenze. Vorher war das die
größte Masse, die ein entartetes Elektronengas am Einsturz hindern kann. Jetzt ist das die Grenzmasse, die als Ganzes
kollabieren kann. Gebiete innerhalb dieses Teils des Cores können durch Schall- und Druckwellen erreicht werden und
somit Dichteschwankungen ausgleichen.

\subsubsection{Aufbau einer Schockfront}
Der Einsturz des Cores geht weiter. Irgendwann hat die Materie im Zentrum Kerndichte und darüber 
($\rho_{\mathrm{Kern}}2.7\, \cdot 10^{14} \, \text{g}\cdot\text{cm}$) erreicht, weil von oben immer neues Material
einfällt. Kernmaterie ist zwar schwer komprimierbar, aber es ist möglich. Die inneren Schichten stehen dadurch unter 
einer großen Spannung wie eine zusammengedrückte Feder. Die Implosionsfront kommt mit einer gewaltigen 
Erschütterung zum Stillstand und wird zu einer Explosionswelle, die durch die Entspannung des Cores ausgelöst wird. 
Die Geschwindigkeit, mit der eine Coreschale 
einstürzt, ist proportional zu ihrer Entfernung zum Sternzentrum. Man sagt deshalb, der Einsturz sei homolog. 
Die Dichte nimmt nach außen hin ab, die Schallgeschwindigkeit ebenfalls, da die Kompressibilität mit kleiner 
werdender Dichte zunimmt. Im Core gibt es eine Grenze, wo die Einfallgeschwindigkeit gleich der 
Schallgeschwindigkeit ist. Diese markiert die Grenze des homologen Cores und wird sonischer Punkt oder Schallpunkt 
genannt. Eine Störung innerhalb des homologen Cores hat keinen 
Einfluß auf Gebiete außerhalb dieses Punktes, denn eine Schallwelle, die am Schallpunkt durch die einfallende Materie
nach außen läuft, bleibt in bezug auf das Sternzentrum in Ruhe, da dieses Coreelement mit Schallgeschwindigkeit 
einstürzt. Demzufolge kann eine Störung im Inneren den Außenbereich nicht erreichen.

Wenn die Implosionsfront zum Stillstand gekommen ist, entstehen durch die Erschütterung Schallwellen, die nach außen
laufen, mit größer werdender Entfernung zum Zentrum langsamer werden und am Schallpunkt schließlich stehen bleiben.
Auf die Materie im Zentrum, die ja Kerndichte erreicht hat, fällt immer mehr Material und erzeugt weitere 
Schallwellen, die sich am Schallpunkt sammeln. Dort baut sich eine Druckfront auf. Diese bremst auffallende Materie,
wodurch ein Sprung in der Schallgeschwindigkeit entsteht. Diese Unstetigkeit verursacht eine Schockwelle.

Nachdem das Core Kerndichte erreicht hat und durch weiteres einfallendes Material weiter komprimiert worden ist,
entspannt sich das Core und sendet weitere Schallwellen aus, die sich mit der Schockfront am Schallpunkt vereinen.

\subsubsection{Schockwellen}
An dieser Stelle soll der Unterschied zwischen einer Schall- und einer Schockwelle geklärt werden. Schallwellen 
durchlaufen ein Medium mit Schallgeschwindigkeit und verändern die Eigenschaften des Mediums nicht. Es wird vor 
allen Dingen nicht zerstört. Eine Schockwelle pflanzt sich mit Überschallgeschwindigkeit fort und ändert dabei 
Druck, Dichte, Entropie des durchlaufenen Mediums.

\subsubsection{Neutrinokühlung}
Die bei der Supernova SN 1987A beobachteten Neutrinos wurden in einer Zeitspanne von etwa $10\,\text{s}$ emittiert.
Nach etwa einer Millisekunde werden die Neutrinos freigesetzt, die beim Elektroneneinfang entstehen. Dies ist der
prompte Neutrinoausbruch. Es bleiben aber noch genügend Leptonen (Neutrinos und Elektronen) im Inneren des Core 
gefangen, wo sich die Schockwelle gebildet hat. Der äußere Teil des Cores fällt in der nächsten Sekunde in sich 
zusammen. Seine Bindungsenergie wird in Form von Neutrinos emittiert. Mittlerweile konnte die Schockwelle weiter nach 
außen laufen und die äußeren Schichten des Sterns wegreißen. Das Core (Radius etwa $30 \,\text{km}$) kann als Stern im 
Stern betrachtet werden. Er kontrahiert weiter und wird durch die Emission von Neutrinos aller Generationen gekühlt. 
Durch die Emission von Elektronneutrinos gehen die Leptonen bis auf einen kleinen Rest verloren. Nach 
$5 - 10\,\text{s}$ ist diese Phase abgeschlossen (sogenanntes Helmholtzkühlen, s. \cite{Stars:96}). Danach setzt 
die späte Kühlungsphase des Neutronensterns ein.

Der wesentliche Punkt ist die Dauer der Neutrinokühlung von $\geqq 10\, \text{s}$, die im Vergleich zur 
Zeitdauer des Kollaps sehr lang ist. Die Neutrinos können aus dem Core nicht frei herausströmen, denn ihre mittlere
Weglänge $\lambda_\nu$ ist kleiner als der Coreradius. Sie müssen aus dem Core langsam herausdiffundieren.
Dieses Neutrinosignal ist 1987 nachgewiesen worden (IMB, Kamiokande).
Ein zusätzlicher Kühlmechanismus darf deshalb nicht zu effektiv sein. Für die mittlere Weglänge gilt nämlich
$\lambda_\nu = 1/(n \sigma) \propto 1/(n T^2)$, wobei $n$ die Teilchendichte, $\sigma$ den totalen Wirkungsquerschnitt
und $T$ die Temperatur bezeichnen. Wäre ein Kühlmechanismus effektiver, würde die Coretemperatur rasch sinken und die 
mittlere freie Weglänge der Neutrinos größer werden. Ein kürzeres Neutrinosignal wäre die Folge. Das steht aber im
Widerspruch zu den Beobachtungen.

\subsubsection{Weitere Verlauf}
Das Geschehen danach ist unklar. Im einfachsten Szenario läuft die Schockwelle durch alle Schichten des Sterns in
Richtung Oberfläche: Der Stern explodiert, man kann eine Supernova beobachten.
Der größte Teil der Hülle wird weggesprengt, übrig bleibt ein kleiner, extrem dichter Reststern, der zu einem 
Neutronenstern kondensiert. Die Schockwelle setzt auf ihrem Weg nach außen die gefangenen Neutrinos frei, die den 
Stern dabei kühlen.

Computersimulationen zeigen aber ein anderes Bild. Die Schockwelle läuft nach $100-200\, \mathrm{km}$ aus,
weil sie auf ihrem Weg die Atomkerne in kleinere Bestandteile (z. B. Protonen) zerlegt und dabei an Energie 
verliert. Druck und Temperatur werden stark vermindert. Die Protonen fangen Elektronen ein, die dabei entstehenden 
Neutrinos entkommen und führen Energie ab. Wegen der vielen Zwischenfälle, die der Schockwelle in dieser Region 
passieren können, wird dieser Teil des Sterns ``Minenfeld'' genannt. Es gibt kein befriedigendes Modell,
das die Explosion zutreffend beschreibt.

Das soeben beschriebene Szenario ist ein mögliches unter vielen. Simulationen haben gezeigt, daß der Mechanismus 
stark von der Masse und der Kompressibilität der Kernmaterie abhängt. Das letztgenannte Modell trifft wohl auf 
Sterne mit einer Anfangsmasse von $12-18\, M_{\odot}$ zu.

\subsubsection{Schicksal des Reststerns}
Zum Schluß noch ein paar Bemerkungen zu dem Schicksal des Cores.
Die Gravitation kann den Elektronendruck überwinden. Sogar Kernmaterie kann den weiteren Einsturz nicht immer 
verhindern. Die Grenzmasse für einen erkalteten Neutronenstern liegt bei etwa $1.8 \, M_{\odot}$. Die Argumente sind
dieselben wie bei der Chandrasekhar-Grenze. Die nicht-relativistisch entarteten Neutronen können den weiteren 
Kollaps bis zu dieser Grenze verhindern.
Größere Massen kollabieren wahrscheinlich zu einem Schwarzen Loch.
Unmittelbar nach der Explosion ist der Reststern in seinem Inneren etwa $10^{11} \, \mathrm{K}$ heiß. Diese 
Temperatur reicht aus, um größere Massen als die $1.8 \, M_{\odot}$ zu stabilisieren. Die Kühlung erfolgt durch 
Neutrinos. Diese führen hundert Mal mehr Energie als die Explosion ab, in Zahlen $\approx 10^{53} \, \mathrm{erg}$, 
was etwa einem Zehntel der Gesamtmasse des Sterns entspricht.

Auf der Erde können diese Modelle durch Nachweis der Neutrinos und der Häufigkeit der chemischen Elemente überprüft 
werden.

\chapter{Neutralinokühlung der Supernova}

\section{Die Emissivität des Elektron-Positron-Kanals}

In diesem Kapitel werden die Produktionsraten für Neutralinos durch die verschiedenen Mechanismen berechnet.

\subsection{$e^+ e^-$-Annihilation und Selektronaustausch}
Wir nehmen an, daß folgende $e^+ e^-$-Prozesse zur Kühlung der Supernova beitragen können:
\begin{eqnarray}
e^+ e^- \longrightarrow Z^0 \longrightarrow  \chi^0_1 \chi^0_1 \\
e^+ e^- \overset{\tilde{e}_{L/R}}{\longrightarrow} \chi^0_1 \chi^0_1 
\end{eqnarray}
Der erste Prozeß ist nur dann möglich, wenn zum leichtesten Neutralino Higgsino-Anteile beigemischt sind,
weil nur die Higgsinos an das $Z^0$ koppeln.
Die Feynman-Graphen für diese fünf Prozesse sind in erster Ordnung:

\unitlength=1mm
\vspace{10mm}
\begin{fmffile}{diagramm3}

\begin{table}[htbp]
\begin{center}
\begin{tabular}{c c c}
\vspace{8mm}
\begin{fmfgraph*}(28,21)
\fmfpen{thick}
\fmfleft{ia1,ia2} \fmfright{oa1,oa2}
\fmf{fermion}{oa1,va1,ia1} \fmf{fermion}{ia2,va2,oa2}
\fmf{dashes_arrow,label=$\tilde{e}_{L/R}$}{va2,va1} \fmfdot{va1,va2}
\fmflabel{$e^+, p_1$}{ia1}
\fmflabel{$e^-, p_2$}{ia2}
\fmflabel{$\chi^0_1, p_3$}{oa1}
\fmflabel{$\chi^0_1, p_4$}{oa2}
\end{fmfgraph*}

\hspace{20mm}

\begin{fmfgraph*}(28,21)
\fmfpen{thick}
\fmfbottom{ina1,vna1,ina2} \fmftop{ona1,vna2,ona2}
\fmfforce{(0.0w,0.0h)}{ina1}
\fmfforce{(1.w,0.0h)}{ina2}
\fmfforce{(0.0w,1.h)}{ona1}
\fmfforce{(1.0w,1.0h)}{ona2}
\fmfforce{(0.37w,0.2h)}{vna1}
\fmfforce{(0.37w,0.8h)}{vna2}
\fmf{fermion}{ona2,vna1,ina1} \fmf{fermion}{ona1,vna2,ina2}
\fmf{dashes_arrow,label=$\tilde{e}_{L/R}$}{vna2,vna1} \fmfdot{vna1,vna2}
\fmflabel{$e^+, p_1$}{ina1}
\fmflabel{$e^-, p_2$}{ona1}
\fmflabel{$\chi^0_1, p_4$}{ona2}
\fmflabel{$\chi^0_1, p_3$}{ina2}
\end{fmfgraph*}

\hspace{23mm}

\begin{fmfgraph*}(28,21)
\fmfpen{thick}
\fmftop{ima1,ima2} \fmfbottom{oma1,oma2}
\fmf{fermion}{oma1,vma1,ima1} \fmf{fermion}{ima2,vma2,oma2}
\fmf{boson,label=$Z^0$}{vma1,vma2} \fmfdot{vma1,vma2}
\fmflabel{$e^-, p_2$}{ima1}
\fmflabel{$e^+, p_1$}{oma1}
\fmflabel{$\chi^0_1, p_4$}{ima2}
\fmflabel{$\chi^0_1, p_3$}{oma2}
\end{fmfgraph*}

\end{tabular}
\end{center}
\end{table}
\end{fmffile}

Mit den Feynmanregeln des Standardmodells (ich beziehe mich auf \cite{element:84}) und des MSSM aus
\cite{super:85} und \cite{higgs:86} kann man den totalen Wirkungsquerschnitt berechnen. Bartl et al.
haben in \cite{production:86} diesen Wirkungsquerschnitt in allgemeinerer Form für zwei verschiedene
Neutralinos als Endzustand berechnet. Da die anderen Neutralinos zu große Massen haben
als daß sie thermisch angeregt werden können, beschränke ich mich auf den Sonderfall, daß zwei leichteste
Neutralinos auslaufen. 
Für die ausführliche Rechnung verweise ich auf den Anhang, hier nenne ich nur die wesentlichen Ergebnisse.

Es ergeben sich folgende Amplituden:
\begin{eqnarray}
- \ie M_1 & = &  \frac{g^2}{\cos^2\theta_W} D_{Z^0}(s) \vv(p_1) \gamma^\mu (c_L P_L + c_R P_R) u(p_2)
                                      \uu(p_3) \gamma_\mu (L_1 P_L + R_1 P_R) v(p_4) \label{eq:mmma} \\
- \ie M_2 & = & -g^2|l_1|^2 D_{\tilde{e}}(t) \vv(p_1) P_R v(p_3) \uu(p_4) P_L u(p_1) \label{eq:mmmb}\\
- \ie M_3 & = & -g^2|l_1|^2 D_{\tilde{e}}(u) \vv(p_1) P_R v(p_4) \uu(p_3) P_L u(p_2)\cdot (-1)\label{eq:mmmc} \\
- \ie M_4 & = & -g^2|r_1|^2 D_{\tilde{e}}(t) \vv(p_1) P_L v(p_3) \uu(p_4) P_R u(p_2)\label{eq:mmmd} \\
- \ie M_5 & = & -g^2|r_1|^2 D_{\tilde{e}}(u) \vv(p_1) P_L v(p_4) \uu(p_3) P_R u(p_2) \cdot (-1)\label{eq:mmme}
 \enspace .
\end{eqnarray}

Summation über die Ausgangsspins und Mittelung über die Eingangsspins liefert den differentiellen 
Wirkungsquerschnitt, ausgewertet im Schwerpunktssystem:
\begin{eqnarray}
\diff{\sigma}{\Omega} =  \frac{g^4}{128 \pi^2 s} \sqrt{1 - \frac{4 M^2}{s}} \Biggl(
                         (t - M^2)^2 + (u - M^2)^2 - 2M^2s \Biggr) \times \\
                         \Biggl[ \frac{c_L^2 L_1^2 + c_R^2L_1^2}{ M^4_{Z^0}\cos^4\theta_W } +
                  \frac{L_1 \bigl( c_L|l_1|^2 - c_R |r_1|^2 \bigr)}{ M^2_{Z^0} M_{\tilde{e}}^2\cos^2\theta_W} +
                  \frac{|l_1|^4+|r_1|^4}{4 M_{\tilde{e}}^4} \Biggr] \enspace ,
\end{eqnarray}
die auftretenden Konstanten bedeuten:
\begin{eqnarray}
c_L & = &\frac{1}{2}(c_V + c_A); \enspace c_R = \frac{1}{2}(c_V - c_A); \enspace c_V = T_3 - 2 e_{\ell} \sw[2];
\enspace c_A = T_3 \\
L_1 & = & \frac{1}{2}\left( N_{14}^{}N_{14}^* -  N_{13}^{}N_{13}^* \right) \cos(2\beta) -
      \frac{1}{2}\left( N_{13}^{}N_{14}^* -  N_{13}^{}N_{14}^* \right) \sin(2\beta)   \\
R_1 & = & -L_1\\
l_1 & = & \sqrt{2} \left[e_{\ell} \sw N_{11} + \frac{1}{\cw}
                \left(\frac{1}{2} - e_e \sw[2] \right)N_{12}\right]\\
r_1 & = & \sqrt{2} \left[e_{\ell} \sw N_{11}^* - e_{\ell} \frac{\sw[2]}{\cw} N_{12}^* \right]\\
D_{Z^{0}}(s) & = & \frac{\ie}{M^2_{Z^{0}}};\quad D_{\tilde{e}}(x)  =  \frac{-\ie}{M^2_{\tilde{e}}} \quad .
\end{eqnarray}
Nach Integration über die Winkel erhält man den totalen Wirkunsquerschnitt:
\begin{eqnarray}
\sigma =  \frac{2 \pi \alpha^2 s}{3 \sw[4]} 
                \Biggl[ \frac{c_L^2 L_1^2 + c_R^2L_1^2}{ M^4_{Z^0}\cos^4\theta_W } +
                \frac{L_1 \bigl( c_L|l_1|^2 - c_R |r_1|^2 \bigr)}{ M^2_{Z^0} M_{\tilde{e}}^2\cos^2\theta_W} +
                        \frac{|l_1|^4+|r_1|^4}{4 M_{\tilde{e}}^2} \Biggr] 
                        \left(\sqrt{1 - \frac{4 m^2}{s}} \right)^{3/2}  \, .
\end{eqnarray}
Weitere Details und Erklärungen zu dieser Rechnung befinden sich im Anhang.
\subsection{Definition der Emissivität}
Die Emissivität $\epsilon$ gibt den Energieverlust pro Zeit und Volumen an. Sie kann analog zum
Wirkungsquerschnitt aus dem Betragsquadrat des Matrixelements unter Verwendung Fermis Goldener Regel
 berechnet werden. Sie ist für einen 
$2 \rightarrow 2$-Prozeß gegeben durch:

\begin{eqnarray}
\epsilon & = &(2\pi)^4 \int \frac{\dif{p}_1^4}{(2\pi)^2 2E_1} \frac{\dif{p}_2^4}{(2\pi)^2 2E_2}
                       \frac{\dif{p}_3^4}{(2\pi)^2 2E_3} \frac{\dif{p}_4^4}{(2\pi)^2 2E_4}
	              (E_3 + E_4)f_1 f_2 (1-f_3)(1-f_4) \notag \\ 
    &  &           \phantom{(2\pi \int)^4}\delta^4({p}_1 + {p}_2 - {p}_3 - {p}_4)
	            |\overline{\mathcal{M}}|^2 \\
f_i & = & \frac{1}{\mathrm{e}^{(p_i \pm \mu_i)/T} + 1} \quad \text{(alle (Anti-)Teilchen sind Fermionen)}.
\end{eqnarray}

Aus der Formel ist der Zusammenhang zwischen Emissivität und Wirkungsquerschnitt ersichtlich:
\begin{eqnarray}
\epsilon & = & \frac{1}{(2 \pi)^6}\int \dif {p}_3 \dif {p}_4 
                          (E_3 + E_4) f_1 f_2 (1 - f_3)(1 - f_4)\,
	       v_{\text{M\o{}l}} \,\sigma \quad . \label{eq:emissivitaet}
\end{eqnarray} 
 
An dieser Stelle sei noch darauf hingewiesen, daß die beiden einlaufenden Teilchen im Sternsystem nicht
kollinear sind.
Deswegen muß für $|\mathbf{v}_1 - \mathbf{v}_2|$ die M\o{}ller-Geschwindigkeit angesetzt werden 
(siehe \cite{Landau:71}), die als Funktion der Teilchengeschwindigkeiten folgende Form hat:
\begin{eqnarray}
{v}_{\text{M\o{}l}} = \sqrt{(\mathbf{v}_1 - \mathbf{v}_2)^2 -
                              (\mathbf{v}_1 \times \mathbf{v}_2)^2} \quad .
\end{eqnarray}
Um die Emissivität ausrechnen zu können, muß man die Temperatur und das chemische Potential kennen. Wenn die
Temperaturen hoch und damit die Entartung niedrig ist, so sind $f_3$ und $f_4$ 
Null. Die Faktoren $1 - f_3$, $1 - f_4$ geben die Phasenraum\-unterdrückung an, da Fermionen nur einen freien
Zustand pro Teilchen besetzen dürfen. $f_1$ und $f_2$ geben die Teilchendichte an. 

\subsection{Emissivität des Elektron-Positron-Prozesses}

Der totale Wirkungsquerschnitt des Prozesses $e^+ e^- \rightarrow \chi^0 \chi^0$ wurde schon ausgerechnet. Das
Ergebnis kann jetzt für die Berechnung der Emissivität verwendet werden. 
Die Impulse der einlaufenden Teilchen (Elektron und Positron) werden wie folgt parametrisiert:
\begin{eqnarray}
p_1^\mu & = & (p_1, 0, 0, p_1),  \quad p_2^\mu = (p_2, 0, p_2 \sin \theta, p_2 \cos \theta) \label{eq:para}\\
\theta & = & \angle(\mathbf{p_1}, \mathbf{p_2}) \quad
\end{eqnarray} 
Einige Winkelintegrationen in Formel (\ref{eq:emissivitaet}) können schon ausgeführt werden. Man erhält als
Resultat, wenn man $E_{1,2} = p_{1,2}$ (masselose Elektronen bzw. Positronen) und $E_1 + E_2 = E_3 + E_4$
(Energieerhaltung) benutzt (siehe \cite{anomaly:00}):
\begin{eqnarray}
\epsilon & = & \frac{1}{2 \pi^4} \int \dif p_1 \dif p_2 \,\dif \!\cos \theta 
               \frac{p_1^2}{\mathrm{e}^{(p_1 - \mu)/T}+1}\:
               \frac{p_2^2}{\mathrm{e}^{(p_2 + \mu)/T}+1} (p_1 + p_2)\,v_{\text{M\o{}l}} \,\sigma \quad .
              \label{eq:emm}
\end{eqnarray}
Unter Verwendung von Gl.(\ref{eq:para}) und 
$\mathbf{v} = \mathbf{p}/E$ erhält man für die Möller-Geschwindigkeit:
\begin{equation}
v_{\text{M\o{}l}} =|\mathbf{v}_1 - \mathbf{v}_2|_{\text{M\o{}l}} = \left| \frac{\mathbf{p}_1}{E_1} - \frac{\mathbf{p}_2}{E_2}
                                                         \right | =
                \left| \frac{\mathbf{p}_1}{p^0} - \frac{\mathbf{p}_2}{p^0} \right | = (1 - \cos\theta) \quad .
\end{equation}

\subsection{Die abgestrahlte Gesamtenergie}

Integration der Emissivität über das Sternvolumen und die Abstrahldauer liefert die Leistung beziehungsweise 
die Energie dieses Prozesses:

\begin{eqnarray}
E & = & \int_{\Delta t} \dif t\int_{\Delta V} \dif V \epsilon\big( M_\chi, \rho, T, \eta)\big); \\
\Delta t & : & \text{Abstrahldauer} \quad \Delta V :  \text{Sternvolumen} 
\end{eqnarray} 
Weil in \cite{birthneutronstar:86} Temperatur und Entartung (siehe die entsprechenden Abb. aus \cite{birthneutronstar:86}) 
nicht in Abhängigkeit des Radius, sondern der 
eingeschlossenen Baryonenmasse angegeben sind, muß $\dif V$ auf $\dif m$ umgerechnet werden.
Das Dichteprofil aus \cite{birthneutronstar:86} 
zeigt, daß die Dichte des kollabierenden Cores nach  $250$ ms etwa konstant ist.
Damit ist die Umrechnung einfach: $\dif V = \dif \frac{m}{\rho} = \frac{\dif m}{\rho}$.
Für die Abstrahldauer nehme ich Zeitspannen zwischen $0.005 \,\text{s}$ und $2$ s an, weil die Zeitspannen,
in denen der Kollaps stattfindet, diesen Größenordnungen entspricht. Für die anderen Parameter fixiere ich
sie bei $1\,\mathrm{s}$. 

\subsection{Temperatur und chemisches Potential}

Ein Ziel meiner Arbeit ist es, für die Temperatur und das chemische Potential eine realistische Verteilung
anzunehmen. Burrows und Lattimer haben in \cite{birthneutronstar:86} verschiedene Parameter einer
Supernova untersucht, unter anderem haben sie Profile für die Temperatur, chemische Potential 
und Dichte zu verschiedenen Zeiten berechnet.
Diese Profile verwende ich, um die Emissivität in Abhängigkeit des Sternradius zu berechnen. Ich benutze die
Profile zur Zeit $t = 0.5\, \mathrm{s}$, weil diese in der Mitte der ersten Sekunde liegen.
Aus den Funktionsgraphen werden etwa ein halbes Dutzend 
charakteristische Punkte (z. B. Extrema) abgelesen,
und diese dann durch kubische Splines miteinander verbunden. Um ein gutes glattes Resultat zu erzielen,
reichen die
wenigen Punkte aus. Ein Problem bei diesen Simulationen ist die Unkenntnis der Anfangsdaten. Die Simulation 
von Burrows und Lattimer beginnt mit dem Kollaps des Sterns. Die Simulationen können aber noch nicht den
Prozeß des Kollabierens erfolgreich berechnen. Die Schockwelle bleibt im Core stecken. Deshalb sind die Anfangsdaten nur
unzureichend bekannt. Nach einer gewissen Zeit ($\mathcal{O}(1)\, \mathrm{s}$) steht das Supernovasystem aber 
im Diffusionsgleichgewicht, so daß die genaue Kenntnis der Anfangsdaten für die weitere Entwicklung nicht nötig ist.
Man erkennt aus den Graphen, daß bei $r \approx M_{\odot}$ die Temperatur
hoch und die Entartung gering ist. Die numerische Auswertung der Emissivität zeigt, daß die meiste Energie aus 
einer Kugelschale mit diesem Radius herrührt. 

Der Radius ist als eingeschlossene Baryonenmasse angegeben. Über die Coredichte kann dann die
Massenabhängigkeit in eine Radiusabhängigkeit umgewandelt werden. 
Aus der Größe des Sterns $R = 13 \mathrm{km}$ und seiner Masse $M = 1.4 \,M_{\odot}$ ergibt sich $\rho$ zu
\begin{equation}
\rho = \frac{m}{V} = \frac{m}{\frac{4}{3}\pi R^3} \approx 3 \cdot 10^{14}\, \mathrm{g}\,\mathrm{cm}^3 \quad;
\end{equation}
das Core hat also Kerndichte erreicht.




Alle zur Berechnung der Emissivität notwendigen Größen sind jetzt beisammen.  
Die drei verbleibenden Integrationen in Gl. (\ref{eq:emm}) werden numerisch ausgeführt. Dazu verwende ich die verallgemeinerte 
Gaußsche Integrationsmethode, bei der der Integrand $f(x)$ an den Nullstellen orthogonaler Polynome ausgewertet wird
und diese Werte gewichtet addiert werden. Die zu verwendenden Polynome hängen von der Gestalt des Integranden
ab:
\begin{eqnarray}
\int_a^b k(x)f(x)\dif x & \approx & \sum_{i = 1}^n c_i f(x_i),\\
&& k(x):\quad \text{nichtnegative, meßbare Gewichtsfunktion},\\
&& c_i :\quad \text{Gewichtsfaktoren} \quad.
\end{eqnarray}
Die Intervallgrenzen können auch uneigentlich sein.
Die $x_i$ sind Nullstellen von Polynomen, diese Polynome bilden bezüglich des Skalarprodukts
\begin{eqnarray}
 (f,g) = \int_a^b k(x)f(x)g(x) \dif x 
\end{eqnarray}
eine Orthonormalbasis.
Die genaue Theorie dazu findet man z. B. in \cite{num:94}. 

Für dieses Problem geeignete Polynome sind für die Impulsintegrationen die Laguerre-Polynome, da die
Fermi-Dirac-Verteilungen sich wie $\mathcal{O}(\mathrm{e}^{-x})$ verhalten, für die Winkelintegration wähle 
ich die Legendre-Polynome. Die Nullstellen $x_i$ und Gewichtsfaktoren $c_i$ findet man in \cite{zeroes:66} oder in
\cite{math:70} tabelliert. Dieses Verfahren hat den Vorteil, daß man mit wenigen Schritten (und damit in
kürzerer Zeit gegenüber Keplerscher Faßregel oder ähnlichem) ein genaues Endergebnis erhält, nämlich
die Emissivität in Abhängigkeit von der Masse und der Kopplung, das heißt, den
Mischungsparametern und der Squarkmasse.

\subsection{Das Resultat}
\begin{figure}[h]
\centering
\vspace{5mm}
\scalebox{0.6}{\includegraphics{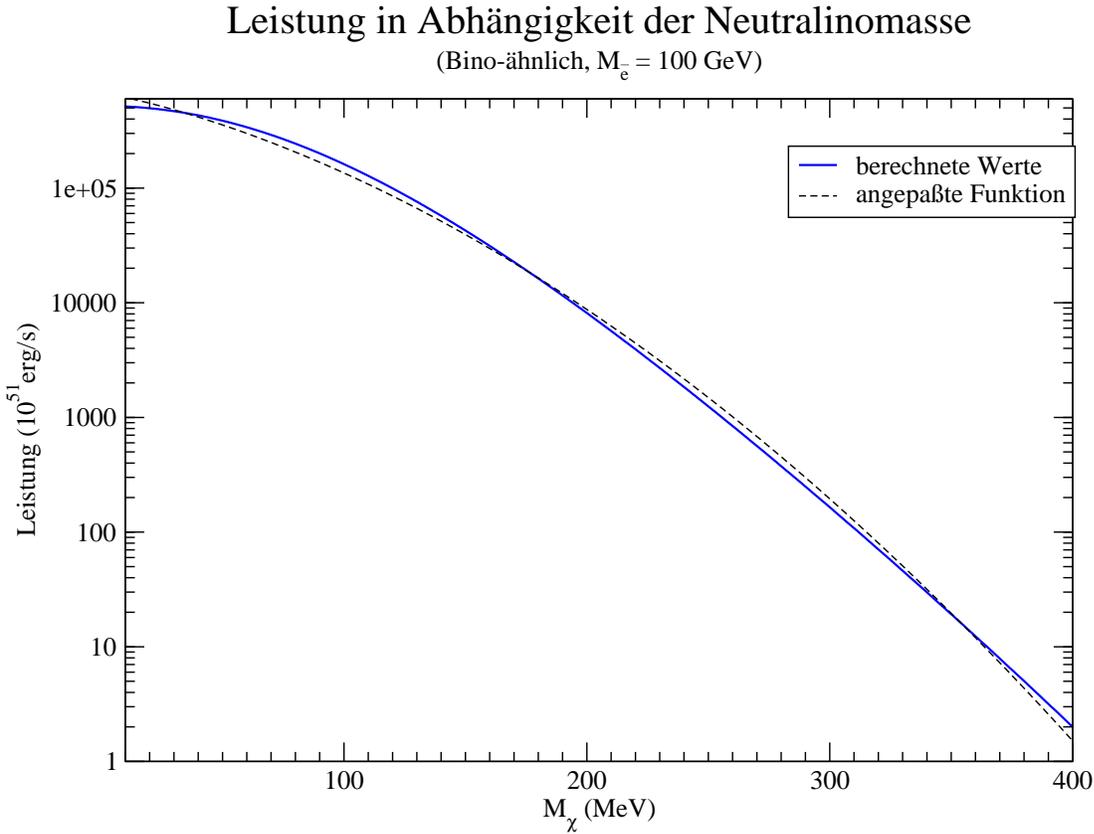}}
\caption[Massenabhängigkeit der Leistung]{Leistung in Abhängigkeit von der Masse für ein Bino-Neutralino, 
                                          $M_{\tilde{e}} = 100$ GeV}
\label{fig:bino}
\end{figure}

In Abbildung \ref{fig:bino} ist die Leistung in Abhängigkeit der Neutralinomasse exemplarisch für ein Bino 
mit einer Selektronmasse von $100$ GeV dargestellt.

Zur besseren Handhabung der Funktion habe ich sie durch eine geeignete analytische Funktion approximiert. 
Folgender Ausdruck ist eine mögliche Näherung:

\begin{eqnarray}
P(M_\chi) = 6.47 \cdot 10^5 \left(\frac{M_\chi}{\mathrm{MeV}}\right)^{0.03}
              \exp\left(-0.0115\,\frac{M_\chi}{\mathrm{MeV}} 
                       - 0.000053\left(\frac{M_\chi}{\mathrm{MeV}}\right)^2 \right)\, 10^{51}\,\mathrm{erg/s} \, .
\end{eqnarray}
Die Näherung ist als gestrichelte Funktion ebenfalls in Abb. \ref{fig:bino} zu sehen. Diese Näherungsformel hat bis
$M_\chi=370\,\text{MeV}$ einen maximalen Fehler von $20\%$, darüber hinaus ist der Fehler so groß, daß die Formel nicht 
mehr brauchbar ist. 

\subsection{Frühere Resultate}
Ellis et al. haben in \cite{lowmass:88} die Möglichkeit von Photinos mit einer Masse von etwa $100\,\text{eV}$ 
diskutiert. Squark- und Selektronmassen in Bereich der $W$-Masse zerstören eventuell den durch Neutrinobeobachtungen
erlaubten Bereich. Neutralinos sind dann genauso wie Neutrinos gefangen, während Sfermionmassen von 
$2.5\,\text{TeV}$ die Photinos frei herausströmen lassen.

Kachelrieß hat in \cite{anomaly:00} die Möglichkeit eines $34\,\text{MeV}$-Neutralinos untersucht und gefunden,
daß die relative Emissivität (Emissivität normiert auf die zentrale Emissivität) nur schwach unterdrückt ist.

Dicus et al. haben in \cite{gravitino:97} eine unter Massenschranke für ein superleichtes Gravitino mittels
Supernova-Kühlung ausgerechnet. Sie haben als maximale Leistung $10^{52}\,\text{erg/s}$ angenommen und damit
die Gravitino-Emission nach oben begrenzt.

\subsection{Diskussion der Parameter}
\subsubsection{Abschätzung der abgestrahlten Energie}
Um eine untere Schranke für die Neutralinos zu gewinnen, braucht man eine obere Schranke für die maximal zulässige
abführbare Energie, ohne daß das Neutrinosignal gestört wird. 
Einfache Überlegungen führen zu folgender Formel für die
Bindungsenergie, die beim Einsturz einer kugelsymmetrischen Massenverteilung frei wird (aus \cite{Stars:96}):
\begin{eqnarray}
E_b = \frac{3}{5} \frac{G M^2}{r} = 1.6\cdot10^{53}\,\mathrm{erg} \left(\frac{M}{M_{\odot}}\right)^2
                                        \left(\frac{10 \mathrm{km}}{r}\right) \quad.
\end{eqnarray}

\begin{figure}
\centering
\scalebox{0.6}{\includegraphics{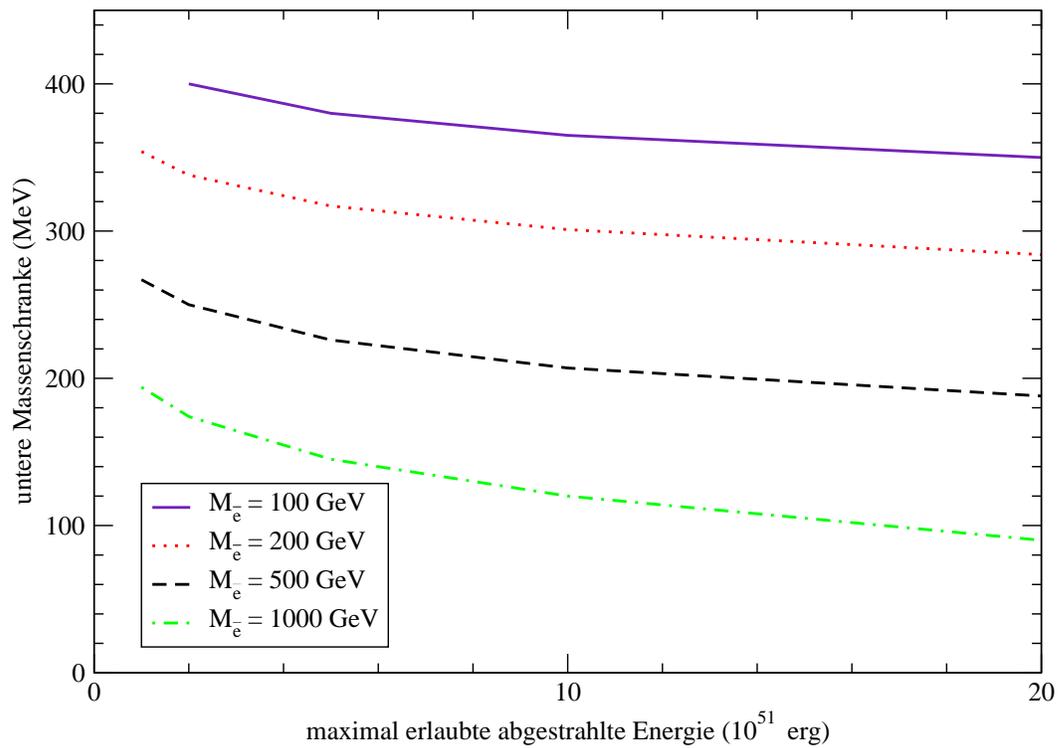}}
\caption[Energieschranke]{Abhängigkeit der Massenschranke von der maximal zulässigen abgestrahlten Energie}
\label{fig:schranke}
\end{figure}

Nimmt man einen Endradius des Cores von $13$ km wie Burrows und Lattimer in \cite{birthneutronstar:86} an, so
 erhält man als freiwerdende Bindungsenergie $3\cdot 10^{53}$ erg. Diese Energie wird in Form 
von Photonen und Neutrinos abgegeben. Obwohl eine Supernova eine ganze Galaxie überstrahlen kann,
ist die abgestrahlte Energie in 
Form von Photonen gegenüber den Neutrinos vernachlässigbar. Es werden etwa $2.4 - 3.0 \cdot 10^{53}$ erg als
Neutrinos abgestrahlt, das entspricht etwa $50000$ Erdmassen oder $0.15$ Sonnenmassen (\cite{sn:90})! Da es
sechs Neutrinoarten gibt, führt jede im Mittel $4 - 5\cdot 10^{52}$ erg ab. Damit kann ich die maximale Energie, die 
die Neutralinos abführen können, zu $E = 10^{52}$ erg mit einer Unsicherheit von der Größenordnung
$\mathcal{O}(1)$ festlegen. In Abb. \ref{fig:schranke} wird die Abhängigkeit der unteren Massenschranke von der
erlaubten abgestrahlten Energie für verschiedene Selektronenmassen dargestellt.

Für größere Energieschranken fällt die untere Massenschranke für das Neutralino bei konstanter
Selektronenmasse schwach monoton, wie man das auch erwartet.
Die Selektronenmasse hat einen erheblichen Einfluß. So liegt die untere Massengrenze bei einem leichten Selektron ($100\,\text{GeV}$)
bei $350 \ldots 400\,\text{MeV}$, bei einem schweren Selektron ($1000 \,\text{GeV}$) ist das Neutralino leichter: 
$100 \ldots 200\, \text{MeV}$.

\subsubsection{Die Zeitskala}
\begin{figure}
\vspace{5mm}
\centering
\scalebox{0.6}{\includegraphics{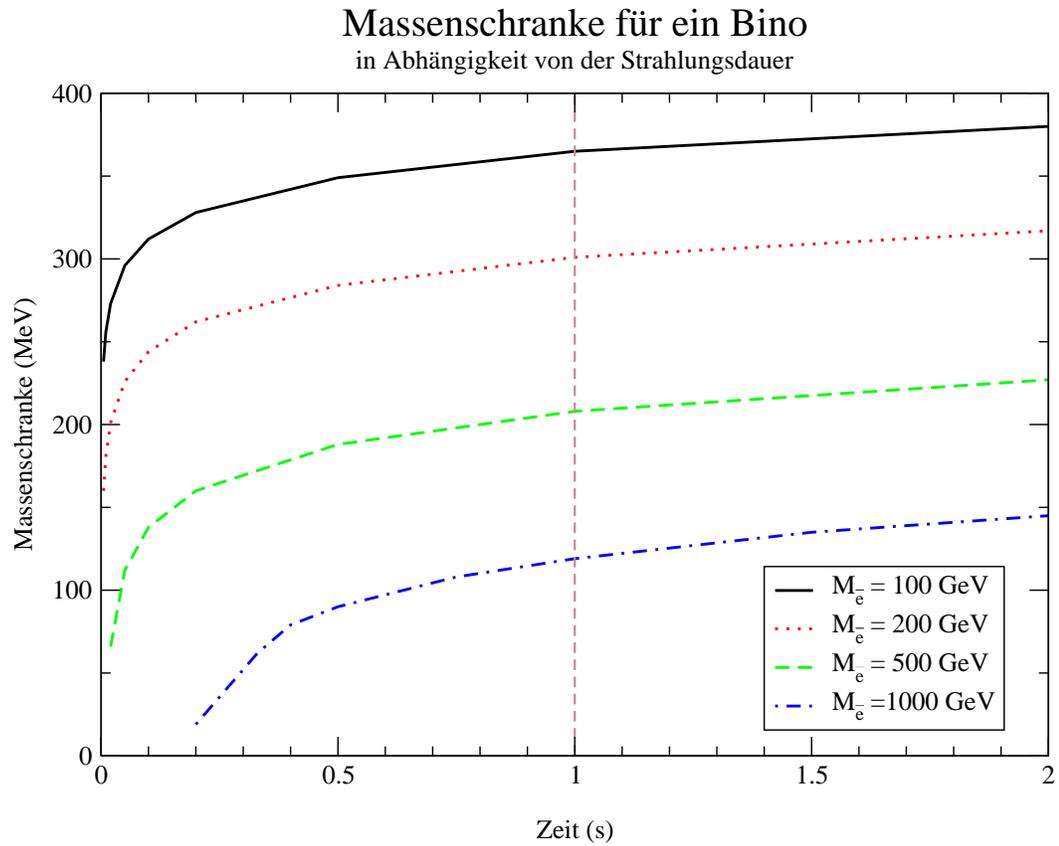}}
\vspace{5mm}
\caption{Abhängigkeit der Massenschranke von der Dauer der Energieabstrahlung}
\label{fig:dauerabst}
\end{figure}

Weiterhin muß man abschätzen, in welcher Zeit die Energie abgeführt wird. Der Kollaps findet auf einer 
Zeitskala von Bruchteilen einer Sekunde bis eine Sekunde statt.
Innerhalb einer Sekunde werden $10^{52}\,\mathrm{erg}$ abgeführt, damit hätte ein Bino (siehe Abb.
\ref{fig:bino}) eine untere Massenschranke von etwa $360 \, \mathrm{MeV}$. Vergrößert man die Abstrahlzeit 
(die maximal abgestrahlte Energie wird konstant bei $10^{52} \, \text{erg}$ gehalten) für verschiedene Selektronenmassen, so wird 
die untere Massenschranke für das Bino kleiner, wie man aus Abb. \ref{fig:dauerabst} ablesen kann.
Bei einem Selektron mit einer Masse von $1\,\text{TeV}$ ist hier keine Aussage möglich. 
Für sehr kleine Zeiten hängt die Massenschranke empfindlich von der gewählten Abstrahlzeit ab, weil in diesem Bereich die Kurven
steil verlaufen.

\subsubsection{Die Mischungsparameter}
\begin{figure}[!h]
\centering
\scalebox{0.5}{\includegraphics{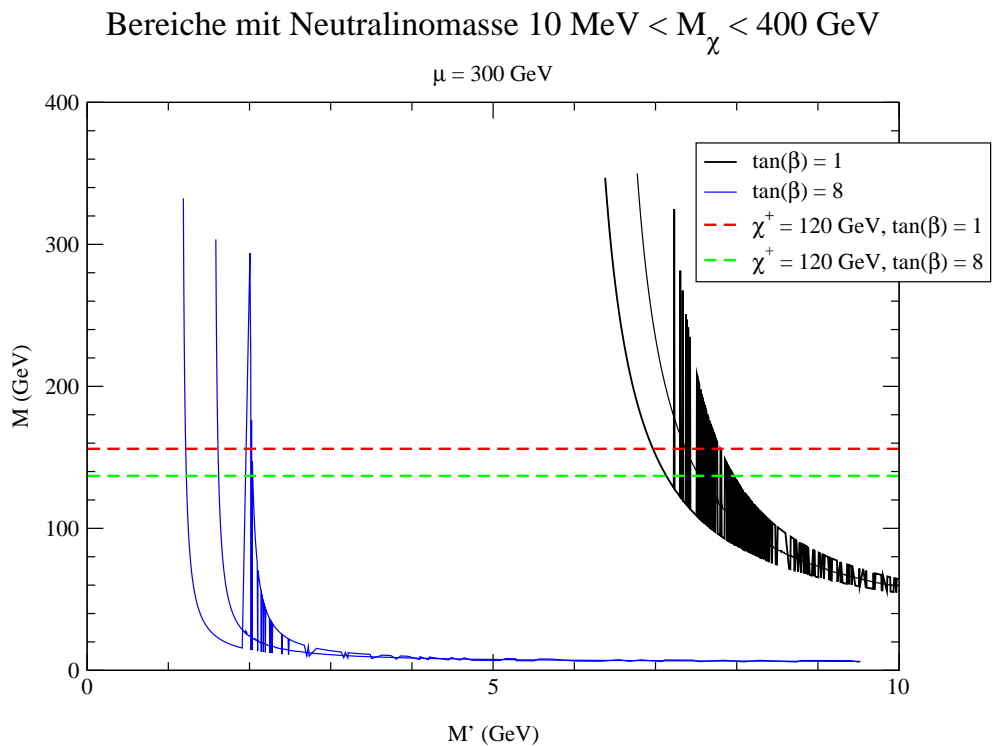}}
\caption[Bereiche mit Neutralino zwischen 10 und 400 MeV]{Bereiche mit Neutralino zwischen 10 und 400 MeV,
           oberhalb der gestrichelten Linie ist die Masse  des Chargino größer als $120$ GeV}
\label{fig:bereichn}
\end{figure}

Betrachtet man die Abhängigkeit von den Mischungsparametern, so muß man im $M-M'-\mu-\tan\beta$-Parameterraum
nach Gebieten suchen, die folgende Bedingungen erfüllen:
\begin{itemize}
\item $10\, \mathrm{MeV} \le M_{\chi^0} \le 400\,\mathrm{MeV}$ 
\item $M_{\chi^{\pm}} \ge 120\,\mathrm{GeV}$ (siehe einführendes Kapitel)
\item $\sqrt{N_{13}^2 + N_{14}^2} < 0.5 = (0.08)^{1/4}$ \\
Diese Bedingung folgt aus dem Fehler der unsichtbaren Zerfallsbreite des $Z^0$-Bosons, siehe \cite{karmen:00} 
\end{itemize} 
Eine ähnliche Suche im Parameterraum wurde in \cite{karmen:00} durchgeführt, allerdings mit einer engeren
Massenbreite und einer höheren Charginomasse. Für $\mu \le 0\, \mathrm{GeV}$ gibt es keinen zulässigen
Eigenwert im Parameterraum. Den Parameter $M_2$ muß man in sehr kleinen Schritten erhöhen, um das enge Tal
zulässiger Werte nicht zu verfehlen. Da der kleinste Eigenwert von Interesse ist und ``kleinster Eigenwert'' 
einer Matrix, die von einer Reihe von Parametern abhängt, nicht unbedingt eine differenzierbare Funktion ist, kann der
zulässige Bereich etwas sprunghaft verlaufen, wie die Abb. \ref{fig:bereichn} zeigt. 

Für eine Selektronmasse in der Nähe der $Z^0$-Masse ändert sich die Gesamtkopplung bei einer Variation der 
Mischungsparameter wenig, das heißt, man kann
von der Massenschranke nicht auf mögliche Mischungsparameter schließen. In diesem Szenario ist der 
$Z^0$-Beitrag äußerst klein. Das ändert sich, wenn die Selektronmasse hoch wird. Dann wird der Selektron\-anteil
klein gegenüber dem $Z^0$-Anteil.
Für zwei weitere mögliche Neutralinomischungen sind auf den folgenden Abbildungen die Massenschranken
dargestellt. Die Energieschranke beträgt auch hier $10^{52}\,\text{erg}$.

\begin{figure}[!h]
\centering
\scalebox{0.6}{\includegraphics{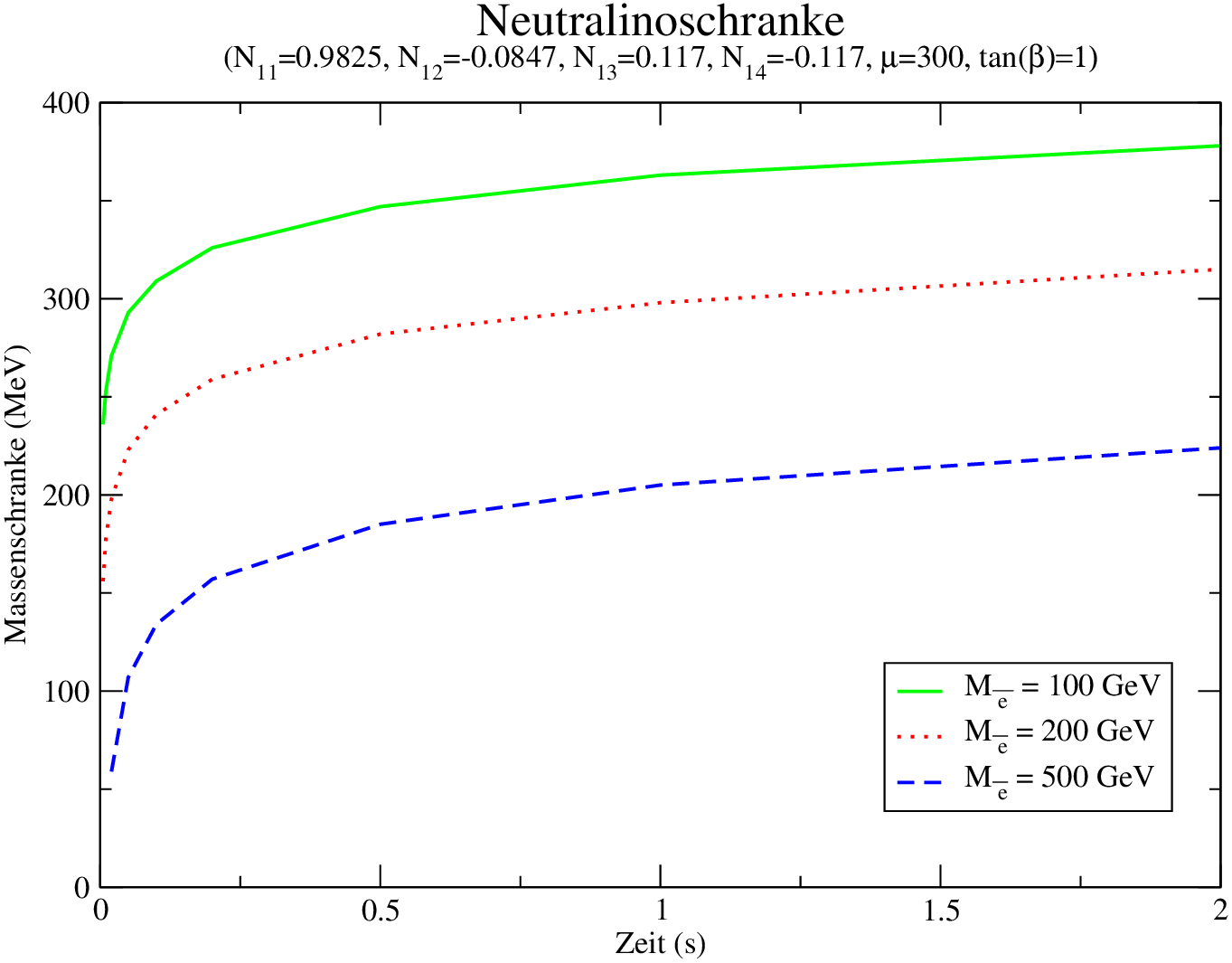}}
\caption{Szenario 1}
\label{fig:neutralino1}  
\end{figure}

\begin{figure}[!ht]
\centering
\scalebox{0.6}{\includegraphics{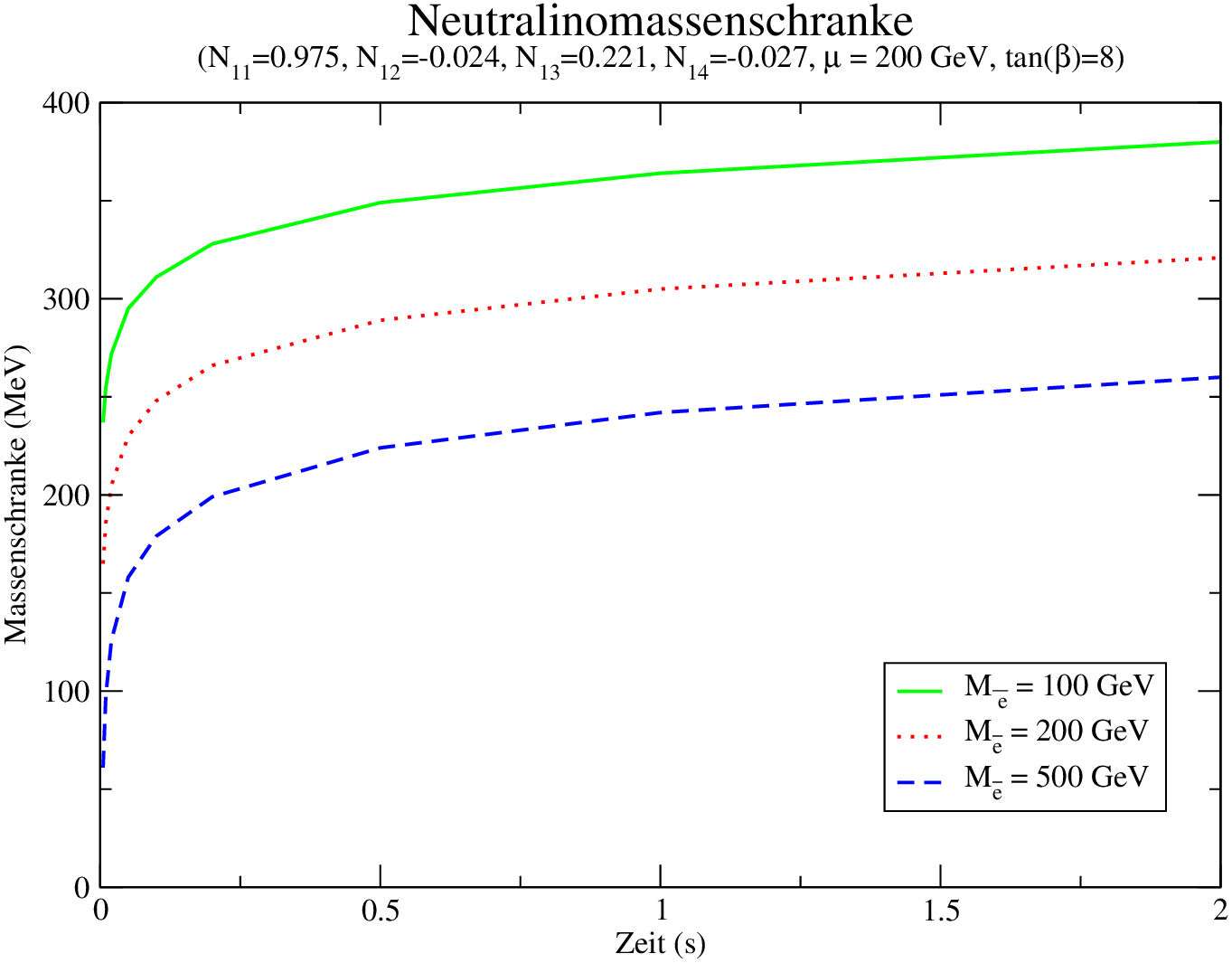}}
\caption{Szenario 2}
\label{fig:neutralino2}  
\end{figure}

Der Einfluß der Mischungsparameter $N_{13}$ und $N_{14}$ ist bei hohen Selektronmassen deutlich zu erkennen.
Bei Szenario 1 sind die letztgenannten Parameter gleich groß: $N_{13} = N_{14} = 0.117$, 
$\sqrt{N_{13}^2 + N_{14}^2} = 0.165$, im zweiten Fall ist $N_{13} = 0.221, N_{14} = 0.027, 
\sqrt{N_{13}^2 + N_{14}^2} = 0.223$ (Dies ist der größte Wert, den ich für $\sqrt{N_{13}^2 + N_{14}^2}$ 
gefunden habe.). Im ersten Fall tritt partiell Auslöschung der Parameter auf, im zweiten
Fall dagegen nicht. Außerdem spielt ein weiterer Effekt eine Rolle. Die Kopplungskonstante ist eine Summe aus $Z^0$-, 
Selektron- und Interferenzterm. Die $Z^0$-Masse ist bekannt, die Selektronmasse nicht. Bei hohen Selektronmassen 
(TeV-Bereich) spielt der Selektronterm gegenüber dem $Z^0$-Term keine Rolle mehr, da die Massen mit der inversen vierten 
Potenz eingehen, die Mischungsparameter jedoch nur quadratisch. Man erwartet also, daß die untere Massenschranke nicht 
mehr von der Selektronmasse abhängt, wenn das Selektron entsprechend schwer ist.
Der Effekt der Mischungsparameter ist in den Abbildungen \ref{fig:neutralino1} und \ref{fig:neutralino2}
bei einer Selektronenmasse von $500\,\text{GeV}$ deutlich zu erkennen, der andere dagegen nicht.
Zwischen Szenario 1 (Abb. \ref{fig:neutralino1}) und einem reinen Bino (Abb. \ref{fig:bino}, mit $t = 1\,\text{s}$)
ist kein Unterschied zu erkennen, da der Effekt der Mischungsparameter sich herauskürzt.

\subsubsection{Selektronenmasse}
\begin{figure}[!ht]
\centering
\scalebox{0.6}{\includegraphics{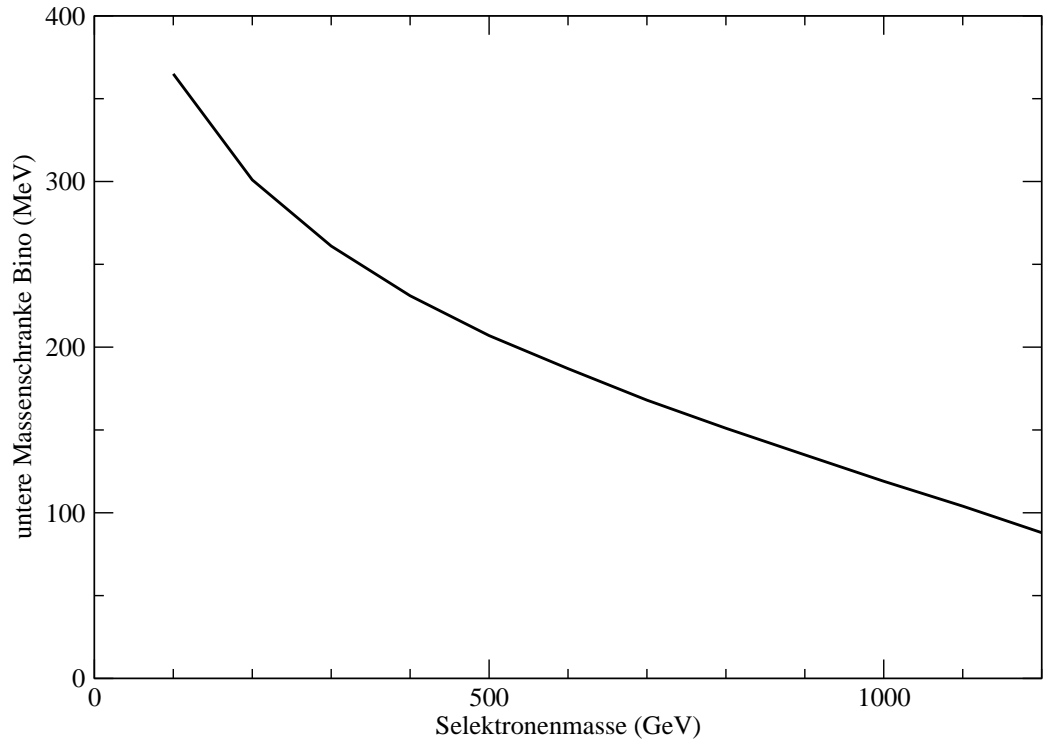}}
\caption[Abhängigkeit von der Selektronenmasse]{Abhängigkeit der unteren Massenschranke des Bino von der 
                                                Selektronenmasse}
\label{fig:smasse}  
\end{figure}

Bisher ist die Selektronenmasse als dritter Freiheitsgrad bei den bisher diskutierten Parametern betrachtet worden. 
Nun werden diese fixiert (Energieschranke $10^{52}\,\text{erg}$, Bino) und die untere Massenschranke nur in Abhängigkeit 
der Selektronenmasse diskutiert, siehe Abb. \ref{fig:smasse}. Wie zu erwarten war, verkleinert sich die untere 
Massenschranke bei größer werdender Selektronenmasse. Für noch größere Selektronenmassen ist meine Rechnung vermutlich 
nicht mehr brauchbar. Die untere Massenschranke variiert zwischen $38\,\text{MeV}$ ($M_{\tilde{e}} = 1400\,\text{GeV}$) 
und $365\,\text{MeV}$ ($M_{\tilde{e}} = 100\,\text{GeV}$). 
 
\clearpage
\newpage

\section{Die Nukleon-Nukleon-Bremsstrahlung}

\subsection{Die Mechanismen der Nukleon-Nukleon-Bremsstrahlung}

Es gibt einen weiteren Mechanismus, der Neutralinos erzeugen kann: Nukleon-\-Nukleon-\-Brems\-strahlung.
Um die möglichen Reaktionen zu erhalten, nimmt man die Diagramme der $e^+ e^-$-Prozesse und baut sie in jedes
Beinchen der Nukleon-Nukleon-Streuung ein (Natürlich ist Elektron durch Quark und Selektron durch Squark zu ersetzen.).
Die abgestrahlte Energie in Form von Neutralinos wird als weich betrachtet, obwohl dies für hohe Massen nicht zutrifft.
Aber für eine Größenordnungsabschätzung ist diese Methode trotzdem angemessen. 
Die Diagramme für die Kopplung der Neutralinos an die Quarks sind in Tabelle \ref{tab:nukleon} aufgelistet.
Später wird von den Quarks zu Nukleonen übergegangen.

\unitlength=1mm
\begin{fmffile}{diagramm4}

\begin{table}[htbp]
\begin{center}
\begin{tabular}{c c}
\vspace{8mm}

\begin{fmfgraph*}(64,48)
\fmfpen{thick}
\fmfpoly{smooth,shade}{g1,g2}
\fmftop{a1,a2}
\fmfforce{(.57w,.8h)}{g2}
\fmfforce{(.57w,.48h)}{g1}
\fmfforce{(.05w,.85h)}{a1}
\fmfforce{(.95w,.85h)}{a2}
\fmfbottom{b1,b2,b3,b4}
\fmfforce{(.05w,.42h)}{b1}
\fmfforce{(.2w,.2h)}{b2}
\fmfforce{(.4w,.2h)}{b3}
\fmfforce{(.95w,.42h)}{b4}
\fmf{quark}{a1,g2} \fmf{quark}{g2,a2}
\fmf{quark}{b1,c1,b2} \fmf{quark}{b3,c2,g1} \fmf{quark}{g1,b4}
\fmfforce{(.22w,.44h)}{c1}
\fmfforce{(.38w,.46h)}{c2}
\fmfdot{c1} \fmfdot{c2} 
\fmf{dashes_arrow,label=$\tilde{q}_{L/R}$}{c1,c2}
\fmflabel{$q$}{a1} \fmflabel{$q$}{a2}
\fmflabel{$q$}{b1} \fmflabel{$q$}{b4}
\fmflabel{$\chi$}{b2} \fmflabel{$\chi$}{b3}
\end{fmfgraph*}
\hspace{1cm}
&
\begin{fmfgraph*}(64,48)
\fmfpen{thick}
\fmfpoly{smooth,shade}{h1,h2}
\fmftop{d1,d2}
\fmfforce{(.57w,.8h)}{h2}
\fmfforce{(.57w,.48h)}{h1}
\fmfforce{(.05w,.85h)}{d1}
\fmfforce{(.95w,.85h)}{d2}
\fmfbottom{e1,e2,e3,e4}
\fmfforce{(.05w,.42h)}{e1}
\fmfforce{(.13w,.05h)}{e2}
\fmfforce{(.47w,.05h)}{e3}
\fmfforce{(.95w,.42h)}{e4}
\fmf{quark}{d1,h2} \fmf{quark}{h2,d2}
\fmf{quark}{e1,f1,h1,e4}
\fmfdot{f1}
\fmflabel{$q$}{d1} \fmflabel{$q$}{d2}
\fmflabel{$q$}{e1} \fmflabel{$q$}{e4}
\fmffreeze
\fmfforce{(.31w,.22h)}{f2} \fmfdot{f2}
\fmf{boson,label=$Z^0$}{f1,f2}
\fmf{quark}{e2,f2,e3}
\fmflabel{$\chi$}{e2} \fmflabel{$\chi$}{e3}
\end{fmfgraph*}
\\
\begin{fmfgraph*}(64,48)
\fmfpen{thick}
\fmfpoly{smooth,shade}{ga1,ga2}
\fmftop{aa1,aa2}
\fmfforce{(.57w,.8h)}{ga2}
\fmfforce{(.57w,.48h)}{ga1}
\fmfforce{(.05w,.85h)}{aa1}
\fmfforce{(.95w,.85h)}{aa2}
\fmfbottom{ba1,ba2,ba3,ba4}
\fmfforce{(.05w,.42h)}{ba1}
\fmfforce{(.17w,.17h)}{ba2}
\fmfforce{(.43w,.17h)}{ba3}
\fmfforce{(.95w,.42h)}{ba4}
\fmf{quark}{aa1,ga2} \fmf{quark}{ga2,aa2}
\fmf{quark}{ba1,ca1,ba3} \fmf{quark}{ba2,ca2,ga1} \fmf{quark}{ga1,ba4}
\fmfforce{(.22w,.44h)}{ca1}
\fmfforce{(.38w,.46h)}{ca2}
\fmfdot{ca1} \fmfdot{ca2} 
\fmf{dashes_arrow,label=$\tilde{q}_{L/R}$}{ca1,ca2}
\fmflabel{$q$}{aa1} \fmflabel{$q$}{aa2}
\fmflabel{$q$}{ba1} \fmflabel{$q$}{ba4}
\fmflabel{$\chi$}{ba2} \fmflabel{$\chi$}{ba3}
\end{fmfgraph*}
\hspace{1cm}
&
\\
\end{tabular}
\caption{Nukleon-Nukleon-Bremsstrahlungsprozesse}
\label{tab:nukleon}
\end{center}
\end{table}

\end{fmffile}

Die Feynmanregeln sind im Anhang \ref{app:sigma} zusammengestellt. Die Matrixelemente haben die Form {$P_R \otimes P_L$} bzw.
$P_L \otimes P_R$, falls ein Squark ausgetauscht wird, und {$(c_L P_L + c_R P_R) \otimes (L_1 P_L + R_1 P_R)$},
falls ein $Z^0$-Boson abgestrahlt wird. Weil die Sterntemperatur im Vergleich zur $Z^0$- und zur Squarkmasse
klein ist (30 MeV gegenüber $\approx$ 100 GeV), werden die Impulse in den Propagatoren vernachlässigt.
Dadurch entsteht eine effektive Vierpunktwechselwirkung. Anschließend werden die Ströme durch eine Fierztransformation
umgeordnet, so daß hadronischer Strom und Neutralinostrom miteinander wechselwirken. Unter Verwendung der 
$SU(6)$-Wellenfunktionen werden die Quarks in die Neutronen eingebunden. Dann wird durch den Neutralen Strom normiert, um 
nichtperturbative Effekte der Starken Wechselwirkung zu berücksichtigen. Dieser Prozeß ist in Tab. \ref{tab:trafo} 
graphisch dargestellt.
Die Neutralinos können an allen vier äußeren Beinchen des Graphen abgestrahlt werden.
 
\subsection{Der Neutralinotensor}
\subsubsection{Näherungen}
Um die Bremsstrahlungsrate zu berechnen, nimmt man die Prozesse ohne Strahlung und fügt nacheinander an jedes
Beinchen die Strahlung an. 
Die Nukleonen werden nichtrelativistisch behandelt, die Neutralinos dagegen relativistisch. 
Durch die nichtrelativistische Näherung für die Nukleonen vereinfachen sich die möglichen Kopplungsstrukturen, es gibt in diesem Fall 
nur zwei: den Einheitsoperator $\boldsymbol{1}$ und den Spinoperator $\boldsymbol{\sigma}$ 
(mehr dazu siehe Anhang \ref{app:fierz}). Da die Kopplungen denen der Neutrinos sehr ähnlich sind und Neutrinobremsstrahlung in 
\cite{axion:00} schon gerechnet wurde, die nukleonische Seite also schon bekannt gewesen ist, 
ist es notwendig, durch eine Fierztransformation Quark- und Neutralinostrom voneinander zu trennen. 
Friman und Maxwell haben in \cite{pion:79} für
die Streuung einen Pion-Austausch zugrunde gelegt. C. Hanhart et al. beziehen dagegen die Produktionsrate
weicher Strahlung (die Neutralinostrahlung wird ja näherungsweise als weich behandelt) auf die \textit{on-shell}
$NN$-Streuamplitude (s. \cite{axion:00}). 
Dazu wird die Nukleon-Nukleon-Streu-$T$-Matrix $T_{NN}$ in Partialwellen zerlegt. 
Die Partialwellen können in Beziehung zu den Phasenverschiebungen aus den Nukleon-Nukleon-Streudaten gesetzt werden.
In dem betrachteten Fall ist nur elastische Streuung kinematisch erlaubt.

Für die Berechnung des Neutralinotensors können die Ergebnisse aus dem letzten Abschnitt teilweise übernommen werden.

\begin{fmffile}{diagramm6}
\begin{table}
\begin{center}
\begin{tabular}{c c c c c c c}
\vspace{8mm}
\begin{fmfgraph*}(30,48)
\fmfpen{thick}
\fmfbottom{fb1,fb2}
\fmftop{ft1,ft2}
\fmf{quark}{fb1,fv1,ft1}
\fmf{quark}{ft2,fv2,fb2}
\fmf{dashes_arrow,label=$M_{\tilde{q}}$}{fv1,fv2}
\fmfdot{fv1,fv2}
\fmflabel{$q$}{fb1} \fmflabel{${q}$}{fb2}
\fmflabel{$\chi$}{ft1} \fmflabel{$\chi$}{ft2}
\end{fmfgraph*}
\hspace{15mm}
&
\begin{fmfgraph*}(10,48)
\fmfpen{thick}
\fmfleft{fl}
\fmflabel{$\boldsymbol{\xrightarrow{\lim M_{\tilde{q}} = \infty}}$}{fl}
\end{fmfgraph*}
&
\hspace{-15mm}
\begin{fmfgraph*}(30,48)
\fmfpen{thick}
\fmfbottom{ffb1,ffv1,ffb2}
\fmftop{fft1,ffv2,fft2}
\fmfforce{(.05w,0.1h)}{ffb1}
\fmfforce{(.55w,0.1h)}{ffb2}
\fmfforce{(.05w,0.9h)}{fft1}
\fmfforce{(.55w,0.9h)}{fft2}
\fmfforce{(.26w,.5h)}{ffv1}
\fmfforce{(.34w,.5h)}{ffv2}
\fmf{quark}{ffb1,ffv1,fft1}
\fmf{quark}{fft2,ffv2,ffb2}
\fmf{dashes}{ffv1,ffv2}
\fmfdot{ffv1,ffv2}
\fmflabel{$q$}{ffb1} \fmflabel{${q}$}{ffb2}
\fmflabel{$\chi$}{fft1} \fmflabel{$\chi$}{fft2}
\end{fmfgraph*}
&
\begin{fmfgraph*}(10,48)
\fmfpen{thick}
\fmfleft{fl}
\fmflabel{$\boldsymbol{\xrightarrow{\text{Fierz-Trafo}}}$}{fl}
\end{fmfgraph*}
&
\hspace{-15mm}
\begin{fmfgraph*}(30,48)
\fmfpen{thick}
\fmfbottom{afb1,afv1,afb2}
\fmftop{aft1,afv2,aft2}
\fmfforce{(.05w,0.1h)}{afb1}
\fmfforce{(.55w,0.1h)}{afb2}
\fmfforce{(.05w,0.9h)}{aft1}
\fmfforce{(.55w,0.9h)}{aft2}
\fmfforce{(.3w,.48h)}{afv1}
\fmfforce{(.3w,.52h)}{afv2}
\fmf{quark}{afb1,afv1,afb2}
\fmf{quark}{aft2,afv2,aft1}
\fmf{dashes}{afv1,afv2}
\fmfdot{afv1,afv2}
\fmflabel{$q$}{afb1} \fmflabel{${q}$}{afb2}
\fmflabel{$\chi$}{aft1} \fmflabel{$\chi$}{aft2}
\end{fmfgraph*}
&
\hspace{0mm}
\begin{fmfgraph*}(10,48)
\fmfpen{thick}
\fmfleft{fl}
\fmflabel{$\boldsymbol{\xrightarrow{SU(6)}}$}{fl}
\end{fmfgraph*}
&
\hspace{-15mm}
\begin{fmfgraph*}(30,48)
\fmfpen{thick}
\fmfbottom{xk1,xk2,xk3,xk4,xk5,xk6,xvv}
\fmftop{xw3,xw4,xuu}
\fmfforce{(.1w,0.10h)}{xk1}
\fmfforce{(.9w,0.10h)}{xk6}
\fmfforce{(.16w,0.10h)}{xk2}
\fmfforce{(.21w,0.10h)}{xk3}
\fmfforce{(.78w,0.10h)}{xk4}
\fmfforce{(.84w,0.10h)}{xk5}
\fmfforce{(.10w,0.9h)}{xw3}
\fmfforce{(.90w,0.9h)}{xw4}
\fmfforce{(.5w,.48h)}{xvv}
\fmfforce{(.5w,.52h)}{xuu}
\fmf{quark}{xk1,xvv,xk6}
\fmf{quark}{xw4,xuu,xw3}
\fmf{dashes}{xvv,xuu}
\fmf{quark,left=0.55,tension=0.5}{xk2,xk5}
\fmf{quark,left=0.40,tension=0.3}{xk3,xk4}
\fmfdot{xuu,xvv}
\fmflabel{$\chi$}{xw3}
\fmflabel{$\chi$}{xw4}
\fmflabel{$n$}{xk3}
\fmflabel{$n$}{xk4}
\end{fmfgraph*}

\end{tabular}
\caption{Wirkung der Transformationen}
\label{tab:trafo}
\end{center}
\end{table}
\end{fmffile}

Durch die Fierztransformation zerfallen die Betragsquadrate der Matrixelemente in einen hadronischen Tensor und einen 
Neutralinotensor: $\M = \mathcal{H}^{\mu\nu} \mathcal{N}_{\mu\nu}$.  
Der Vektorstrom trägt in führender Ordnung nicht bei, weil $\gamma^0 \rightarrow \boldsymbol{1}$ und 
$\boldsymbol{\gamma} \rightarrow 0$ in nichtrelativistischer Näherung übergehen und der Kommutator mit dem 
Identitätsoperator verschwindet. Denn die Neutralinos können vor oder nach der Streuung der Neutronen
abgestrahlt werden. In dem einen Fall ist die Energie des Zwischenzustandes positiv, im anderen Fall negativ.
Die Summe beider Möglichkeiten ergibt den Kommutator zwischen der Streu-$T$-Matrix und dem Operator der
Kopplungsstruktur.  
Der einzige Beitrag stammt von den räumlichen Komponenten des axialen Stroms. Demzufolge sind auch nur die
räumlichen Komponenten des Neutralinotensors von Interesse. 
Da der hadronische Tensor von C. Hanhart et al. in \cite{axion:00} schon berechnet worden ist, beschränke ich mich auf den Neutralinotensor.
\subsubsection{Von den Quarks zu Neutronen}
Um aus der Kopplung der Quarks an die Neutralinos die Kopplung der Neutronen an letztere zu gewinnen,
benutzen wir als einfachstes Modell die $SU(6)$-Wellenfunktionen (s. \cite{nachtmann:86}) und als Normierung den 
Neutralen Strom, so daß Renormierungseffekte der Starken Wechselwirkung berücksichtigt werden. Dies ist nur eine grobe 
Näherung und vernachlässigt die Dynamik der Quarks innerhalb des Nukleons. Da es sich hier um eine 
Größenordnungsabschätzung handelt, ist dieses Modell aber gerechtfertigt.
\begin{eqnarray}
W_{nn\chi\chi} =  (W_{nn\nu\overline{\nu}})_{NC}
               \frac{W_{q\overline{q}\chi\chi}^{SU(6)}}{(W_{q\overline{q}\nu\overline{\nu}})_{NC}^{SU(6)}} \quad.
\label{eq:qq}
\end{eqnarray}
Der effektive Vertex der $q\overline{q}\chi\chi$-Kopplung sieht nun so aus:
\begin{eqnarray}
W_{q\overline{q}\chi\chi} & = & \frac{e^2}{8 M_{\tilde{q}}} 
               \gamma_\mu \gamma_5 (l_1^2\otimes \gamma^\mu P_L - r_1^2 \otimes \gamma^\mu P_R) \label{eq:effkopp}\\
l_1^2 & = & \kappa(Y_L^2 + \Delta^2 + 2\tau_3 Y_L \Delta); \quad r_1^2 = \kappa^2 Y_R^2 \label{eq:faktoren}\\
\kappa^2 & = & (N'_{j1} - N'_{j2} \tw)^2; \quad 
\Delta = \ctw \left(\frac{\sw N'_{j1} + \cw N'_{j2}}{\cw N'_{j1} - \sw N'_{j2}}\right)\\
\tau:&& \text{3-Komponente des schwachen Isospins}
\end{eqnarray}

\begin{table}
\begin{center}
\begin{tabular}{c c c}
\toprule
& $T_3$ & Y \\
\midrule
$\begin{pmatrix}u_L\\[2mm] d_L \end{pmatrix}$ &
$\begin{matrix} \phantom{-}\frac{1}{2}\\[2mm] -\frac{1}{2}  \end{matrix}$ & 
$\begin{matrix} \phantom{-}\frac{1}{6}\\[2mm] \phantom{-}\frac{1}{6}  \end{matrix}$\\[6mm]
$u_R$ & $\phantom{-} 0$ & $\phantom{-}\frac{2}{3}$ \\[4mm]
$d_R$ & $\phantom{-} 0$ & $-\frac{1}{3}$ \\[2mm]
\bottomrule
\end{tabular}
\end{center}
\caption[Isospin und Hyperladung]{Isospin und Hyperladung von up- und down-Quark} \label{tab:hyper}
\end{table}
\noindent 
Der Erwartungswert der Hyperladung wird durch den schwachen Isospin und die Identität ausgedrückt:
\begin{eqnarray}
T_3^L & = &\frac{1}{2}\tau_3; \enspace T_3^R = 0; \enspace Y_L = \frac{1}{6}\boldsymbol{1}, \enspace 
       Y_R = \frac{1}{2}\left(\tau_3 + \frac{1}{3}\right) \\
Y_R^2 & = & \frac{1}{4} \left(1 + \frac{1}{9} + \frac{2}{3}\tau_3 \right) =
            \frac{1}{6} \left( \frac{5}{3} + \tau_3 \right) \\
l_1^2 & = & \kappa^2 \left(\frac{1}{36} + \Delta^2 + \frac{1}{3}\tau_3 \Delta \right) \\
r_1^2 & = & \kappa^2 \left(\frac{5}{18} + \frac{1}{6}\tau_3 \right) 
\end{eqnarray}
Die $SU(6)$-Wellenfunktion des Neutrons ist gegeben durch 
\begin{eqnarray}
n(1/2,1/2) & \propto & \frac{1}{\sqrt{18}}
                \left( 2(d^{\uparrow}d^{\uparrow}u^{\downarrow} +
                         d^{\uparrow}u^{\downarrow}d^{\uparrow} +
                         u^{\downarrow} d^{\uparrow}d^{\uparrow})\right.\\
        &&\quad\quad   \left. -(d^{\uparrow}d^{\downarrow}u^{\uparrow} +
                           d^{\downarrow}d^{\uparrow}u^{\uparrow} +
                           d^{\downarrow}u^{\uparrow}d^{\downarrow}+
                           d^{\downarrow}u^{\uparrow}d^{\uparrow} +
                           u^{\uparrow}d^{\uparrow}d^{\downarrow}+
                           u^{\uparrow}d^{\downarrow}d^{\uparrow})  
                         \right) \quad.
\end{eqnarray} 
In nichtrelativistischer Näherung geht $\gamma_\mu \gamma_5$ in Gl. (\ref{eq:effkopp}) aus dem Quarkstrom
in $\boldsymbol{\sigma}$ über (s. \ref{app:fierz}). Gemäß dem Wigner-Eckart-Theorem braucht man nur den Fall $\sigma_3$ 
auszuwerten:
\begin{eqnarray}
\langle n|\sigma_3|n \rangle = 1, \quad \langle n|\tau_3 \sigma_3|n \rangle = - \frac{5}{3} 
\end{eqnarray}  
und dies in (\ref{eq:faktoren}) einsetzen:
\begin{eqnarray}
\langle n|\overline{q} \Gamma^{(3)}_{\chi\chi}q|n\rangle^{(SU(6))} & = &
      \frac{e^2}{8 M_{\tilde{q}}^2}\kappa^2\left[\left
       (\frac{1}{36} + \Delta^2 -\frac{5}{9}\Delta \right)\gamma^\mu P_L
      -\left(\frac{5}{18}-\frac{1}{6} \cdot \frac{5}{3}\right)\gamma^\mu P_R \right] \notag \\
 & = & \frac{e^2}{8 M_{\tilde{q}}^2}\kappa^2\left[\left
       (\frac{1}{36} + \Delta^2 -\frac{5}{9}\Delta \right)\gamma^\mu P_L \right] \quad.
\end{eqnarray}

Die Normierung berechnet man analog zu
\begin{eqnarray}
\langle n|\overline{q} \Gamma^{(3)}_{NC} q|n\rangle^{(SU(6))} & = &
\frac{5e^2}{24\cw[2]\sw[2]M^2_{Z^{0}}}\gamma^\mu P_L \quad .
\end{eqnarray} 
weil der Vertex die Struktur $\tau_3 \sigma_3 \otimes \gamma^\mu P_L$ hat.
Einsetzen der letzten Resultate in die Ausgangsgleichung (\ref{eq:qq}) führt zu
\begin{eqnarray}
W^i_{nn\chi\chi} = \frac{C_A G_F}{\sqrt{8}} \cdot \frac{1}{5}\kappa^2 \sw[2] \cw[2]
                          \left(\frac{M_{Z^{0}}}{M_{\tilde{q}}} \right)^2
                    \left(\frac{1}{12} + 3 \Delta^2 + \frac{5}{3}\Delta \right)\gamma^i P_L \quad .
\end{eqnarray}
Zur besseren Übersicht definiert man einen Faktor $\beta$, der die Kopplungskonstanten enthält:
\begin{equation}
\sqrt{\beta} = \frac{\sqrt{2}}{5}\kappa^2 \sw[2] \cw[2]  \left(\frac{M_{Z^{0}}}{M_{\tilde{q}}} \right)^2
                    \left( \frac{1}{12} + 3 \Delta^2 + \frac{5}{3}\Delta \right) \quad.
\end{equation}
Bildung des Betragsquadrats und Kontraktion mit den Spinoren liefert
\begin{eqnarray}
\mathcal{N}^{ij} =  \frac{C_A^2 G_F^2}{8} \cdot \underbrace{\frac{2}{25}\kappa^4 \sw[4] \cw[4]
                          \left(\frac{M_{Z^{0}}}{M_{\tilde{q}}} \right)^4
                   \left(\frac{1}{12} + 3 \Delta^2 - \frac{5}{3}\Delta \right)^2}_{\beta} 
                  (\omega_1\omega_2 + M^2)\delta^{ij}\enspace,
\end{eqnarray} 
wobei berücksichtigt worden ist, daß die linear auftretenden Impulse bei der Winkelintegration verschwinden, so daß
nur die Energie- und die Massenterme beitragen. Die $Z^0$-Abstrahlung wird hier vernachlässigt.

\subsection{Die Emissivität}
Der Energieverlust in ein Phasenraumelement ist gegeben durch
\begin{eqnarray}
\dif E & = & (\omega_1 + \omega_2) \frac{\dif \mathbf{k}_1}{(2\pi)^3 2\omega_1} 
                              \frac{\dif \mathbf{k}_2}{(2\pi)^3 2\omega_2}|\overline{\M}|^2 \label{eq:phase} \\
& & \omega_1,\, \omega_2: \enspace \text{Energie der abgestrahlten Teilchen}\hspace{3cm} \notag\\
& & \mathbf{k}_1,\,\mathbf{k}_1: \enspace \text{3-Impulse der abgestrahlten Teilchen} \quad .\notag 
\end{eqnarray}

\begin{fmffile}{diagramm7}

\begin{table}[htbp]
\begin{center}
\begin{tabular}{c c c}
\vspace{8mm}

\begin{fmfgraph*}(48,36)
\fmfpen{thick}
\fmfpoly{smooth,shade}{gbx1,gbx2}
\fmftop{abx1,abx2}
\fmfforce{(.57w,.6h)}{gbx2}
\fmfforce{(.57w,.28h)}{gbx1}
\fmfforce{(.05w,.65h)}{abx1}
\fmfforce{(.95w,.65h)}{abx2}
\fmfbottom{bbx1,bbx2,bbx3,bbx4}
\fmfforce{(.05w,.22h)}{bbx1}
\fmfforce{(.2w,.85h)}{bbx2}
\fmfforce{(.4w,.85h)}{bbx3}
\fmfforce{(.95w,.22h)}{bbx4}
\fmf{quark}{abx1,cbx1,gbx2,abx2}
\fmf{quark}{bbx1,gbx1,bbx4} 
\fmfdot{cbx1} 
\fmflabel{$n$}{abx1} \fmflabel{$n$}{abx2}
\fmflabel{$n$}{bbx1} \fmflabel{$n$}{bbx4}
\fmflabel{$\underset{\phantom{a}}{\hat{O}}\phantom{-}$}{cbx1}
\fmffreeze
\fmf{quark}{bbx2,cbx1,bbx3}
\fmflabel{$\chi$}{bbx2} \fmflabel{$\chi$}{bbx3}
\end{fmfgraph*}
\hspace{1.0cm}
&
\hspace{-10mm}
\begin{fmfgraph*}(10,36)
\fmfpen{thick}
\fmfleft{xx1}
\fmfforce{(0.5w,.5h)}{xx1}
\fmflabel{\Large{$\boldsymbol{+}$}}{xx1}
\end{fmfgraph*}
\hspace{1cm}
&
\begin{fmfgraph*}(48,36)
\fmfpen{thick}
\fmfpoly{smooth,shade}{gcx1,gcx2}
\fmftop{acx1,acx2}
\fmfforce{(.43w,.6h)}{gcx2}
\fmfforce{(.43w,.28h)}{gcx1}
\fmfforce{(.05w,.65h)}{acx1}
\fmfforce{(.95w,.65h)}{acx2}
\fmfbottom{bcx1,bcx2,bcx3,bcx4}
\fmfforce{(.05w,.22h)}{bcx1}
\fmfforce{(.6w,.85h)}{bcx2}
\fmfforce{(.8w,.85h)}{bcx3}
\fmfforce{(.95w,.22h)}{bcx4}
\fmf{quark}{acx1,gcx2,ccx1,acx2}
\fmf{quark}{bcx1,gcx1,bcx4} 
\fmfdot{ccx1} 
\fmflabel{$n$}{acx1} \fmflabel{$n$}{acx2}
\fmflabel{$n$}{bcx1} \fmflabel{$n$}{bcx4}
\fmflabel{$\phantom{-}\underset{\phantom{a}}{\hat{O}}$}{ccx1}
\fmffreeze
\fmf{quark}{bcx2,ccx1,bcx3}
\fmflabel{$\chi$}{bcx2} \fmflabel{$\chi$}{bcx3}
\end{fmfgraph*}
\\
&\hspace{-5mm}{\Large $\boldsymbol{= \frac{1}{\omega}\langle [T_{NN}, \hat{O} ]\rangle}$} &\\
\vspace{5mm}
\end{tabular}
\caption[Summe zweier Diagramme]{Summe der Diagramme mit Strahlung vor und nach der Streuung}
\label{tab:summedia}
\end{center}
\end{table} 

\end{fmffile}

Wenn man die beiden Diagramme aus Tab. \ref{tab:summedia} aufsummiert, so erhält man 
$\frac{1}{\omega}\langle[T_{NN},\hat{O}]\rangle$,
da in dem einem Fall die Energie vorher, im anderen Fall nachher abgestrahlt wird. Daraus folgt, daß der 
Identitätsoperator, der ja aus dem Vektorstromanteil der Kopplung stammt, nicht beiträgt. 

In Gl. (\ref{eq:phase}) fügen wir $1 = \int \dif\,\omega \delta(\omega - \omega_1 - \omega_2)$ und erhalten:
\begin{eqnarray}
\diff{E}{\omega} & = & \left( \omega \frac{\dif \mathbf{k}_1}{(2\pi)^3 2 \omega_1}
         \frac{\dif \mathbf{k}_2}{(2\pi)^3 2 \omega_2}\delta(\omega - \omega_1 - \omega_2) N_{ij} \right) H^{ij}\\
                 & = & \frac{\omega}{(2\pi)^4} \dif \omega_1 \dif \omega_2 \delta
        (\omega - \omega_1 - \omega_2) k_1 k_2 n(\omega_1, \omega_2) \frac{16}{\omega^2}H^{ij}\\
n(\omega_1, \omega_2)& = & \frac{G_F^2 C_A^2}{8}\beta \left(\omega_1 \omega_2 + M_{\chi}^2 \right)\\
H^i_i & = & \sum_{M_s,M_{s'}}\left| \langle M_{s'} \mathbf{p}'|[S_i, T_{NN}]|M_{s} \mathbf{p} \rangle \right|^2 \\
S_i  & = & \frac{\sigma^1_i + \sigma^2_i}{2}
\end{eqnarray}
Austauschdiagramme sind mit einem Faktor $4$ berücksichtigt. 
\noindent
Die totale Emissivität eines $2\rightarrow 4$-Prozesses hat die Form
\begin{eqnarray}
\epsilon & = &  \int\! \dif \omega \!\int \left
                       (\prod_{i=1,2}\frac{\dif \mathbf{p}_i \dif \mathbf{p}'_i}{(2\pi)^6} \right)
                        S f(E_1) f(E_2)(1 - f(E_1'))(1 - f(E_2')) \times \\ 
         & & \quad   \times (2\pi)^4 \delta^{(4)}(p_1 + p_2 - k - p_1'- p_2')\diff{E}{\omega} \label{totem}\\
f(E) & = &  \frac{1}{\exp(\frac{E - \mu}{T}) + 1}, \quad S = \frac{1}{8} \quad 
             \text{statistischer Faktor \hspace{2cm}}
\end{eqnarray}  
Weiterhin wird der $3$er-Impuls der Strahlung $\mathbf{k}$ gegenüber dem Impuls der Nukleonen in den 
$\delta$-Distributionen vernachlässigt:
\begin{eqnarray}
p_N & \propto &\sqrt{2 M_N T},\quad \text{aus der nichtrelativistischen Energie}\quad  E = \frac{p_N^2}{2 M_N} = T \\
E & \approx &p_{\chi} \propto T; \quad \text{die Neutralinos sind relativistisch} \\
\frac{p_{\chi}}{p_N}& = &\frac{\sqrt{T}}{\sqrt{2M_N}} \approx 0.12 \enspace \text{mit} 
\quad T \approx 30\,\mathrm{MeV},\enspace M_N \approx 1000 \,\mathrm{MeV} \quad.
\end{eqnarray} 
Dies vereinfacht die Integrationen, die dann nur noch den Neutralinotensor betreffen. Dabei wird die sphärische 
Symmetrie und Energie-Impuls-Erhaltung ausgenutzt.
 
Im nächsten Schritt definieren wir Gesamt- und Relativimpulse sowie dimensionslose Variablen:
\begin{eqnarray}
\mathbf{P} = \mathbf{p_1 + p_2 = p'_1 + p'_2 }, \enspace \mathbf{p} = \frac{1}{2}\mathbf{(p_1 - p_2 )},
                                                \enspace \mathbf{p'} = \frac{1}{2}\mathbf{(p'_1 - p'_2 )}, \\
u^2 = \frac{\mathbf{p^2}}{2MT},\enspace u^{\prime 2} = \frac{\mathbf{p'^2}}{2MT},\enspace 
\lambda^2 = \frac{\mathbf{P^2}}{8MT},\enspace y = \frac{\mu - M}{T}, \enspace \hat{M}_{\chi} = \frac{M_\chi}{T} \quad. 
\end{eqnarray}
$T$ ist die Temperatur des Cores und damit die Temperatur des Neutronengases, $M$ die Masse eines Neutrons.
Die $z$-Achse wird entlang von $\mathbf{P}$ gewählt.
Außerdem können einige Winkelintegrationen schon ausgeführt werden. Bei hohen Temperaturen wird die 
Winkelabhängigkeit in den Fermi-Funktionen schwach, im Grenzfall völliger Entartung verschwindet diese Abhängigkeit.
Da das Neutronengas nahezu entartet ist, können die Fermi-Funktionen durch ihre winkelgemittelten Funktionen ersetzt 
werden:
\begin{eqnarray}
\int_{-1}^1 \dif \cos\gamma \left[ e^{u^2+\lambda^2+2u\lambda \cos\gamma}+1\right]^{-1}
\left[ e^{u^2+\lambda^2-2u\lambda \cos\gamma}+1\right]^{-1} = \\
\frac{e^{-(u^2+\lambda ^2)}}{2\lambda u \sinh (u^2+\lambda ^2 -y)}
\ln \left( \frac{\cosh \left( \frac{1}{2}((u+\lambda )^2-y)\right)}
{\cosh \left( \frac{1}{2}((u-\lambda )^2-y)\right)} \right) \quad.
\end{eqnarray}
Genauso wird ein winkelgemittelter hadronischer Tensor eingeführt:
\begin{eqnarray}
\overline{\mathcal{H}}_{ii} = \int \dif \Omega \dif \Omega' \mathcal{H}_{ii} \quad ,
\end{eqnarray}
diese Integrationen können analytisch ausgeführt werden, da die $T$-Matrix als Partialwellenzerlegung vorliegt.
Übrig bleibt eine vierdimensionale Integration:
\begin{eqnarray}
\epsilon & = & \beta \frac{2 C_A^2 G_F^2}{(2 \pi)^{11}}(2 M_\chi)^{4.5}T^{4.5}M_\chi^{4} 
\int_{\hat{M_{\chi}}}^{\infty}\dif \delta e^{-\delta} h(\delta/{\hat{M_{\chi}}})
         \int_\delta^{\infty} \dif \sigma F(\delta,\sigma) \overline{\mathcal{H}}(T\sigma)_{ii},
\end{eqnarray}
mit den Abkürzungen
\begin{eqnarray}
\delta & = & \frac{p^2-p'^2}{2MT}, \quad \sigma = \frac{p^2 + p'^2}{2MT}, \\
h(x) & = & \frac{(x-1)^3}{x}[(x^2 - 1) I_0(x) - (x - 1)^2 I_2(x)], \\
I_k(x) & = & \int_{-1}^1 \dif t t^k \sqrt{1-t^2}\sqrt{\left(\frac{x + 1}{x - 1} \right)^2 -t^2}, \\
F(\delta,\sigma) & = & \int_0^{\infty} \dif \lambda L\left(\sqrt{\frac{1}{2}(\sigma + \delta)},\lambda \right)
                                                    L\left(\sqrt{\frac{1}{2}(\sigma - \delta)},\lambda \right),\\
L(a,\lambda) & = & \frac{1}{\sinh(a^2 + \lambda^2 - y)}
                 \ln\left(\frac{\cosh(\frac{1}{2}((\lambda+a)^2-y))}{\cosh(\frac{1}{2}((\lambda+a)^2-y))} \right)
                 \quad .
\end{eqnarray} 
Die Funktionen $I_k(x)$ lassen sich durch Taylorentwicklung der Wurzelterme bis zur dritten Ordnung in $(t/x)^2$  
recht genau analytisch berechnen. Die Entwicklung hat sogar bei $x=1$ nur einen Fehler von $5 \%$ auf das exakte
Ergebnis. Alle anderen Integrationen werden numerisch mittels verallgemeinerter Gauß-Integration ausgeführt.
Als Entartung $\eta$ folgt für $T = 30\,\text{MeV}$ und Kerndichte  ein Wert von $1.3$.
\subsubsection{Das Endergebnis}
\begin{figure}
\centering
\scalebox{0.5}{\includegraphics{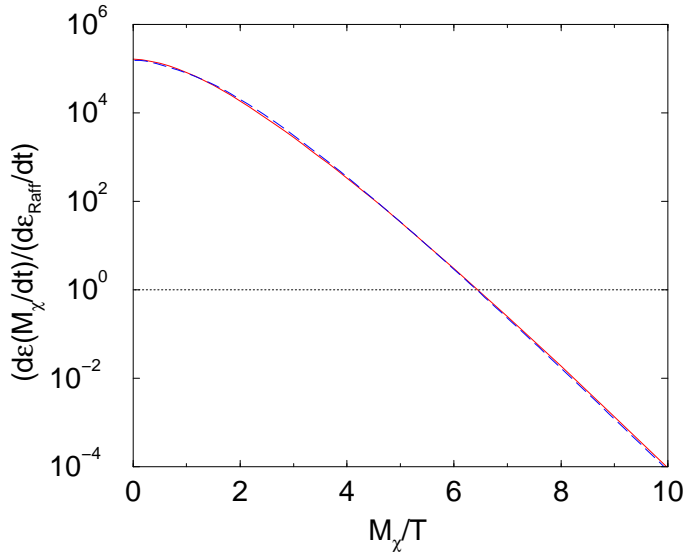}}
\caption{Emissivität der Bremsstrahlung}
\label{fig:ergebnisbrems}
\end{figure}

Das Endergebnis läßt sich mit hoher Genauigkeit durch
\begin{eqnarray}
\epsilon = \alpha (1.5 + 6 \hat{M}_\chi + 8.6 \hat{M}_\chi^3 )e^{-3\hat{M}_\chi}
\times10^{24}\frac{\mathrm{erg}}{\mathrm{g}\cdot \mathrm{s}},\enspace T_{\text{Core}} = 30 \,\text{MeV}
\end{eqnarray} 
approximieren (s. Abb. \ref{fig:ergebnisbrems}). 

Wegen der \textit{soft-radiation}-Näherung ist dieses Ergebnis für alle Felder gültig, die an den axialen Strom koppeln.
Man muß nur das entsprechende $\beta$ bestimmen.

\subsubsection{Massenschranke für ein Bino}

Mit diesem Resultat kann eine untere Massenschranke für ein Bino in Abhängigkeit der Squarkmasse  berechnet werden, 
siehe Tab. \ref{tab:bremsschranke}.

\begin{table}
\begin{center}
\begin{tabular}{c c c}
\toprule
Temperatur & Squarkmasse & Massenschranke\\
 (MeV)     &  (GeV)      &    (MeV)      \\
\midrule
30         & 100         & 147           \\
30         & 200         & 110           \\
30         & 500         &  49            \\
\bottomrule 
\end{tabular}
\caption{Untere Massenschranke}
\label{tab:bremsschranke}
\end{center}
\end{table}

Dieses Ergebnis widerspricht nicht der Existenz eines leichten Neutralinos mit einer Masse von $34 \,\mathrm{MeV}$,
wie es z. B. in \cite{karmen:00} vorgeschlagen wird, wenn die Squarkmasse in der Größenordnung von einigen TeV liegt.
Im Vergleich zu den unteren Massenschranken, die man aus dem Elektron-Positron-Kanal erhält, sind die letzteren etwas 
kleiner, liegen aber in derselben Größenordnung.

\chapter{Diskussion}

In diesem Kapitel möchte ich die Ergebnisse und die in sie einfließenden Annahmen aus dem vorhergehenden 
Abschnitt eingehender untersuchen.
Die beiden Verfahren sollen miteinander verglichen werden, und die Neutralinoemissivität soll der 
Neutrinoemissivität gegenüber gestellt werden.

\section{Diskussion des Elektron-Positron-Kanals}

In die Berechnung der Emissivität des Elektron-Positron-Kanals sind eine Reihe von Eingangsdaten eingegangen,die im
folgenden noch etwas ausführlicher diskutiert werden:
\begin{enumerate}
\item Temperatur- und Entartungsprofil einer Supernova
\item Coredichte 
\item Dauer der Neutralinoemission und Menge der abgestrahlten Energie
\item Selektronmasse
\item Mischungsparameter
\item Näherungen
\end{enumerate}

\subsubsection{Temperatur- und Entartungsprofil einer Supernova}
Das Temperatur- und Entartungsprofil einer Supernova stammt aus \cite{birthneutronstar:86}.
Die Autoren haben eine Supernova simuliert, dazu mußten sie Anfangs- und Randwerte annehmen, um
die Differentialgleichungen zu lösen. Über die Anfangswerte einer Supernova-Explosion weiß man wenig.
Kleine Änderungen der Anfangsdaten führen zur gleichen Entwicklung nach etwa einer halben Sekunde.
Die Supernova befindet sich nach dieser Zeit im Diffusionsgleichgewicht. Als Temperatur- und Entartungsprofil
habe ich die Profile nach einer halben Sekunde genommen. Man kann somit vermuten, daß dies vernünftige Profile sind.
Für meine Rechnungen habe ich die Profile über die Zeit als konstant angenommen (es liegen auch nicht mehr Daten 
im relevanten Zeitintervall vor), betrachtet man deren Abbildungen, 
so stellt man fest, daß die Änderungen nach einer weiteren halben Sekunde nicht 
allzu groß sind, vor einer halben Sekunde allerdings ist die Änderung, aber auch die Unsicherheit groß.
Diese Profile sind ungefähre Mittelwerte für die erste Sekunde.

\subsubsection{Coredichte}
Da die Verteilungen in Abhängigkeit der eingeschlossenen Baryonenmasse angegeben sind, muß die Volumenintegration
in eine Massenintegration transformiert werden. Dafür wird die Coredichte benötigt. Da das Core Kerndichte erreicht
hat und diese Dichte bekannt ist, stellt dies kein Problem dar. Die Dichte wird konstant gehalten, was nur nicht ganz richtig
ist. Weil aber die Emissivität aus einer dünnen Kugelschale stammt, macht das nichts aus.

\subsubsection{Dauer der Neutralinoemission und Menge der abgestrahlten Energie}
Die abgestrahlte Gesamtenergie kann einfach aus einer kollabierenden, kugelsymmetrischen Massenverteilung
abgeschätzt werden. Das meiste davon wird in Form von Neutrinos abgestrahlt. Ein Teil allerdings soll in Form 
von Neutralinos abgeführt werden. Da anzunehmen ist, daß die Neutralinos nicht wie die Neutrinos gefangen werden,
können erstere die Supernova sofort verlassen. Wir wissen aber, daß der Hauptteil der Energie von Neutrinos 
abgeführt wird. Also dürfen die Neutralinos nur eine kleine, im Verhältnis zu den Neutrinos vernachlässigbare Störung
darstellen. Die Neutrinos führen einige $10^{53} \, \mathrm{erg}$ ab, die Neutralinos sollen maximal etwa 
$10^{52} \, \mathrm{erg}$ abführen. Da die Unsicherheit bezüglich dieser Zahl groß ist, habe ich 
verschiedene Zeiten gewählt, um aus der Leistung $P = \int \dif M_{\text{Stern}} \epsilon$ die Energie 
$E = P \Delta t$ zu berechnen. Vergleicht man dies mit dem Kriterium von Raffelt (\cite{Stars:96}), der eine 
Höchstemissivität von $10^{19} \, \mathrm{erg}/(\mathrm{g}\mathrm{s})$ für zulässig hält, ohne daß das 
Neutrinosignal gestört wird, so ist meine Schranke um einen Faktor Drei niedriger 
($10^{19}\, \mathrm{erg}/(\mathrm{g}\mathrm{s}) \cdot 
2,8\cdot 10^{33}\,\mathrm{g} \cdot 1\,\mathrm{s} = 2.8\cdot 10^{52} \, \mathrm{erg}$).
Bei diesem Vergleich muß man vorsichtig sein, weil dieses Kriterium auf diesen Prozeß nicht anwendbar ist.
Die Neutralinos werden im Gegensatz zu den Neutrinos nicht verzögert emittiert. Dies ist aber eine Voraussetzung
zur Anwendbarkeit des Raffelt-Kriteriums.  
Da ich von einer Emissionsdauer von etwa einer Sekunde ausgehe, habe ich die Profile nach einer halben Sekunde 
gewählt, so daß diese Profile ein zeitliches Mittel darstellen.  

\subsubsection{Selektronmasse}
Über die Selektronmasse wissen wir nur, daß sie mindestens im $100 \, \mathrm{GeV}$-Bereich liegen muß. Ich habe 
drei verschiedene Werte für die Selektronmasse gewählt und eine Massenschranke
berechnet. Wird die Selektronenmasse zu groß, wird die Berechnung einer unteren Massenschranke nicht mehr möglich.
Ein Vorteil der Rechnung ist, daß die Selektronmasse und die Mischungsparameter (s. u.) in einem Faktor stecken, 
der nicht in die Integration eingeht. 

Wenn die Selektronmasse im TeV-Bereich liegt, steht ein leichtes Neutralino ($30 \,\mathrm{MeV}$, \cite{karmen:00}) 
nicht im Widerspruch zu Supernova-Ereignissen. 

\subsubsection{Mischungsparameter}
Ebenso unbekannt sind die Mischungsparameter $N_{1j}$ (Mischungsmatrix) der Neutralinos. Aus den Experimenten 
ergeben sich aber Schranken an diese Parameter (siehe vorheriges Kapitel). Aus der Supernova lassen sich keine 
Folgerungen darüber ziehen. Der 
Einfluß auf die Kopplung ist gering, solange die Sfermionmassen im Bereich der $Z^0$-Masse liegen. Bei 
Sfermionmassen im TeV-Bereich wird der Einfluß der $N_{13/4}$ größer.

\subsubsection{Näherungen}
Bei der Berechnung des Wirkungsquerschnitts sind eine Reihe von Näherungen gemacht worden: Die Elektronmasse wurde zu Null 
gesetzt, die Impulsüberträge in den Propagatoren  
vernachlässigt und die Massen von Selektron(L) und Selektron(R) als gleich angenommen. Die ersten beiden Näherungen 
sind durch die Kleinheit der Größen gerechtfertigt. Die letzte Näherung ist eine Vereinfachung, die mit der Supergravitation
gerechtfertigt werden kann.   

\section{Die Nukleon-Nukleon-Bremsstrahlung}

Einige Annahmen, die für den Elektron-Positron-Kanal gemacht werden müssen, fallen hier weg, da bei der 
Nukleon-Nukleon-Bremsstrahlung Emissivitäten mit dem Raffelt-Kriterium verglichen werden. 
Die Partialwellen der Nukleon-Nukleon-Streuung sind dem Experiment entnommen. Die Streuung wird realistischer
behandelt. Für die Näherungen gilt dasselbe wie für den Elektron-Positron-Kanal.

Die Unbekannten stecken in dem Faktor $\beta$: die Mischungsparameter und die Squarkmasse (der $Z^0$-Kanal wurde 
in dieser Rechnung vernachlässigt). $\beta$ verhält sich wie 
\begin{eqnarray}
\beta = \left(\frac{M_{Z^{0}}}{M_{\tilde{q}}} \right)^4 \times \mathcal{O}(1) \quad .
\end{eqnarray} 
(Die $Z^0$-Massenabhängigkeit stammt aus der Normierung durch den Neutralen Strom). 
Auch hier ist die wesentliche Abhängigkeit durch die Squarkmasse gegeben. Ein leichtes (auch masseloses) Neutralino 
ist möglich, wenn die Squarkmasse im TeV-Bereich liegt. 

\subsubsection{Soft-radiation-Näherung}

Die abgestrahlten Neutralinos wurden als soft behandelt. Dies ist schon für leichte Neutralinos ($\approx 30\,\text{MeV}$) problematisch.
Die Nukleon-Nukleon-Streuung spielt sich auf einer Längenskala ab, die einer Pionmasse ($\approx 100\,\text{MeV}$) entspricht.
Der Entwicklungsparameter der Näherung $\frac{\M_\chi}{m_\pi}$ ist also mit $\approx \frac{1}{3}$ schon für ein leichtes Neutralino nicht 
besonders klein. Damit ist diese Näherung nicht gerechtfertigt. Aber es gibt nichts Besseres.
Außerdem wurden Effekte vernachlässigt, die durch das Medium Kernmaterie verursacht werden.

\section{Vergleich mit der Neutrinoluminosität} 
    

In Abb. 2 aus \cite{sn:90} ist die Neutrinoluminosität in den ersten Millisekunden nach dem Kollaps dargestellt. 
Der Peak bei $113\,\mathrm{ms}$ stammt von den durch die Schockwelle ausgetriebenen Elektron-Neutrinos. Die 
Luminosität erreicht $6\cdot 10^{53}\, \mathrm{erg/s}$, die im Peak abgestrahlte Energie beträgt aber nur 
$2 \cdot 10^{51}\,  \mathrm{erg}$, das Spektrum ist ein Einfangspektrum, kein (pseudo-)thermisches und damit auch 
kein Schwarzer Strahler. Zu derselben Zeit setzt abrupt die Emission
von Myon-Neutrinos und Elektron-Antineutrinos ein, die kleinere Luminositäten erreichen. 
Nach dem Durchlaufen der Schockwelle ist das Spektrum ein pseudothermisches.

Da das Core für Neutralinos durchsichtig ist, kann man solche Peaks in der Luminosität für die Neutralinos nicht 
erwarten. Die Emissivität wird vielmehr durch das Zusammenspiel von Core-Temperatur und dem chemischen Potential von
Elektronen und Positronen getrieben. Die Emissivität findet in meinem Modell einer Schale bei einer eingeschlossenen
Masse von einer Sonnenmasse statt. An der Stelle ist die Temperatur hoch und das chemische Potential der Elektronen 
niedrig.   

In \cite{anomaly:00} wurde die Neutralinoemissivität bezogen auf das Corezentrum für drei verschiedene Entartungen 
berechnet (das hat den Vorteil, daß Mischungsparameter, Temperatur, Selektronmasse nicht gebraucht werden). Für ein 
Neutralino mit einer Masse von $34 \, \mathrm{MeV}$ ist die Emissivität kaum unterdrückt. Kachelrieß fordert in 
\cite{anomaly:00} eine deutliche Unterdrückung ($M_\chi /T > 10$), folglich muß entweder die Temperatur niedrig sein 
oder die Masse des Neutralino hoch.  
\paragraph{Anmerkung} Abb. 1 in \cite{anomaly:00} konnte von mir nicht exakt reproduziert werden. Für größere 
$x = M_\chi /T$-Werte habe ich höhere relative Emissivitäten erhalten: Bei $x=2$ ergibt sich $R = \epsilon(x)/\epsilon(0) \approx 0.7$
statt $R \approx 0.6$ wie bei Kachelrieß.

\appendix
\chapter{Der totale Wirkungsquerschnitt
 $\sigma(\lowercase{e}^+\lowercase{e}^- \rightarrow \chi^0_1 \chi^0_1)$} \label{app:sigma}
In diesem Abschnitt soll die Berechnung des totalen Wirkungschnitts des Prozesses
$e^+e^- \rightarrow \chi^0_1 \chi^0_1$ dargestellt werden. Dabei stütze ich mich auf die allgemeinere Rechnung
von A. Bartl et al. (\cite{production:86}).
\section{Beteiligte Diagramme}
Folgende Prozesse tragen in niedrigster Ordnung zum Wirkungsquerschnitt bei:
\unitlength=1mm
\vspace{10mm}
\begin{fmffile}{diagramm1}

\begin{table}[htbp]
\begin{center}
\begin{tabular}{c c c}
\vspace{8mm}
\begin{fmfgraph*}(28,21)
\fmfpen{thick}
\fmfleft{i1,i2} \fmfright{o1,o2}
\fmf{fermion}{o1,v1,i1} \fmf{fermion}{i2,v2,o2}
\fmf{dashes,label=$\tilde{e}_{L/R}$}{v1,v2} \fmfdot{v1,v2}
\fmflabel{$e^+, p_1$}{i1}
\fmflabel{$e^-, p_2$}{i2}
\fmflabel{$\chi^0_1, k_1$}{o1}
\fmflabel{$\chi^0_1, k_2$}{o2}
\end{fmfgraph*}

\hspace{20mm}

\begin{fmfgraph*}(28,21)
\fmfpen{thick}
\fmfleft{in1,in2} \fmfright{on1,on2}
\fmf{fermion}{on1,vn1,in1} \fmf{fermion}{in2,vn2,on2}
\fmf{dashes,label=$\tilde{e}_{L/R}$}{vn1,vn2} \fmfdot{vn1,vn2}
\fmflabel{$e^+, p_1$}{in1}
\fmflabel{$e^-, p_2$}{in2}
\fmflabel{$\chi^0_1, k_1$}{on2}
\fmflabel{$\chi^0_1, k_2$}{on1}
\end{fmfgraph*}

\hspace{20mm}

\begin{fmfgraph*}(28,21)
\fmfpen{thick}
\fmftop{im1,im2} \fmfbottom{om1,om2}
\fmf{fermion}{om1,vm1,im1} \fmf{fermion}{im2,vm2,om2}
\fmf{boson,label=$Z^0$}{vm1,vm2} \fmfdot{vm1,vm2}
\fmflabel{$e^+, p_1$}{im1}
\fmflabel{$e^-, p_2$}{om1}
\fmflabel{$\chi^0_1, k_1$}{im2}
\fmflabel{$\chi^0_1, k_2$}{om2}
\end{fmfgraph*}

\end{tabular}
\end{center}
\end{table}
\end{fmffile}

\section{Feynmanregeln}
Die Feynmanregeln für die Vertizes des Standardmodells sind \cite{element:84} entnommen, bei den Regeln für
die SUSY-Teilchen beziehe ich mich auf \cite{super:85} und \cite{higgs:86}. Die Elektronmassen werden
vernachlässigt sowie die Impulsüberträge gegenüber den Massen des $Z^0$-Bosons und des Selektrons. Weiter sollen
links-chirales und rechts-chirales Selektron (d. h. das dazugehörige Standardmodellelektron ist links- bzw. 
rechts-chiral) in ihren Massen entartet sein. Als Neutralinobasis wird die Photinobasis verwendet.

\vspace{8mm}

\begin{fmffile}{diagramm2}

\begin{table}[htbp]
\begin{center}
\begin{tabular}[h]{c l}
\toprule
Graph des Vertex & \hspace{25mm}Vertexfaktor \\
\midrule
& \\[5mm]
\begin{fmfgraph*}(28,21)
\fmfpen{thick}
\fmfleft{ii1,oi1}
\fmfright{vi2}
\fmf{fermion}{oi1,vi1,ii1}
\fmf{scalar, label=$\tilde{\ell}_L$}{vi2,vi1}
\fmfdot{vi1}
\fmflabel{$\ell$}{ii1}
\fmflabel{$\chi^0_1$}{oi1}
\end{fmfgraph*} & 
\begin{large}
$- \text{i} \sqrt{2} [e e_{\ell} N_{11} + \frac{g}{\cw}(\frac{1}{2} - e_{\ell} \sw[2])N_{12}] P_R$
\end{large}
 \\
&  \\[5mm]
 
\begin{fmfgraph*}(28,21)
\fmfpen{thick}
\fmfleft{ig1,og1}
\fmfright{vg2}
\fmf{fermion}{og1,vg1,ig1}
\fmf{scalar, label=$\tilde{\ell}_R$}{vg2,vg1}
\fmfdot{vg1}
\fmflabel{$\ell$}{ig1}
\fmflabel{$\chi^0_1$}{og1}
\end{fmfgraph*} & 
\begin{large}
$\text{i}  \sqrt{2} [e e_{\ell} N_{11}^* - e_{\ell}\frac{g \sw[2]}{\cw} N_{12}^*]P_L$
\end{large}
 \\
& \\[5mm]
\begin{fmfgraph*}(28,21)
\fmfpen{thick}
\fmfright{ir1,or1}
\fmfleft{vr2}
\fmf{fermion}{ir1,vr1,or1}
\fmf{scalar, label=$\tilde{\ell}_L$}{vr1,vr2}
\fmfdot{vr1}
\fmflabel{$\ell$}{ir1}
\fmflabel{$\chi^0_1$}{or1}
\end{fmfgraph*} & 

\begin{large}
$- \text{i} \sqrt{2} [e e_{\ell} N_{11} + \frac{g}{\cw}(\frac{1}{2} -e_{\ell}  \sin^2\theta_W)N_{12}] P_L$
\end{large}
 \\
& \\[5mm]
\begin{fmfgraph*}(28,21)
\fmfpen{thick}
\fmfright{it1,ot1}
\fmfleft{vt2}
\fmf{fermion}{it1,vt1,ot1}
\fmf{scalar, label=$\tilde{\ell}_R$}{vt1,vt2}
\fmfdot{vt1}
\fmflabel{$\ell$}{it1}
\fmflabel{$\chi^0_1$}{ot1}
\end{fmfgraph*} &
\begin{large}
$\text{i}  \sqrt{2} [e e_{\ell} N_{11}^* -e_{\ell} \frac{g \sw[2]}{\cw[2]} N_{12}^*]P_R$
\end{large}
 \\ 
& \\[5mm]
\begin{fmfgraph*}(21,28)
\fmfpen{thick}
\fmfbottom{ib1,ib2}
\fmftop{vb2}
\fmf{fermion}{ib2,vb1,ib1}
\fmf{dashes, label=$Z^0$}{vb1,vb2}
\fmfdot{vb1}
\fmflabel{$e^-$}{ib2}
\fmflabel{$e^-$}{ib1}
\end{fmfgraph*} &
\begin{large}
$- \frac{\text{i}g}{\cos\theta_W} \gamma^{\mu}(c_L P_L + c_R P_R)$
\end{large}
 \\ 
& \\[5mm]
\begin{fmfgraph*}(21,28)
\fmfpen{thick}
\fmftop{if1,if2}
\fmfbottom{vf2}
\fmf{fermion}{if2,vf1,if1}
\fmf{dashes, label=$Z^0$}{vf1,vf2}
\fmfdot{vf1}
\fmflabel{$\chi^0_1$}{if2}
\fmflabel{$\chi^0_1$}{if1}
\end{fmfgraph*} &

\begin{large}
$\frac{\text{i}g}{\cw} \gamma^{\mu}(L_1 P_L + R_1 P_R)$
\end{large}
 \\ 
& \\[2mm]
\bottomrule
\end{tabular}
\caption{Vertexfaktoren} \label{tab:vertizes}
\end{center}
\end{table}
\end{fmffile}

Die Propagatoren sehen wie folgt aus ($\ell$: Fermion mit schwachem Isospin $-1/2$):
\begin{eqnarray}
\tilde{\ell}\text{-Propagator}& = &  \frac{\text{i}}{p^2 - M^2_{\tilde{\ell}}} 
\approx - \frac{\text{i}}{M^2_{\tilde{\ell}}} \\
Z^0\text{-Propagator}& = & \frac{-\text{i}(g_{\mu\nu} - p_\mu p_\nu / M^2_{Z^0})}{p^2 - M^2_{Z^0}}
  \approx \frac{{\text{i}}g_{\mu\nu}}{M^2_{Z^0}} \enspace .
\end{eqnarray}
Für den $Z^0$-Propagator wurde die Landau-Eichung benutzt.
Die Vertizes stehen in Tabelle \ref{tab:vertizes}. Die dabei auftretenden Konstanten bedeuten:
\begin{eqnarray}
c_L & = &\frac{1}{2}(c_V + c_A); \enspace c_R = \frac{1}{2}(c_V - c_A); \enspace c_V = T_3 - 2 e_{\ell} \sw[2];
\enspace c_A = T_3 \\
L_1 & = & \frac{1}{2}\left( N_{14}^{}N_{14}^* -  N_{13}^{}N_{13}^* \right) \cos(2\beta) -
      \frac{1}{2}\left( N_{13}^{}N_{14}^* -  N_{13}^{}N_{14}^* \right) \sin(2\beta)  = -R_1 \enspace .
\end{eqnarray}
Zur Abkürzung vereinbaren wir:
\begin{eqnarray}
l_1 & = & \sqrt{2} \left[e_{\ell} \sw N_{11} + \frac{1}{\cw}
                \left(\frac{1}{2} - e_e \sw[2] \right)N_{12}\right]\,,\\
r_1 & = & \sqrt{2} \left[e_{\ell} \sw N_{11}^* - e_{\ell} \frac{\sw[2]}{\cw} N_{12}^* \right] \quad .
\end{eqnarray}

\section{Die Übergangsmatrixelemente}
Die Anwendung der Feynmanregeln liefert fünf Übergangsmatrixelemente:
\begin{eqnarray}
- \ie \M_1 & = &  \frac{g^2}{\cos^2\theta_W} D_{Z^0}(s) \vv(p_1) \gamma^\mu (c_L P_L + c_R P_R) u(p_2)
                                                       \uu(p_3) \gamma_\mu (L_1 P_L + R_1 P_R) v(p_4), \notag\\
\label{eq:mma} \\
- \ie \M_2 & = & -g^2|l_1|^2 D_{\tilde{e}}(t) \vv(p_1) P_R v(p_3) \uu(p_4) P_L u(p_2), \label{eq:mmb}\\
- \ie \M_3 & = & -g^2|l_1|^2 D_{\tilde{e}}(u) \vv(p_1) P_R v(p_4) \uu(p_3) P_L u(p_2)\cdot (-1),\label{eq:mmc} \\
- \ie \M_4 & = & -g^2|r_1|^2 D_{\tilde{e}}(t) \vv(p_1) P_L v(p_3) \uu(p_4) P_R u(p_2),\label{eq:mmd} \\
- \ie \M_5 & = & -g^2|r_1|^2 D_{\tilde{e}}(u) \vv(p_1) P_L v(p_4) \uu(p_3) P_R u(p_2)\cdot(-1),\label{eq:mme} \enspace 
 \\ \text{mit} & &
  D_{Z^0}(s) = \frac{\ie}{M_{Z^0}^2},\quad  D_{\tilde{e}}(x) = - \frac{\ie}{M^2_{\tilde{e}}} \quad .
\end{eqnarray}
Da die Ausgangszustände identische Fermionen sind (Neutralinos sind Majorana-Fermionen), erhalten die Amplituden für
$M_3$ und $M_5$ das zusätzliche, angeschriebene Minuszeichen.

\section{Der differentielle Wirkungsquerschnitt}
\subsection{Berechnung der Betragsquadrate}
Der nächste Schritt ist die Berechnung von
\begin{eqnarray} 
\sum_{\sigma(\text{out})}|\M|^2 &  = & \sum_{\sigma(\text{out})}|\M_1 + \M_2 + \M_3 + \M_4 + \M_5|^2  \\
                         & = &\sum_{\sigma(\text{out})}
	\left( \sum_{i=1}^5 |\M_i|^2 + 2 \,\real \sum_{i<j} \M_i \M_j^+\right) \quad .
\end{eqnarray}
Hierbei bedeutet $\mink{a}{b} = a_\mu b^\mu$. Die Summation über die auslaufenden Spins wird ab sofort nicht mehr
mitgeführt und implizit vorausgesetzt.
\begin{eqnarray}
\M_1\M_1^+ & = &  \phantom{-}\frac{16 g^4}{M_{Z^0}^4 \cos^4 \theta_W} \Big(
  c_L^2 L_1^2 \mink{p_1}{p_4}\mink{p_2}{p_3} + c_R^2 L_1^2 \mink{p_1}{p_3} \mink{p_2}{p_4}, \\ 
& & \phantom{ \frac{16 g^4}{M_{Z^0}^4 \cos^4 \theta_W}}
 + c_L^2 L_1 R_1 M^2 \mink{p_1}{p_2} + c_R^2 L_1 R_1 M^2 \mink{p_1}{p_2}  \nonumber \\
& & \phantom{ \frac{16 g^4}{M_{Z^0}^4 \cos^4 \theta_W}}
   + c_R^2 R_1^2 \mink{p_1}{p_4}\mink{p_2}{p_3} + c_L^2 R_1^2 \mink{p_1}{p_3} \mink{p_2}{p_4} \Big)\label{eq:aa} \nonumber \\
\M_1\M_2^+  & = & - \frac{g^4|l_1|^2}{M_{Z^0}^2 M_{\tilde{e}^2} \cos^2\theta_W}
                  \Big(4c_L L_1 M^2 \mink{p_1}{p_2} + 8 c_L R_1\mink{p_1}{p_3} \mink{p_2}{p_4}  \Big),\label{eq:ab} \\[2mm]
\M_1\M_3^+  & = & - \frac{g^4|l_1|^2}{M_{Z^0}^2 M_{\tilde{e}^2} \cos^2\theta_W}
                  \Big(4c_L L_1 M^2 \mink{p_1}{p_2} + 8 c_L R_1\mink{p_1}{p_4} \mink{p_2}{p_3}  \Big),\label{eq:ac} \\[2mm]
\M_1\M_4^+  & = & {-}\frac{g^4|r_1|^2}{M_{Z^0}^2 M_{\tilde{e}^2} \cos^2\theta_W}
               \Big(4c_R R_1 M^2 \mink{p_1}{p_2} + 8 c_R L_1\mink{p_1}{p_3} \mink{p_2}{p_4} \Big),\label{eq:ad}\\[2mm]
\M_1\M_5^+  & = &  {-}\frac{g^4|r_1|^2}{M_{Z^0}^2 M_{\tilde{e}^2} \cos^2\theta_W}
               \Big(4c_R R_1 M^2 \mink{p_1}{p_2} + 8 c_R L_1\mink{p_1}{p_4} \mink{p_2}{p_3} \Big),\label{eq:ae}\\[2mm]
\M_2\M_2^+  & = &  \phantom{-}\frac{g^4|l_1|^4}{M^4_{\tilde{e}}} 4 \mink{p_1}{p_3} \mink{p_2}{p_4},\label{eq:bb} \\[1mm]
\M_2\M_3^+  & = &          {-}\frac{g^4|l_1|^4}{M^4_{\tilde{e}}} 2 M^2 \mink{p_1}{p_2}, \label{eq:bc}\\[1mm]
\M_2\M_4^+  & = &  \phantom{-} 0, \\[1mm]
\M_2\M_5^+  & = &  \phantom{-}0, \\[1mm]
\M_3\M_3^+  & = &  \phantom{-}\frac{g^4|l_1|^4}{M^4_{\tilde{e}}} 4 \mink{p_1}{p_4} \mink{p_2}{p_3},\label{eq:cc} \\[1mm]
\M_3\M_4^+  & = &  \phantom{-} 0, \\[1mm]
\M_3\M_5^+  & = &  \phantom{-}0, \\[1mm]
\M_4\M_4^+  & = &  \phantom{-}\frac{g^4|r_1|^4}{M^4_{\tilde{e}}} 4 \mink{p_1}{p_3} \mink{p_2}{p_4},\label{eq:dd} \\[1mm]
\M_4\M_5^+  & = & - \frac{g^4|r_1|^4}{M^4_{\tilde{e}}} 2 M^2 \mink{p_1}{p_2},\label{eq:de} \\[1mm]
\M_5\M_5^+  & = & \phantom{-} \frac{g^4|r_1|^4}{M^4_{\tilde{e}}} 4 \mink{p_1}{p_4} \mink{p_2}{p_3}\quad.\label{eq:ee}
\end{eqnarray} 
\vfill
\subsection{Weitere Vollständigkeitsrelationen}
Für die Berechnung des Betragsquadrates des Gesamtmatrixelements sind die folgenden Voll\-ständig\-keits\-relationen nützlich.
\begin{alignat}{2}
&\sum_s u(\mathbf{p},s) \uu(\mathbf{p},s) = \slashed{p} + m ,&\quad 
&\sum_s v(\mathbf{p},s) \vv(\mathbf{p},s) = \slashed{p} - m  , \\
&\sum_s u(\mathbf{p},s) v^T(\mathbf{p},s) = (\slashed{p} + m)C^T ,&\quad 
&\sum_s \uu^T(\mathbf{p},s) \vv(\mathbf{p},s) = C^{-1}(\slashed{p} - m), \\
&\sum_s \vv^T(\mathbf{p},s) \uu(\mathbf{p},s) = C^{-1} (\slashed{p} + m) ,&\quad 
&\sum_s v(\mathbf{p},s) u^T(\mathbf{p},s) = (\slashed{p} - m)C^T \quad.
\end{alignat}
Diese Regeln werden sofort einsichtig, wenn man
\begin{eqnarray}
u(\mathbf{p},s) = C\vv^T(\mathbf{p},s); \quad v(\mathbf{p},s) = C\uu^T(\mathbf{p},s) 
\end{eqnarray}
in die bekannten Vollständigkeitsrelationen einsetzt.
Eine weitere hilfreiche Identität ist
\begin{eqnarray}
{\gamma^\mu}^T = - C^{-1}\gamma^\mu C \quad.
\end{eqnarray}
\subsection{Die Kinematik}
Der totale Wirkungsquerschnitt ist ein Lorentzskalar. Also kann man ein beliebiges Bezugssystem wählen,
um den totalen Wirkungsquerschnitt zu berechnen. Ich wähle als Bezugssystem das Schwerpunktsystem,
weil dann die Rechnung einfacher ist. 
\begin{figure}
\centering 
\scalebox{0.7}{\includegraphics{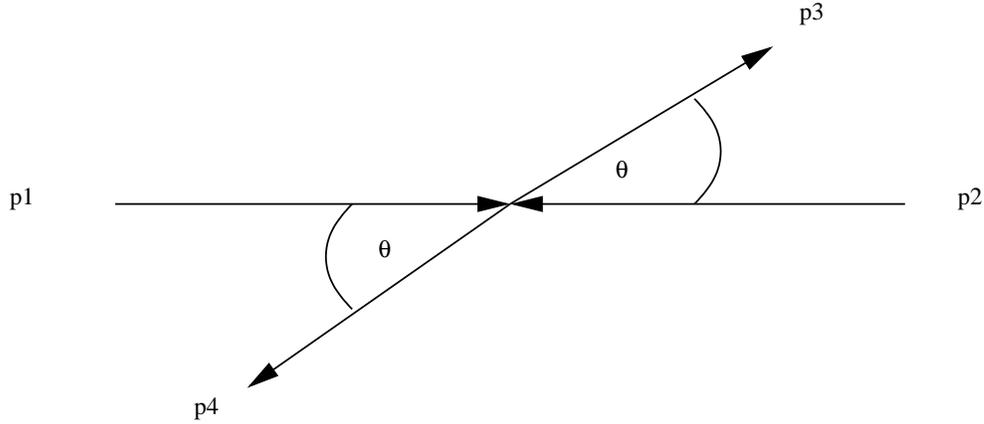}} 
\caption{Impulse im Schwerpunktsystem}
\label{fig:kin}
\end{figure}

Als Parametrisierung der Impulse (genauer: $4$er-Impulse) wähle ich (siehe Abbildung \ref{fig:kin})
\begin{alignat}{2}
p_1 & =  \begin{pmatrix} E, & 0, & 0, & \phantom{-}E,\end{pmatrix} & \qquad,
p_3 & =  \begin{pmatrix} \omega, & 0, &\phantom{-} k\sin\theta, &\phantom{-} k\cos\theta, \end{pmatrix}\enspace, \\
p_2 & =  \begin{pmatrix} E, & 0, & 0, & -E,\end{pmatrix} & \qquad,
p_4 & =  \begin{pmatrix} \omega, & 0, & -k\sin\theta, & -k\cos\theta, \end{pmatrix} \quad .
\end{alignat}
Hierbei sind $k = |\boldsymbol{p}_3| = |\boldsymbol{p}_4|$ der Betrag des $3$er-Impulses, $\omega$ die Energie
der auslaufenden Teilchen, $E$ die Energie und der Impuls des einlaufenden Elektrons und Positrons (Sie werden
als masselos betrachtet.).
Damit folgt für die Mandelstam-Variablen:
\begin{eqnarray}
s & = & (p_1 + p_2)^2 = 4E^2 \label{eq:sman}\enspace, \\
t & = & (p_1 - p_3)^2 = -\frac{1}{2}\left(s - \sqrt{s(s - 4M^2)}\cos\theta \right) + M^2 \label{eq:tman}\enspace, \\
u & = & (p_1 - p_4)^2 = -\frac{1}{2}\left(s + \sqrt{s(s - 4M^2)}\cos\theta \right) + M^2 \label{eq:uman} \quad .
\end{eqnarray}
Die Minkowski-Produkte in den Gleichungen (\ref{eq:aa})-(\ref{eq:ee}) werden mit diesen Beziehungen durch die Mandelstam-Variablen
ausgedrückt:
\begin{eqnarray}
\mink{p_1}{p_2} & = & \phantom{-}\frac{1}{2}s \phantom{- M^2}\enspace, \\
\mink{p_1}{p_3} & = & \mink{p_2}{p_4} = -\frac{1}{2}\big(t - M^2 \big) \enspace,\\
\mink{p_1}{p_4} & = & \mink{p_2}{p_3 }= -\frac{1}{2}\big(u - M^2 \big) \enspace,\\
\mink{p_3}{p_4} & = & \phantom{-}\frac{1}{2}s - M^2 \quad.
\end{eqnarray} 

\subsection{Der differentielle Wirkungsquerschnitt}
Der differentielle Wirkungsquerschnitt im Schwerpunktsystem für eine $2\rightarrow2$-Streuung ist allgemein gegeben durch 
(siehe z. B. \cite{element:84}):
\begin{equation}
\diff{\sigma}{\Omega} = \frac{1}{64 \pi^2 s} \frac{p_f}{p_i} |\M|^2 \quad .
\end{equation}
Außerdem muß noch über die Eingangsspins gemittelt werden. Da die auslaufenden Neutralinos identisch sind, ist der 
Phasenraum nur halb so groß. Einsetzen liefert:
\begin{eqnarray}
\diff{\sigma}{\Omega} =  \frac{g^4}{128 \pi^2 s} \sqrt{1 - \frac{4 M^2}{s}} \Biggl(
                          (t - M^2)^2 + (u - M^2)^2 - 2M^2s \Biggr) \times \\
                          \Biggl[ \frac{c_L^2 L_1^2 + c_R^2L_1^2}{ M^4_{Z^0}\cos^4\theta_W } +
                          \frac{L_1 \bigl( c_L|l_1|^2 - c_R |r_1|^2 \bigr)}{ M^2_{Z^0} M_{\tilde{e}}^2\cos^2\theta_W} +
                             \frac{|l_1|^4+|r_1|^4}{4 M_{\tilde{e}}^2} \Biggr]\quad. 
\end{eqnarray}

\subsection{Der totale Wirkungsquerschnitt}

Der totale Wirkungsquerschnitt ist gegeben durch
$\sigma = \int_O \diff{\sigma}{\Omega} \, \dif{\Omega}$.
Weil wir es in unserem Falle mit $2\rightarrow2$-Streuung zu tun haben, hängt der 
differentielle Wirkungsquerschnitt nur vom Winkel  zwischen einlaufendem und auslaufendem
Teilchen ab. Für die Integration ist nur der Term $(t-M^2)^2 + (u-M^2)^2-2M^2 s$ relevant, die restlichen Terme
sind für die Integration unwichtige Vorfaktoren. Setzt man die Gleichungen (\ref{eq:tman}) und (\ref{eq:uman}) für die
Mandelstam-Variablen ein, so vereinfacht sich der differentielle Wirkungsquerschnitt  zu  
\begin{eqnarray}
\diff{\sigma}{\Omega}  =  \frac{g^4 s(1+\cos^2\theta)}{128 \pi^2} 
                             \Biggl[ \frac{c_L^2 L_1^2 + c_R^2L_1^2}{ M^4_{Z^0}\cw[4] } +
                            \frac{L_1 \bigl( c_L|l_1|^2 - c_R |r_1|^2 \bigr)}{ M^2_{Z^0} M_{\tilde{e}}^2 \cw[2]} +
                             \frac{|l_1|^4+|r_1|^4}{4 M_{\tilde{e}}^4} \Biggr]\times \notag \\ 
                          \left(1 - \frac{4 m^2}{s} \right)^{3/2} 
\end{eqnarray}
Der Wirkungsquerschnitt $\sigma$ hat dann folgende Gestalt, wenn man $\int_O (1+\cos^2\theta)\dif{\phi}\,
\dif{\cos \theta} = \frac{16}{3} \pi$ benutzt:
\begin{eqnarray}
\sigma =  \frac{\pi \alpha^2}{3 \sw[4]} 
                       \Biggl[ \frac{c_L^2 L_1^2 + c_R^2L_1^2}{ M^4_{Z^0}\cos^4\theta_W } +
                       \frac{L_1 \bigl( c_L|l_1|^2 - c_R |r_1|^2 \bigr)}{ M^2_{Z^0} M_{\tilde{e}}^2\cos^2\theta_W} +
                       \frac{|l_1|^4+|r_1|^4}{4 M_{\tilde{e}}^4} \Biggr] \times \notag \\
                       \left(1 - \frac{4 m^2}{s} \right)^{3/2} s .
\end{eqnarray}

\chapter{Fierz-Transformation}\label{app:fierz}
\section{Fierz-Transformation}
\subsection{Allgemeine Transformationsformel}
Ziel dieser Transformation ist es, Strom-Strom-Wechselwirkungen der Form 
\begin{equation}
\M_{AB} \propto (\vv_1\, \Gamma_A\, v_2)\,(\uu_3\, \Gamma_B \, u_4), \quad
\Gamma_k \in \{1,\gamma_5, \gamma_\mu, \gamma_5\gamma_\mu, 
\sigma_{\mu\nu} = \ie/8(\gamma_\mu \gamma_\nu - \gamma_\nu \gamma_\mu ) \}
\end{equation}
wie sie zum Beispiel in der Reaktion $e^+e^- \rightarrow \overline{\nu} \nu$ auftreten, in der Form
\begin{equation}
\M_{AB} \propto \sum_{C,D} c_{ABCD}(\vv_1\, \Gamma_C\,u_4)\,(\uu_3\, \Gamma_D\, v_2), \quad \label{eq:fierzf}
\end{equation}
zu schreiben (Die Summe läuft über alle Lorentz-Kovarianten). Dann zerfallen auch die Interferenzterme mehrerer
Matrixelemente in Produkte zweier Tensoren mit gleicher Stromstruktur,
was sonst nicht immer der Fall ist. Dadurch können Rechnungen einfacher werden.
Man entwickelt $(\vv_1\, \Gamma_A\, v_2)\,(\uu_3\, \Gamma_B \, u_4)$ nach einem vollständigen Satz von Matrizen, nämlich
den $16\,\, \Gamma$-Matrizen. Außerdem benutzt man, daß diese orthogonal zueinander bzgl. der Spur als
Skalarprodukt und (anti-)hermitesch sind. 

\begin{fmffile}{diagramm8}
\begin{table}
\begin{center}
\begin{tabular}{c c c}
\vspace{8mm}
\hspace{15mm}
\begin{fmfgraph*}(36,48)
\fmfpen{thick}
\fmfbottom{fbb1,fbb2}
\fmftop{ftb1,ftb2}
\fmf{quark}{fbb1,fvb1,ftb1}
\fmf{quark}{ftb2,fvb2,fbb2}
\fmf{phantom}{fvb1,fvb2}
\fmfdot{fvb1,fvb2}
\fmflabel{$q$}{fbb1} \fmflabel{$\overline{q}$}{fbb2}
\fmflabel{$\chi$}{ftb1} \fmflabel{$\chi$}{ftb2}
\end{fmfgraph*}
\hspace{25mm}
&
\begin{fmfgraph*}(12,48)
\fmfpen{thick}
\fmfleft{flb}
\fmflabel{$\boldsymbol{\xrightarrow{\text{Fierz-Transformation}}}$}{flb}
\end{fmfgraph*}
&
\hspace{-15mm}
\begin{fmfgraph*}(36,48)
\fmfpen{thick}
\fmfbottom{afbb1,afbb2}
\fmftop{aftb1,aftb2}
\fmf{quark}{afbb1,afvb1,afbb2}
\fmf{quark}{aftb2,afvb2,aftb1}
\fmf{phantom}{afvb1,afvb2}
\fmfdot{afvb1,afvb2}
\fmflabel{$q$}{afbb1} \fmflabel{$\overline{q}$}{afbb2}
\fmflabel{$\chi$}{aftb1} \fmflabel{$\chi$}{aftb2}
\end{fmfgraph*}

\end{tabular}
\caption{Wirkung der Fierz-Transformation}
\label{tab:fierz}
\end{center}
\end{table}
\end{fmffile}

Die Entwicklungskoeffizienten ergeben sich wie folgt:
\begin{eqnarray}
& &             (\Gamma_A)_{\alpha\beta}(\Gamma_B)_{\gamma\delta} =
                 \sum c_{ABCD}(\Gamma_C)_{\alpha\delta}(\Gamma_D)_{\gamma\beta} \notag \\
&\Rightarrow &(\Gamma_A)_{\alpha\beta}(\Gamma_D)_{\beta\gamma}(\Gamma_B)_{\gamma\delta}(\Gamma_C)_{\delta\alpha}
   =c_{ABCD}(\Gamma_C)_{\alpha\delta}(\Gamma_C)_{\delta\alpha}(\Gamma_D)_{\beta\gamma}(\Gamma_D)_{\gamma\beta}
\notag\\
&\Leftrightarrow& \Tr(\Gamma_A\Gamma_D\Gamma_B\Gamma_C) = c_{ABCD}\Tr(\Gamma_C^2)\Tr(\Gamma_D^2)\notag\\
&\Rightarrow & c_{ABCD} = \frac{\Tr(\Gamma_A\Gamma_D\Gamma_B\Gamma_C)}{\Tr(\Gamma_C^2)\Tr(\Gamma_D^2)} 
\label{eq:spur} \quad.
\end{eqnarray}

Die Gramsche Matrix $G_{AB} = (\Tr(\Gamma_A\Gamma_B))$ lautet (Reihenfolge: Skalar, Pseudoskalar, Vektor, 
Axialvektor, Tensor):
\begin{eqnarray}
G = 4 \cdot
\begin{pmatrix}
 1 & 0 & 0          & 0            & 0 \\
 0 & 1 & 0          & 0            & 0 \\
 0 & 0 & g_{\mu\nu} & 0            & 0 \\
 0 & 0 & 0          & -g_{\mu\nu}  & 0 \\
 0 & 0 & 0          & 0            & (g_{\mu\alpha}g_{\nu\beta} - g_{\nu\alpha}g_{\mu\beta})\\ 
\end{pmatrix} \quad.
\end{eqnarray}

\subsection{Anwendung auf $q\overline{q}\rightarrow \chi^0\chi^0$}

Die Fierz-Transformation wird auf das Matrixelement $ \M = \uu(p_1)P_R v(p_4) \uu(p_3)P_L u(p_2)$ angewandt,
was der Struktur $\frac{1}{2}(1 + \gamma_5) \otimes \frac{1}{2}(1 - \gamma_5)$ entspricht. Es ist
\begin{eqnarray}
\frac{1}{2}(1 + \gamma_5) \otimes \frac{1}{2}(1 - \gamma_5) & = & 
\frac{1}{4}(1 \otimes 1 - 1 \otimes \gamma_5 + \gamma_5 \otimes 1 - \gamma_5 \otimes \gamma_5)\label{eq:melement} \quad.
\end{eqnarray} 
Anwendung von Formel \ref{eq:spur} auf die einzelnen Terme ergibt:
\begin{eqnarray}
1 \otimes 1         & = & \frac{1}{4}\left(1 \otimes 1 + \gamma_5 \otimes \gamma_5 + \gamma_\mu \otimes \gamma^\mu - 
   \gamma_5 \gamma_\mu \otimes \gamma_5 \gamma^\mu +\frac{1}{2} \sigma_{\mu\nu} \otimes \sigma^{\mu\nu}\right), \\
1 \otimes \gamma_5  & = & \frac{1}{4}\left(1 \otimes \gamma_5 +  \gamma_\mu \otimes \gamma_5 \gamma^\mu + 
\gamma_5 \otimes 1 - \gamma_5 \gamma_\mu \otimes  \gamma^\mu \phantom{\frac{1}{2}}\right), \\
\gamma_5 \otimes 1  & = &  \frac{1}{4}\left(1 \otimes \gamma_5 -  \gamma_\mu \otimes \gamma_5 \gamma^\mu + 
\gamma_5 \otimes 1 + \gamma_5 \gamma_\mu \otimes  \gamma^\mu \phantom{\frac{1}{2}} \right), \\
\gamma_5 \otimes \gamma_5 & = & \frac{1}{4}\left(1 \otimes 1 + \gamma_5 \otimes \gamma_5 
- \gamma_\mu \otimes \gamma^\mu + \gamma_5 \gamma_\mu \otimes \gamma_5 \gamma^\mu +
\frac{1}{2} \sigma_{\mu\nu} \otimes \sigma^{\mu\nu}\right) \quad .
\end{eqnarray}
Man erhält nach Einsetzen dieses Zwischenresultats in Gl. (\ref{eq:melement}):
\begin{eqnarray}
P_R \otimes P_L = \frac{1}{2}(1 + \gamma_5) \otimes \frac{1}{2}(1 - \gamma_5) 
               = \frac{1}{8}\gamma_\mu(1 + \gamma_5) \otimes \gamma^\mu (1 - \gamma_5)
               = \frac{1}{2}\gamma_\mu P_R \otimes \gamma^\mu P_L
\end{eqnarray}
Dieses Resultat ist aufgrund der Symmetrieeigenschaften der Ausgangsstruktur nicht weiter verwunderlich!

\section{Nichtrelativistischer Grenzfall der Kopplungsstrukturen}

Ein Diracspinor $u^{(s)}(\mathbf{p}), \enspace E> 0 $, oder $v^{(r)}(\mathbf{p}),\enspace E < 0$, hat als Lösung der Dirac-Gleichung
in der Standard-Darstellung die allgemeine Form
\begin{eqnarray}
u^{(s)}(\mathbf{p}) = \sqrt{E + m} 
\begin{pmatrix} \chi^{(s)}\\ \frac{\boldsymbol{\sigma}\cdot\mathbf{p}}{E + m} \chi^{(s)} \end{pmatrix},\quad
v^{(s)}(\mathbf{p}) = \sqrt{|E| + m}
 \begin{pmatrix}  \frac{\boldsymbol{\sigma}\cdot\mathbf{p}}{|E| + m} \chi^{(s)} \\ \chi^{(s)} \end{pmatrix}\\
\chi^{(1)} = \begin{pmatrix}1 \\ 0 \end{pmatrix},\enspace \chi^{(2)} = \begin{pmatrix}0 \\ 1 \end{pmatrix} \quad.
\hspace{7cm}
\end{eqnarray}

Wenn die kinetische Energie gegenüber der Masse vernachlässigbar klein ist, sind die kleinen Komponenten der Diracspinoren
unterdrückt:
\begin{eqnarray}
u^{(s)}(\mathbf{p}) = \sqrt{E + m} 
\begin{pmatrix} \chi^{(s)}\\ \frac{\boldsymbol{\sigma}\cdot\mathbf{p}}{E + m} \chi^{(s)} \end{pmatrix}
\approx \sqrt{2m}\begin{pmatrix} \chi^{(s)}\\ 0 \end{pmatrix} \quad .
\end{eqnarray}
 Dadurch überleben von den fünf Lorentzkovarianten nur solche, die große mit großen Komponenten
verbinden. Am Ende bleiben nur zwei Kopplungsstrukturen übrig.
Dazu verwendet man die $\gamma$-Matrizen in der Dirac-Darstellung:
\begin{eqnarray}
\gamma^0 = \begin{pmatrix}I & \phantom{-}0\\0&-I \end{pmatrix},\enspace 
\boldsymbol{\gamma} = \begin{pmatrix}0 &\boldsymbol{\sigma}\\{\boldsymbol{\sigma}}& 0 \end{pmatrix},\enspace
\gamma^5 = \begin{pmatrix}0 & I\\I& 0 \end{pmatrix} \label{eq:gammamatrix}  
\end{eqnarray}
$\gamma^0$, $\gamma^k \gamma^5$, $\sigma^{ij}$ sind diagonal, alle übrigen Kombinationen mischen kleine mit großen
Komponenten und verschwinden deshalb (vgl. dazu \cite{foundation:95}).
\begin{eqnarray}
\uu u            & \rightarrow & \chi^+ \chi, \\
\uu \gamma^\mu u & \rightarrow & \chi^+ \chi, \enspace \mu = 0; \enspace 0 \enspace \text{sonst},\\
\uu \gamma^5 u   & \rightarrow & 0,\\
\uu \gamma^\mu\gamma^5 u   & \rightarrow & \chi^+ \sigma^i \chi, \enspace \mu = i; \enspace 0 \enspace\text{sonst}, \\
\uu \sigma^{\mu\nu} u   & \rightarrow & \chi^+ \sigma^i \chi, \enspace \mu = j,\enspace \nu = k, ijk, 
\enspace\text{zyklisch}; \enspace 0 \enspace\text{sonst}.
\end{eqnarray}
(Griechische Buchstaben bezeichnen Indizes, die von 0 bis 3 laufen, römische Buchstaben Indizes von 1 bis 3, $u$ ist
ein Diracspinor, $\chi$ ein Weylspinor.) Man sieht, daß nur $I$ und $\boldsymbol{\sigma}$ übrigbleiben.

\chapter{Partialwellenzerlegung}
Hier wird die Partialwellenzerlegung des winkelgemittelten hadronischen Tensors $\overline{\mathcal{H}}^{ii}$ angegeben.
\begin{eqnarray}
\langle S M'_S \mathbf{p}'|T_{NN}|S M_S \mathbf{p}\rangle \!& = & \!
\sum Y_{L'M'_L}(\mathbf{p})Y_{LM_L}(\mathbf{p})^* \mathcal{T}(p,S' M'_S,S M_S,L',\Delta L,M_J)\notag \\
\\
\mathcal{T}(p,S' M'_S,S M_S,L',\Delta L,M_J)\! & = & \! \sum\langle S' M'_S,L'M'_L|J M_J\rangle 
                                                    \langle S' M'_S,L'M'_L|J M_J\rangle T^{JL'LS}(p)\notag \\
\\
T^{JL'LS}(p)\! & = & \!(2\pi)^6\left(\frac{2}{\pi M p}\right)\eu^{\ie \delta_{JL'LS}(p)} \sin \big(\delta_{JL'LS}(p)
\big)\\
&&\text{(falls die Kanäle ungekoppelt sind)}\notag\\
S,\,M_S\!\! &:& \text{Spin und dazugehörende 3-Komponente}\notag\\
L,\,M_L\!\! &:& \text{Drehimpuls und dazugehörende 3-Komponente}\notag\\
J,\,M_J\!\! &:& \text{Gesamtdrehimpuls und dazugehörende 3-Komponente}\notag\\
\delta_{JL'LS}\!\!&:& \text{Phasenverschiebung der Partialwelle}\,JL'LS\notag \quad.
\end{eqnarray}
Eine einfache, aber längliche Rechnung führt zur Beziehung zwischen $\overline{\mathcal{H}}^{ii}$ und $\mathcal{T}$:
\begin{eqnarray}
\overline{\mathcal{H}}^{ii} & \!= \!\!& \sum 2 \Big[(2 - M_S M'_S) \big|\mathcal{T}(p,1 M'_S,1 M_S,L',\Delta L,M_J) \big|^2 \\ 
                        && \quad  - 2 \real \big( \mathcal{T}(p,1 (M'_S-1),1 M_S,L',\Delta L,M_J-1)^*
                                                  \mathcal{T}(p,1 M'_S,1 (M_S+1),L',\Delta L',M_J)\ \big) \Big] \notag
\end{eqnarray} 

\chapter{Liste der verwendeten Symbole}
\begin{table}[htb]
\begin{center}
\begin{tabular}[h]{c l c c}
\toprule
Symbol       &  Bezeichnung         & SI-Einheit                 & natürliche Einheit      \\
\midrule
 $s$         &  Länge               &   m                        &   $ \text{eV}^{-1}$     \\
 $t$         & Zeit                 &   s                        &   $ \text{eV}^{-1}$     \\
 $a$         & Beschleunigung       &$\text{m}\text{s}^{-2}$     &   eV                    \\
 $E$         &    Energie           &   J                        &   eV                    \\
 $F$         &        Kraft         &   N                        &   eV                    \\ 
 $L$         &   Leuchtkraft        &  kg/s                      &   $ \text{eV}^2$        \\
 $M$         &    Masse             &   kg                       &   eV                    \\
 $P$         &     Druck            &$\text{N}\text{m}^{-2}$     &   $ \text{eV}^4$        \\
 $p$, $k$    &    Impuls            &   Ns                       &   eV                    \\
 $R$, $r$    &    Radius            &   m                        &   $ \text{eV}^{-1}$     \\   
 $T$         &  Temperatur          &    K                       &   eV                    \\
 $u$, $e$    &  Energiedichte       & $\text{J}\text{m}^{-3}$    &   $ \text{eV}^4$        \\ 
 $V$         &   Volumen            &  $\text{m}^3$              &   $ \text{eV}^{-3}$     \\
 $v$         & Geschwindigkeit      &$\text{m}\text{s}^{-1}$     &    1                    \\
 $\epsilon$  &  Emissivität         &$\text{J}\text{m}^{-3} \text{s}^{-1}$&$ \text{eV}^5$  \\
 $\rho$      &   Dichte             &$\text{kg}\text{m}^{-3}$    &    $ \text{eV}^4$       \\ 
 $\sigma$    &  Wirkungsquerschnitt &    b                       &  $ \text{eV}^{-2}$      \\
 $P$         &  Leistung            &    W                       &  $\text{eV}^{-2}$       \\
 $\mu$       & chemisches Potential &   J                        & eV                      \\
 $\eta$      & Entartung            &   1                        &  1                      \\
 $w$         & ``chemische Zusammensetzung''& 1                  &  1                       \\
 $N$         &  Teilchenzahl        &         1                  &  1                       \\
 $n$         & Teilchendichte       & $\text{m}^{-3}$            &  $ \text{eV}^3$          \\ 
\bottomrule
\end{tabular}
\caption{Liste physikalischer Größen}
\end{center}
\end{table}

\begin{table}[htb]
\begin{center}
\begin{tabular}[h]{c l  l}
\toprule
Symbol & Konstante            &  Wert \\
\midrule
 $e$         &   Elementarladung    &   $1.60 \cdot 10^{-19}\,$As        \\
 $\alpha$    &  Feinstrukturkonstante& $1/137$ \\
 $G$         & Gravitationskonstante&   $6.67\cdot10^{-11}\,\text{N}\cdot\text{m}^2\text{kg}^{-2}$ \\
 $k$         & Boltzmann-Konstante  &  $1.38\cdot 10^{-23}\,$J/K $=0.862 \cdot10^{-4} \text{eV/K}$ \\   
 $\hbar$     & Planck-Konstante     &  $6.625\cdot10^{-34} \,\text{Js}$ \\
 $c$         & Lichtgeschwindigkeit &  $3\cdot 10^8\,\text{ m/s}$ \\ 
 $M_\odot$   & Sonnenmasse          &  $1.99\cdot 10^{30} \,\text{kg}$ \\
 $m_n$       & Masse Neutron        &  $1.675\cdot 10^{-27} \,\text{kg} = 939.6\,\text{MeV}$\\
 $m_p$       & Masse Proton         &  $1.673\cdot 10^{-27} \,\text{kg} = 938.3\,\text{MeV}$\\
 $\theta_W$  & schwacher Mischungswinkel  &  $\sw[2]= 0.223$ \\
\bottomrule
\end{tabular}
\caption{Liste der Konstanten}
\end{center}
\end{table}

\begin{table}[htb]
\begin{center}
\begin{tabular}[h]{c l  c }
\toprule
Symbol   & Operator            &mögliche Darstellung   \\
\midrule
$C$      & Ladungskonjugation  & $\ie\gamma_2\gamma_0$ \\
$P_{L,R}$& Projektoren         & $\frac{1}{2}(1\mp \gamma_5)$ \\
$a^*$    & komplexe Konjugation von $a$&                 \\
$A^T$    & Transposition von $A$ & \\
$A^+$    & hermitesche Konjugation von $A$ & \\  
\bottomrule
\end{tabular}
\caption{Liste der Operatoren}
\end{center}
\end{table}

\begin{table}[htb]
\begin{center}
\begin{tabular}[h]{c l }
\toprule
Symbol   & Bezeichnung      \\
\midrule
$\mathcal{L}$ & Lagrangedichte \\
$\mathcal{M}$ & Matrixelement \\
$\mathcal{N}_{ij}$ & Neutralinotensor\\
$\mathcal{H}_{ij}$ & Hadrontensor\\
$N$     & unitäre Transformation\\
\bottomrule
\end{tabular}
\caption{sonstige Symbole}
\end{center}
\end{table}

\begin{table}[htb]
\begin{center}
\begin{tabular}[h]{l l}
\toprule
cgs $\rightarrow$ natürliche Einheit & natürliche $\rightarrow$ cgs Einheit \\
\midrule
$1\,\text{cm~} = 5,068\cdot10^4\,1/\text{eV}$&$1\,\text{eV} = 5.068 \cdot10^4 1/\text{cm}$\\
$1\,\text{s~~~} = 1.52\cdot10^{15}\,1/\text{eV}$&$1\,\text{eV} = 1.519\cdot10^{15} 1/\text{s}$\\
$1\,\text{K~~} = 8.62\cdot 10^{-5} \,\text{eV}$ &$1\,\text{eV} =1.160\cdot 10^4\,\text{K}$\\
$1\,\text{erg} = 6.24\cdot 10^{11} \,\text{eV}$ &$1\,\text{eV} =1.602\cdot 10^{-12}\,\text{erg}$\\
$1\,\text{g~~} = 5.61\cdot 10^{32} \,\text{eV}$ &$1\,\text{eV} =1.783\cdot 10^{-33}\,\text{g}$\\
\bottomrule
\end{tabular}
\caption{Umrechnung cgs-Einheiten--natürliche Einheiten}
\end{center}
\end{table}

\backmatter

\newpage
 
\nocite*
\bibliographystyle{alpha}
\bibliography{lit1,lit2,lit3,lit4}  
\end{document}